\documentclass[letterpaper,12pt]{article}
\usepackage{jheppub}

\usepackage{graphics,epsfig}
\usepackage{graphicx}

\usepackage{amsmath}
\usepackage{amssymb}
\usepackage{bbm}
\usepackage{subfigure}
\usepackage{epsfig}
\usepackage{yfonts}
\newcommand{\labell}[1]{\label{#1}}

\newcommand{\be}{\begin{equation}}
\newcommand{\ee}{\end{equation}}
\newcommand{\bea}{\begin{eqnarray}}
\newcommand{\eea}{\end{eqnarray}}
\newcommand{\ba}{\begin{eqnarray}}
\newcommand{\ea}{\end{eqnarray}}

\newcommand{\beq}{\begin{equation}}
\newcommand{\eeq}{\end{equation}}
\newcommand{\beqa}{\begin{eqnarray}}
\newcommand{\eeqa}{\end{eqnarray}}
\newcommand{\beqar}{\begin{eqnarray*}}
\newcommand{\eeqar}{\end{eqnarray*}}

\newcommand{\reef}[1]{(\ref{#1})}

\newcommand{\eg}{{\it e.g.,}\ }
\newcommand{\ie}{{\it i.e.,}\ }

\newcommand{\mt}[1]{\textrm{\tiny #1}}

\newcommand{\veps}{\varepsilon}
\newcommand{\cH}{\mathcal{H}}

\newcommand{\N}{\mathcal{N}}

\newcommand{\la}{\lambda}
\newcommand{\lp}{\ell_{\mt P}}

\newcommand\ra{\rangle}
\renewcommand\la{\langle}

\newcommand{\hr}{\hat\rho}

\newcommand{\modu}{modular\ }
\newcommand{\mv}{v}

\newcommand{\tr}{{\rm tr}}
\newcommand{\Tr}{{\rm Tr}}

\newcommand{\tp}{{\delta\tilde\rho}}
\newcommand{\tH}{{\widetilde{\delta H}}}
\newcommand{\tot}{\textrm{tot}}

\newcommand{\tz}{\tilde{z}}
\newcommand{\zh}{z_\mt{h}}
\newcommand{\vep}{\alpha}
\newcommand{\bV}{\overline{V}}

\def\({\left(} \def\){\right)}
\def\[{\left[} \def\]{\right]}

\arxivnumber{1305.3182}

\title{Relative Entropy and Holography}

\author[a]{David D. Blanco}
\author[a]{Horacio Casini}
\author[b]{Ling-Yan Hung}
\author[c]{Robert C. Myers}

\affiliation[a]{ Centro At\'omico Bariloche,
8400-S.C. de Bariloche, R\'{\i}o Negro, Argentina}
\affiliation[b]{Department of Physics, Harvard University,
Cambridge, MA 02138 USA}
\affiliation[c]{Perimeter Institute for Theoretical Physics,
Waterloo, Ontario N2L 2Y5, Canada}

\emailAdd{blancod@ib.cnea.gov.ar}
\emailAdd{casini@cab.cnea.gov.ar}
\emailAdd{lhung@physics.harvard.edu}
\emailAdd{rmyers@perimeterinstitute.ca}

\abstract{Relative entropy between two states in the same Hilbert space is a fundamental
statistical measure of the distance between these states. Relative entropy is
always positive and increasing with the system size. Interestingly, for two states which are
infinitesimally different to each other, vanishing of relative entropy gives a
powerful equation $\Delta S=\Delta H$ for the first order variation of the
entanglement entropy $\Delta S$ and the expectation value of the \modu
Hamiltonian $\Delta H$. We evaluate relative entropy between the vacuum and
other states for spherical regions in the AdS/CFT framework. We check that the
relevant equations and inequalities hold for a large class of states, giving a
strong support to the holographic entropy formula.  We elaborate on potential uses of
the equation $\Delta S=\Delta H$ for vacuum state tomography and obtain modified versions of the Bekenstein bound.}

\begin{document}

\maketitle

\section{Introduction}

Entanglement entropy has emerged as a remarkable theoretical tool providing new
insights into a variety of topics in physics. For example, in condensed matter
theory, it can be used to distinguish new topological phases or different
critical points \cite{wenx,cardy0}. In the context of quantum field theory
(QFT), entanglement entropy has been proposed as a useful probe of phase
transitions in gauge theories \cite{igor0}. Further, it has provided new
insights on the structure of renormalization group flows \cite{twoD,pcon}. In
particular, it was instrumental in establishing new c-theorems in three and
higher dimensions \cite{threeD,cthem}. Of course, entanglement entropy has long
been proposed as the origin of black hole entropy \cite{sorkin,suss,revueS}.
More recently, considerations of entanglement have led to an exciting new
discussion on the nature of Hawking radiation and black hole evaporation
\cite{amps,samir,sam}. At a more fundamental level, it has been suggested that
entanglement entropy may play an important role in understanding the quantum
structure of spacetime, \eg \cite{mvr,vm,arch}.

Entanglement entropy has also figured in many recent discussions of
gauge/gravity duality. The entanglement entropy in the boundary QFT is
determined with an elegant geometric calculation in the dual gravity theory
\cite{rt1}. In particular, the entanglement entropy between a (spatial) region
$V$ and its complement $\bar V$ in the boundary is computed by
 \be
S(V) = \frac{2\pi}{\lp^{d-1}}\ \mathrel{\mathop {\rm
ext}_{\scriptscriptstyle{\mv\sim V}} {}\!\!} \left[A(\mv)\right]
 \labell{define}
 \ee
where one extremizes over all surfaces $\mv$ in the bulk spacetime which are
homologous to the boundary region $V$. Here, we have adopted the convention
$\lp^{d-1}=8\pi G_\mt{N}$ where $d$ is the spacetime dimension of the boundary.
This prescription \reef{define} was found to pass wide range of consistency
tests, \eg see \cite{rt1,head,EEGB}. However, a derivation was provided for the
special case of a spherical entangling surface in \cite{casini9} and quite
remarkably, \cite{aitor} recently extended this derivation to general (smooth)
entangling surfaces.

Quantum information theory provides a variety of other tools with which we
might refine our understanding of entanglement in holographic theories. For
example, R\'enyi entropies are an infinite family of measures of entanglement
\cite{renyi0,karol}, which in principle provide a full description of the
density matrix spectrum, \eg \cite{cala}. Unfortunately, progress towards
understanding holographic R\'enyi entropies has been more limited
\cite{head,twod,yale}. In particular, a good understanding of R\'enyi entropies
has been developed for a two-dimensional boundary CFT and further, these
quantities are easily computed for a spherical entangling surface in any number
of dimensions \cite{yale}. However, an effective and efficient approach to
calculate holographic R\'enyi entropy for more general situations is still
lacking.

In the present paper, we will consider another quantity known as the {\sl
relative entropy} in the context of holography. The relative entropy between
two states in the same Hilbert space yields a fundamental statistical measure
of the distance between these states. Given two density matrices $\rho_1$ and
$\rho_0$, the relative entropy $S(\rho_1|\rho_0)$ is defined as
 \begin{equation}
S(\rho_1|\rho_0)=\textrm{tr}(\rho_1 \log \rho_1)-\textrm{tr}(\rho_1\log \rho_0)\,.
\labell{RSdef}
 \end{equation}
In general, $S(\rho_1|\rho_0)\ge0$ where it vanishes if and only if the states
are equal. Further, if $\rho_1$ and $\rho_0$ describe reduced states on some
region $V$, the relative entropy always increases with the size of $V$, \ie
$S(\rho_1|\rho_0)$ increases under inclusion (for a review see, \eg
\cite{wehrl,vedral1}). When the set is small, both states should approach the
vacuum state on the operator algebra of the region, and then the relative
entropy tends to zero.

The positivity of $S(\rho_1|\rho_0)$  can be given a physical interpretation in
terms of thermodynamics. If the state $\rho_0$ is thermal with respect to the
Hamiltonian $H$, \ie $\rho_0=\frac{e^{- H/T}}{\textrm{tr}(e^{-H/T})}$, then the
relative entropy with any other state $\rho_1$ can be expressed as
\begin{equation}
S(\rho_1 \vert \rho_0)=\frac1T\, \left(F(\rho_1)-F(\rho_0)\right)\,,\labell{free}
\end{equation}
where $F(\rho)$ is the free energy  given by
\begin{equation}
F(\rho)=\textrm{tr} (\rho\, H )- T\, S(\rho) \,.
 \labell{free2}
\end{equation}
We emphasize that $\rho_1$ can be any other state and need not be thermal.
Hence the temperature used to define $F(\rho_1)$ is that of the initial state
$\rho_0$. Now given the expression in eq~\reef{free}, the positivity of the
relative entropy is equivalent to the fact that the free energy at a fixed
temperature $T$ is minimized by the thermal equilibrium state.

Now consider the reduced density matrices describing states of a QFT on a
region $V$. Since any such density matrix is both Hermitian and positive
semidefinite, it can be expressed as
 \be
\rho=\frac{e^{- H}}{\textrm{tr}(e^{-H})} \labell{important}
 \ee
for some Hermitian operator $H$. The latter is known as the modular Hamiltonian
in the literature on axiomatic quantum field theory, \eg
\cite{haag},\footnote{The precise definition of the modular Hamiltonian for a
region $V$ in algebraic QFT also includes an extension of $H$ in
eq.~(\ref{important}) to the algebra of operators outside the region $V$.
\label{foot}} while it is referred to as the entanglement Hamiltonian in the
condensed matter theory literature, \eg \cite{cmt}. The denominator is included
in the above expression to ensure the normalization $\textrm{tr}(\rho)=1$ and
it could instead be absorbed with an additive constant in $H$. However, it will
be convenient to maintain this form below. While $H$ plays an important role in
addressing certain questions, we emphasize that generically the \modu
Hamiltonian is not a local operator and the evolution generated by $H$ would
not correspond to a local (geometric) flow.

Returning to our considerations of the relative entropy and given
eq.~\reef{important}, formally we can say the state $\rho_0$ is thermal with a
temperature $T=1$. Hence we can apply eq.~\reef{free} to express the relative
entropy as
\begin{equation}
S(\rho_1 \vert \rho_0)= \Delta\langle H\rangle -\Delta S \labell{S1}
\end{equation}
where
\begin{equation}
\Delta\langle H\rangle=\textrm{tr} (\rho_1\, H )- \textrm{tr} (\rho_0\, H )
\qquad{\rm and}\qquad
\Delta S=S(\rho_1)- S(\rho_0)\,.
 \labell{S2}
\end{equation}
Now the positivity of the relative entropy requires\footnote{This inequality
can be regarded as a generalized statement of the Bekenstein bound which holds
for any region in QFT. This is explained in more detail in the appendix
\ref{bound}.}
\begin{equation}
\Delta\langle H\rangle\ge\Delta S\,.\labell{123}
\end{equation}
That is, in comparing two states, the variation of the entanglement entropy is
bounded by the variation of the expectation value of the \modu Hamiltonian.
Much of our analysis in this paper will focus on examining this inequality in a
holographic setting. The holographic prescription \reef{define} allows us to
calculate the necessary entanglement entropies and hence $\Delta S$. Further if
the \modu Hamiltonian is known, we can also evaluate $\Delta \langle H
\rangle$, \eg after evaluating the expectation value of the stress energy $\la
T_{ab}\ra$ using standard methods. Unfortunately there are only few simple
cases where the \modu Hamiltonian is explicitly known, as we describe below.

The cases where the precise form of $H$ is known correspond to special
situations, in which the \modu Hamiltonian (and the corresponding internal time flow generated by $H$) are
local.\footnote{The simplest example is given by considering a global thermal
state, with temperature $T$, and taking $V$ to be the whole space. Then, the
\modu Hamiltonian is simply the ordinary (local) Hamiltonian divided by $T$, as
is evident from eq.~(\ref{important}), and so $H$ simply generates ordinary
time translations.}  Let us enumerate a few of these cases here: One well-known
example is given by the vacuum state in any QFT restricted to the half space
$x>0$. In this case, the \modu Hamiltonian is proportional to $K$, the boost
generator in the $x$ direction \cite{bisognano},
\begin{equation}
H=2 \pi K=2 \pi \int_{x>0} d^{d-1}x\, x \,T_{00}(\vec{x})\,.
\labell{siete}
\end{equation}
In this case, $H$ generates a geometric flow along the boost orbits in the
Rindler wedge. Of course, the density matrix then has a thermal interpretation
with respect to time translations along these orbits \cite{Unruh}. A second
example corresponds to the vacuum of a conformal field theory and a spherical
entangling surface, which yields
\begin{equation}
H=2 \pi \int_{|x|<R} d^{d-1}x\, \frac{R^2-r^2}{2 R}\ T_{00}(\vec{x})\,.
\labell{sphereH}
\end{equation}
This result is easily derived from eq.~\reef{siete} since there is a special
conformal transformation (and translation) which maps the Rindler wedge to the
causal development of the ball $|x|<R$ --- \eg see \cite{haag,Hislop}.
Another
situation where the \modu Hamiltonian is known to be local is the case of a
two-dimensional CFT in a thermal state (with temperature $T$) on the Rindler
wedge \cite{yng}. In this case, the \modu Hamiltonian can be expressed as
\begin{equation}
H=\frac1T \int_{x>0}dx\ \left(1-e^{-2\pi T x}\right) \,T_{00}(\vec{x})\,.
 \labell{2dcftRinT}
\end{equation}

In the following, we focus primarily on the case of a spherical entangling
surface with $H$ given by eq.~\reef{sphereH}. As described above, our strategy
will be to use holographic techniques to calculate both $\Delta \langle
H\rangle$ and $\Delta S$ and to test whether the inequality \reef{123} is
satisfied. We will find that eq.~\reef{123} is always satisfied but further,
that in many of our examples, the inequality is in fact saturated to linear
order in the perturbations of the state. The appearance of an equality in these
cases can be understood because we are examining the relative entropy of two
nearby states. Consider choosing a fixed reference state $\rho_0$ and then
moving through a family of states $\rho_1(\lambda)$ with a parameter $\lambda$
such that $\rho_1(\lambda=0)=\rho_0$. Since the two states coincide for
$\lambda=0$, we have that $S(\rho_1(0)|\rho_0)=0$ but
$S(\rho_1(\lambda)|\rho_0)>0$ for both positive and negative $\lambda$.
Therefore if $S(\rho_1(\lambda)|\rho_0)$ is a smooth function of $\lambda$, its
first derivative must vanish at $\lambda=0$. Of course, this vanishing implies
\begin{equation}
\Delta\langle H\rangle =\Delta S \labell{eeex}
\end{equation}
to first order in $\lambda$ (at $\lambda=0$) --- see further discussion in
 the appendix \ref{review}. In thermodynamical terms of eq. (\ref{free})
 this is the well known equation $dE=T dS$
  holding for nearby equilibrium states.

While the above approach tests the positivity of the relative entropy, we can
also use our holographic results to examine the monotonicity constraint
mentioned below eq.~\reef{RSdef}. That is, the relative entropy should increase
as the radius of the spherical entangling surface increases. Of course, this
property can only be tested in the cases where $\Delta\langle H\rangle\ne\Delta
S$, where we should find
 \beq
\partial_R S(\rho_1|\rho_0)=\partial_R
\left[\Delta\langle H\rangle- \Delta S \right] \ge0\,.
 \labell{include}
 \eeq

The remainder of the paper is organized as follows:  In section \ref{simple},
we test relative entropy bounds and the linear equality (\ref{eeex}) for simple
examples containing black branes in the bulk. In section \ref{general} we
analyze general linear perturbations of the vacuum finding agreement with
eq.~(\ref{eeex}). We also compute quadratic perturbations and find in all our
examples that relative entropy is positive and increasing. In section \ref{two}
we analyze some examples in $d=2$ which allow for exact analytic calculations
of the entropy. We discuss some puzzles about localizations of contributions to
$\Delta \langle H\rangle$ in section \ref{puzzle}. We conclude with a summary
of the results and further comments on section \ref{discuss}. In particular, we
discuss the potential of eq. (\ref{eeex}) to make vacuum state tomography using
entanglement entropy, and argue the results of section \ref{general} are
powerful enough to reconstruct the full density matrix in a sphere from the
minimal area prescription for the entropy, in perfect accord with the CFT
result. Finally, in appendix \ref{review} we review several issues related to
relative entropy, including its relation to the strong subadditivity property
of entanglement entropy, the second law of thermodynamics, and the Bekenstein
bound.

\section{Simple examples testing holographic entanglement entropy}
\labell{simple}

As commented above, our strategy will be to test the inequality \reef{123} in a
holographic setting for the case of a spherical entangling surface, for which
the \modu Hamiltonian \reef{sphereH} is known. The RT prescription \cite{rt1}
allows us to calculate the entanglement entropies and hence $\Delta S$. But in
these cases, we can also evaluate $\Delta \langle H \rangle$ given the
expectation value of the stress energy $\la T_{ab}\ra$. In this section, while
our reference state (defining $\rho_0$) is the vacuum of the CFT, our second
state (defining $\rho_1$) will be the holographic dual of a black hole. This is
a warm-up exercise to give us some insight before proceeding with a more
general analysis in the next section.

The bulk solution dual to the vacuum of the $d$-dimensional boundary CFT is
simply empty AdS$_{d+1}$ space, which we write in the Poincar\'e coordinates:
\begin{equation}
ds^2=\frac{L^2}{z^2}\left(-dt^2
+d\vec{x}^2_{d-1}+dz^2\right)\,.
\labell{adsvac}
\end{equation}
Now we are considering a spherical entangling surface in the boundary theory,
\ie the region $V$ is the ball $\lbrace t=0, r\le R\rbrace$. Now the stress
tensor has vanishing expectation value in the vacuum state and so the
expectation value of the \modu Hamiltonian \reef{sphereH} vanishes for this
state, \ie $\langle H\rangle_0=\tr(\rho_0\,H)=0$. Applying the holographic
prescription \reef{define} to evaluate the entanglement entropy, one finds that
the minimal area surface $v$ is given by \cite{rt1}
\begin{equation}
 z=z_0(r)\equiv\sqrt{R^2-r^2}\,. \labell{sphereb}
\end{equation}
Hence the entanglement entropy takes the form
\begin{equation}
S_0=2\pi\, \frac{A(v)}{\lp^{d-1}}
=2\pi\frac{L^{d-1}}{\lp^{d-1}}\Omega_{d-2}\int_0^R dr\, \frac{r^{d-2}}{z^{d-1}}
\sqrt{1+\partial_r z^{\,2}}\,,
\labell{sphereEE}
\end{equation}
where $\Omega_{d-2}$ denotes the area of a unit $(d-2)$-sphere, \ie
 \be
\Omega_{d-2}=\frac{2\,\pi^{(d-1)/2}}{\Gamma((d-1)/2)}\,.
 \labell{units}
 \ee
We will not need to explicitly evaluate eq.~\reef{sphereEE} for the following,
however, the interested reader may find the result in \cite{rt1,casini9}.

For our second state defining $\rho_1$, we take the holographic dual of a bulk
black brane solution, \ie a planar AdS black hole. In general, the (expectation
value of the) stress tensor dual to a stationary black brane takes the form of
that for an ideal fluid,
 \be
\langle T_{\mu\nu}\rangle= (\veps +P)\, u_\mu u_\nu + P\,\eta_{\mu\nu}\,,
 \labell{idealf}
 \ee
where $\veps$, $P$ and $u_\mu$ correspond to the energy density, pressure and
$d$-velocity of the fluid, respectively. Since the boundary theory is a CFT, we
also have $\langle T^\mu{}_\mu\rangle=0$ which imposes $P=\veps/(d-1)$.

As our first example, we consider a static planar AdS black hole, for which the
metric may be written as
\begin{equation}
ds^2=\frac{L^2}{z^2}\left(-f(z)\,dt^2+d\vec{x}^2_{d-1}
+\frac{dz^2}{f(z)}\right)\quad{\rm with}\ \ f(z)=1-\frac{z^d}{\zh^d}
\,. \labell{static}
\end{equation}
In this case, the dual plasma is at rest, \ie $u^\mu=(1,\vec{0}_{d-1})$, and so
eq.~\reef{idealf} reduces to
 \be
\langle T_{\mu\nu}\rangle= \veps\ {\rm diag} (1,1/(d-1),1/(d-1),\cdots)\,.
 \labell{enerden1a}
 \ee
Now the usual holographic dictionary \cite{construct,nice} gives the energy
density as
 \be
\veps =  \frac{d-1}{2}\,\frac{L^{d-1}}{\lp^{d-1}}\,\frac{1}{\zh^{d}}\,.
 \labell{enerden1}
 \ee
The latter can be interpreted as $\veps= c\, T^d$ using the expression for the
black hole temperature:
\begin{equation}
T=\frac{d}{4 \pi \zh}\,.
 \labell{temper1}
\end{equation}
With these expressions, it is straightforward to evaluate the expectation of
the \modu Hamiltonian \reef{sphereH} for this state,
 \bea
\langle H\rangle_1&=&\pi \Omega_{d-2}\frac{\veps}{R}\,\int_0^R
dr\,r^{d-2}\left(R^2-r^2\right)
 \nonumber\\
 &=& \frac{2 \pi \Omega_{d-2}}{d^2-1}\  R^d\,\veps\,.
 \labell{modH1}
 \eea
Hence we arrive at
 \be
\Delta\langle H\rangle=\langle H\rangle_1-\langle H\rangle_0= \frac{\pi
\Omega_{d-2}}{d+1} \,\frac{L^{d-1}}{\lp^{d-1}}\,\frac{R^d}{\zh^{d}}
 \labell{dmodH1}
 \ee
after replacing $\veps$ using eq.~\reef{enerden1}.

Now to complete our comparison in eq.~\reef{123}, we need to evaluate the
entanglement entropy for a spherical entangling surface in the black brane
background. Applying the holographic prescription \reef{define}, the entropy
functional in this new background becomes
\begin{equation}
S_1=2\pi\frac{L^{d-1}}{\lp^{d-1}}\Omega_{d-2}\int_0^R dr\, \frac{r^{d-2}}{z^{d-1}}
\sqrt{1+\frac{(\partial_r z)^{2}}{f(z)}}\,,
\labell{sphereEEBH}
\end{equation}
where $f(z)$ is the metric function given in eq.~\reef{static}. In principle,
we could extremize the above expression, \ie solve for $z(r)$, and evaluate the
entropy at an arbitrary temperature, but this would require a numerical
evaluation.\footnote{The interested reader is referred to \cite{finiteT} for
various interesting analytic approximations.} To make progress analytically, we
will carry out a perturbative calculation for `small' spheres or low
temperatures, in which we consider the limit $R/\zh\ll1$ (or alternatively,
$RT\ll1$). In this case, the minimal surface is only probing the asymptotic
region of the black brane geometry \reef{static} and so the solution deviates
only slightly from the AdS solution \reef{sphereb}, \ie $z(r)=z_0(r)+\delta
z(r)$. Now since $z_0(r)$ extremizes the entropy functional for the AdS
background in eq.~\reef{sphereEE}, the deviation $\delta z(r)$ will not modify
the result at first order in our perturbative calculation.\footnote{As well as
a bulk term proportional to the equations of motion, the first order variation
by $\delta z(r)$ will also generate a total derivative and so one may worry
that there is a nonvanishing boundary term at the cut-off surface. However, a
careful examination shows that this boundary term actually vanishes. The
simplest approach is to simply define the entangling surface directly at the
cut-off surface and then $\delta z$ vanishes there.} Hence, the leading order
change in the entropy comes from evaluating eq.~\reef{sphereEEBH} with
$z=z_0(r)$ and determining the leading contribution in $R/\zh$. Expanding
eq.~\reef{sphereEEBH} to leading order in $1/\zh^d$ yields
 \beqa
\Delta S&=&\pi\frac{L^{d-1}}{\lp^{d-1}}\Omega_{d-2}\int_0^R dr\,\left.
\frac{r^{d-2}\,z\,(\partial_r z)^{2}}{\zh^d\,\sqrt{1+(\partial_r z)^{2}}}
\right|_{z=z_0(r)}=\pi\frac{L^{d-1}}{\lp^{d-1}}\Omega_{d-2}\int_0^R dr\,
\frac{r^{d}}{\zh^d\,R}
\nonumber\\
&=&\frac{\pi \Omega_{d-2}}{d+1} \, \frac{L^{d-1}}{\lp^{d-1}}
\,\frac{R^d}{\zh^{d}}
 \labell{temp}
 \eeqa
Hence comparing to eq.~\reef{dmodH1}, we see that to leading order
\begin{equation}
\Delta \langle H \rangle=\Delta S
\labell{agree}
\end{equation}
and so we have saturated the inequality in eq.~\reef{123}! Of course, this
equality is perhaps not so surprising given the discussion around
eq.~\reef{eeex}.\footnote{Actually the discussion there does not apply directly
to the present example since one would not consider the energy density of the
fluid dual to the black hole taking negative values. However, one might
consider a state where stress tensor locally takes the form in
eq.~\reef{enerden1} but with $\veps<0$ in a small region around the entangling
sphere.} Here we are looking at a family of density matrices characterized by
the temperature $T$ and our perturbative calculation is evaluating the leading
order change in $\langle H \rangle$ and $S$, which appears linearly at order
$(RT)^d$. Of course, it would be interesting to evaluate both sides of
eq.~\reef{123} at next order in the perturbative expansion, but we leave this
exercise to our general analysis in section \ref{general}. Of course, given the
equality in eq.~\reef{agree}, we can not test the monotonicity inequality
\reef{include} at this order. We should add that calculations similar to those
above has also been done in \cite{tak1}, without any reference to relative
entropy.

\subsection{Boosted black brane} \labell{boost}

We now repeat these calculations for a boosted AdS black brane. That is, the
second state defining $\rho_1$ is a thermal plasma which is uniformly boosted
in a certain direction. Hence this new state $\rho_1$ is characterized by the
temperature $T$ and the velocity $v$. Our calculations will be to leading order
in the temperature and all orders in the velocity.

The stress tensor takes the form given in eq.~\reef{idealf} now with $u^\mu =
(\gamma, \gamma v, \vec{0}_{d-2})$ where $\gamma=1/\sqrt{1-v^2}$, as well as
$P=\veps/(d-1)$. In particular, we have
\begin{equation}
\langle T_{00}\rangle=\veps\left( 1+\frac{d}{d-1}\,\gamma^2\,v^2
\right)\,.
\labell{newstress}
\end{equation}
The corresponding bulk black brane solution is simply derived by applying a
boost along, say, the direction of $x^1\equiv x$ directly to the metric in
eq.~\reef{static}. It is convenient to write the resulting metric as
\begin{equation}
ds^{2}=\frac{L^{2}}{z^{2}}\left[ -dt^2 + dx^2 + \gamma^2\frac{z^{d}}{
\zh^{d}}\left( dt+
v dx\right) ^{2}+d\vec{x}_{d-2}^{2} +\frac{dz^{2}}{1-\frac{z^{d}}{\zh^{d}} }
\right] \,.
 \labell{metrica}
\end{equation}
With the usual holographic approach \cite{construct,nice}, one can verify
eq.~\reef{newstress} with $\veps$ given by eq.~\reef{enerden1}, as before. Now
we wish to evaluate the change in the (expectation value of the) \modu
Hamiltonian \reef{sphereH} for the boosted plasma. Since the energy density is
still uniform the calculation of $\langle H\rangle_1$ is the same as before, up
to the additional overall pre-factor in eq.~\reef{newstress}. Hence, we arrive
at
\begin{equation}
\Delta \langle H^{\prime }\rangle=\Delta \langle H\rangle
\left( 1+\frac{d}{d-1}\,\gamma^2\,v^2
\right) \,,  \labell{energy}
\end{equation}
where $\Delta \langle H\rangle$ is the variation of the \modu Hamiltonian given
in eq.~\reef{dmodH1}.

Now in principle, because the background \reef{metrica} is stationary (but not
static), we must apply the covariant prescription suggested by \cite{station}
to evaluate the holographic entanglement entropy. In fact, the holographic
prescription presented in eq.~\reef{define} already accommodates this
situation. In this new background, we would need to find the extremal surface
with a profile defined by $z=z(x,y)$ and $t=t(x,y)$ where
$y^2\equiv\sum_{i=2}^{d-1} (x^{i})^2$ --- in particular, note that the extremal
surface will not remain on a fixed time slice in the bulk. However, our goal is
to evaluate the change in the entanglement entropy $\Delta S'$ and reasoning as
in the previous section, we deduce that the leading change will be determined
by simply evaluating the area in the new background geometry with the
zero-temperature profile \reef{sphereb}. Hence we can ignore the deviations of
the extremal surface away from the constant time slice in the following.

With a profile $z=z(x,y)$, it is straightforward to show that the entropy in
the boosted background \reef{metrica} takes the form
\begin{equation}
S^\prime_1=2\pi\frac{L^{d-1}}{\lp^{d-1}}\Omega_{d-3}\int_{-R}^R dx
\int_0^{\sqrt{R^2-x^2}}\!\!\!dy\
\frac{y^{d-3}}{z^{d-1}}\,
\left[\left(1+\gamma^2v^2\frac{z^d}{\zh^d}\right)
\left(1+\frac{\partial_y z^{\,2}}{f(z)}\right)+\frac{\partial_x z^{\,2}}{f(z)}
\right]^{1/2}\,,
\labell{sphereEEboost}
\end{equation}
where again $f(z)$ is given in eq.~\reef{static}. While no approximations were
made in evaluating $\Delta \langle H^{\prime }\rangle$ in eq.~\reef{energy}, as
before, in evaluating the change in the entropy, we will work to leading order
in the limit $R/\zh\ll1$. Again, applying the same reasoning as in our previous
calculations, we conclude that the leading order change in the entropy comes
simply from evaluating eq.~\reef{sphereEEboost} with the zero-temperature
profile \reef{sphereb}, \ie $z=z_0(r)=\sqrt{R^2-x^2-y^2}$. We first expand the
above expression to leading order in $1/\zh^d$ and then subtract the zero'th
order contribution \reef{sphereEE}, which yields
\begin{equation}
\Delta S'=\pi\frac{L^{d-1}}{\lp^{d-1}}\Omega_{d-3}\int_{-R}^R dx
\int_0^{\sqrt{R^2-x^2}}\!\!\!dy\
\frac{y^{d-3}\,z}{\zh^d\,\sqrt{1+\partial_r z^{\,2}}}
\left[\partial_r z^{\,2}+\gamma^2v^2\left(1+\partial_y z^{\,2}\right)
\right]\,,
\labell{sphereEEboost1}
\end{equation}
where we have simplified $\partial_x z^{2}+\partial_y z^{2}=\partial_r z^{2}$
in anticipation of substituting $z=z_0(r)$. With this substitution, the first
term in the square brackets will yield precisely the `unboosted' result $\Delta
S$, given in eq.~\reef{temp}. Hence we are left with
 \beqa
\Delta S'&=&\Delta S\,+\,\pi\frac{L^{d-1}}{\lp^{d-1}}\Omega_{d-3}\,\gamma^2v^2
\int_{-R}^R dx \int_0^{\sqrt{R^2-x^2}}\!\!\!dy\ \left. \frac{y^{d-3}\,z
\left(1+\partial_y z^{\,2}\right) }{\zh^d\,\sqrt{1+\partial_r z^{\,2}}}
\right|_{z=z_0(r)}
\nonumber\\
&=&\Delta S\,+\,\pi\frac{L^{d-1}}{\lp^{d-1}}\Omega_{d-3}\,\gamma^2v^2
\int_{-R}^R dx \int_0^{\sqrt{R^2-x^2}}\!\!\!dy\ \frac{y^{d-3}
\left(R^2-x^2\right) }{\zh^d\,R}
 \nonumber\\
&=&\Delta S\,+\,\pi\frac{L^{d-1}}{\lp^{d-1}}\Omega_{d-3}\,\gamma^2v^2
\frac{R^d}{\zh^d}\,\frac{\sqrt{\pi}}{d-2}\,\frac{\Gamma\left(d/2+1\right)}{
\Gamma\left(d/2+3/2\right)}
 \nonumber\\
&=&\Delta S
 \left( 1+\frac{d}{d-1}\,\gamma^2\,v^2
\right)\,,
 \labell{sphereEEboost2}
 \eeqa
where we have used eqs.~\reef{units} and \reef{temp} to produce the simple
expression in the final line.

Recall that we found $\Delta \langle H \rangle=\Delta S$ in the previous
section and hence in comparing to eqs.~\reef{energy} and \reef{sphereEEboost2},
we again find that to leading order
\begin{equation}
\Delta \langle H' \rangle=\Delta S'
\labell{agree1}
\end{equation}
for the boosted plasma. While the expressions appearing in the calculations
above are somewhat more complicated, we may have still anticipated this
equality from the discussion around eq.~\reef{eeex}. In this case, we are
considering  a family of density matrices characterized by the temperature
$T$ and the velocity $v$. While our calculations are valid to all orders in the
velocity, we are only evaluating $\Delta\langle H' \rangle$ and $\Delta S'$ to
leading order in $(RT)^d$.

\subsection{Charged black brane} \labell{charge}

Continuing the analysis of section \ref{simple}, another interesting background
to consider as defining $\rho_1$ is a charged AdS black brane. In this case,
the state in the boundary theory is characterized by the chemical potential
$\mu$, as well as the temperature $T$. Our calculations will be to leading
order in $RT$, however, we allow $\mu/T$ to be order one.

In this case, we consider the bulk gravity action
\begin{equation}
I=\frac{1}{2 \lp^{d-1}}\int d^{d+1}x\, \sqrt{-g}\,
\left(\frac{d(d-1)}{L^2}+R-\frac{L^2}{4}
F_{\mu\nu} F^{\mu\nu}\right)
\end{equation}
with $d\ge3$.\footnote{The normalization of the gauge field term is typically
determined by the microscopic details of the holographic construction --- see
discussion in \cite{chemical}. Here, we simply chose the factor of $L^2$ for
convenience. Further we note that in the case $d=2$, the following solution is
modified by logarithmic terms.} The metric for a planar charged black hole can
be written as
\begin{equation}
ds^{2}=\frac{L^{2}}{z^{2}}\left( -h(z)\, dt^{2}+d\vec{x}
^{2}_{d-1}+\frac{dz^{2}}{h}\right)
\labell{chargemet}
\end{equation}
where
\begin{equation}
h=1-\left(
1+\zh^2\,q^{2}\right) \frac{z^d}{\zh^d}
+q^{2} \frac{z^{2d-2}}{\zh^{2d-4}} \,,
 \labell{mimi}
\end{equation}
and the corresponding gauge potential has only a single nonvanishing component
 \begin{equation}
A_0(z)=\sqrt{\frac{2(d-1)}{d-2}} \,q\,
\left(1-\frac{z^{d-2}}{\zh^{d-2}}\right)\,.
 \labell{pot}
 \end{equation}
Here, $z=\zh$ corresponds to the position of the horizon and $q$ is related to
the charge density carried by the horizon. The temperature of the dual plasma
is given by
 \begin{equation}
T=\frac{d}{4 \pi \zh}\left(1-\frac{d-2}{d}\zh^2\, q^{2}\right)
\labell{temper2}
\end{equation}
and the chemical potential is given by the asymptotic value of the gauge
potential, \ie
 \be
\mu=\lim_{z\to0} A_0 = \sqrt{\frac{2(d-1)}{d-2}} \,q\,.
 \labell{chemu}
 \ee

Since the CFT plasma is at rest, eq.~\reef{idealf} reduces to $\langle
T_{\mu\nu}\rangle= \veps\ {\rm diag} (1,1/(d-1),1/(d-1),\cdots)$ and the usual
holographic prescription yields \cite{construct,nice}
 \be
\veps =  \frac{d-1}{2}\,\frac{L^{d-1}}{\lp^{d-1}}\,\frac{1}{\zh^{d}}\left(
1+\zh^2\,q^{2}\right)\,.
 \labell{enerden2}
 \ee
Now we wish to evaluate the change in the expectation value of the \modu
Hamiltonian produced by going to this new state. Since the energy density is
again uniform, evaluating $\langle H\rangle_1$ is precisely the same
calculation as in eq.~\reef{modH1}, up to the additional overall factor
appearing in eq.~\reef{enerden2}. Hence, we arrive at
\begin{equation}
\Delta \langle H^{\prime\prime }\rangle=\Delta \langle H\rangle\
\left(1+\zh^2\,q^{2}\right) \,,  \labell{energy2}
\end{equation}
where $\Delta \langle H\rangle$ is the result given in eq.~\reef{dmodH1}.

Further since the black brane is static, the extremal surface appearing in the
holographic entanglement entropy \reef{define} again has a spherically
symmetric profile $z=z(r)$ for a spherical entangling surface. Hence with the
metric \reef{chargemet}, the entropy functional becomes
\begin{equation}
S''_1=2\pi\frac{L^{d-1}}{\lp^{d-1}}\Omega_{d-2}\int_0^R dr\, \frac{r^{d-2}}{z^{d-1}}
\sqrt{1+\frac{(\partial_r z)^{2}}{h(z)}}\,,
\labell{sphereEEBH2}
\end{equation}
where $h(z)$ is given in eq.~\reef{mimi}. In proceeding, we again limit our
analysis to a perturbative calculation with $R/\zh\ll1$ but we treat $\zh q =
O(1)$. Further, as before, the leading contribution to the change in the
entropy comes from simply evaluating eq.~\reef{sphereEEBH2} with the vacuum
profile $z=z_0(r)$ and expanding in $R/\zh$. However, we would like to refine
our previous arguments. Here as in the previous examples, the leading changes
to the asymptotic metric are $O(z^d/\zh^d)$ and so we will  find the leading
change in the entropy is $\Delta S''=O(R^d/\zh^d)$. The leading change of the
profile of the extremal surface, $\delta z$, is also controlled by these
leading changes in the metric. However, as we argued before, the entropy is
only changed at quadratic order in $\delta z$ and hence we will find that this
contribution produces a change in the entropy $\Delta S''(\delta
z^2)=O(R^{2d}/\zh^{2d})$ --- see section \ref{quad} for an explicit
calculation. Hence at this point, we note that the next-to-leading order
changes in the above metric \ref{chargemet} are $O(z^{2d-2}/\zh^{2d-2})$ since
we consider $\zh q$ to be order 1. If we calculate with these changes in the
metric and the original profile, there will be an additional contribution to
the change in the entropy at $O(R^{2d-2}/\zh^{2d-2})$ --- this is verified by
our calculation below. This contribution is still lower order in the $R/\zh$
expansion compared to those arising from the change in the profile. Hence it is
legitimate to consider this contribution without concerning ourselves with the
change in the profile of the extremal surface. Therefore we expand
eq.~\reef{sphereEEBH2} as
 \beqa
\Delta S''&=&\pi\frac{L^{d-1}}{\lp^{d-1}}\Omega_{d-2}\int_0^R dr\,\left.
\frac{r^{d-2}\,z\,(\partial_r z)^{2}}{\zh^d\,\sqrt{1+(\partial_r z)^{2}}}
\left[\left(
1+\zh^2\,q^{2}\right)-q^2\frac{z^{d-2}}{\zh^{d-4}}\right]\right|_{z=z_0(r)}
\nonumber\\
&=&\Delta S\,\left( 1+\zh^2\,q^{2}\right)-
\pi\frac{L^{d-1}}{\lp^{d-1}}\Omega_{d-2} \frac{q^2}{\zh^{2d-4}R} \int_0^R
dr\,r^{d}\left(R^2-r^2\right)^{\frac{d-2}{2}}
\nonumber\\
&=&\Delta S\,\left( 1+\zh^2\,q^{2}\right)-
\frac{d-1}{2}\pi^{\frac{d+1}{2}}\frac{\Gamma(d/2)
}{\Gamma\left(d+\frac12\right)} \frac{L^{d-1}}{\lp^{d-1}} (\zh q)^2
\frac{R^{2d-2}}{\zh^{2d-2}}
 \labell{temp2}
 \eeqa
where $\Delta S$ corresponds to the variation given in eq.~\reef{temp}.

Recall that we found $\Delta \langle H \rangle=\Delta S$ in eq.~\reef{agree}.
Hence in comparing to eqs.~\reef{energy2} and \reef{temp2}, we find that the
leading order terms are again equal, however, including the contribution at
$O(R^{2d-2}/z_{0}^{2d-2})$ yields
\begin{equation}
\Delta \langle H'' \rangle>\Delta S''\,.
\labell{agree2}
\end{equation}
Hence we find that adding the chemical potential introduces a next-to-leading
contribution which ensures that the relative entropy is positive. Using the
above expressions, we have
 \be
S(\rho_1|\rho_0)\simeq\frac{\pi}{2}\frac{L^{d-1}}{\lp^{d-1}}\Omega_{d-2}\frac{\Gamma(d/2)
\Gamma((d+1)/2)}{\Gamma\left(d+\frac12\right)} (\zh q)^2
\frac{R^{2d-2}}{\zh^{2d-2}}
 \labell{relate2}
 \ee
Since $S(\rho_1|\rho_0)\propto R^{2d-2}$, we can trivially verify that the
relative entropy also satisfies the monotonicity property \reef{include}, \ie
$\partial_RS(\rho_1|\rho_0)>0$. Using eqs.~\reef{temper2} and \reef{chemu}, it
is straightforward to re-express the right-hand side as a function of the
temperature and chemical potential. While the full expression is not
particularly insightful, note that in the regime $1\gg\mu/T\gg RT$, we have
$S(\rho_1|\rho_0)\sim (RT)^{2d-2}(\mu/T)^2$ and so, in particular, we observe
that this nonvanishing contribution begins at quadratic order in the chemical
potential.

In closing, we note that the result in eq.~\reef{temp2} was generated by a
first-order deformation in the asymptotic metric, however, the latter is
produced by the back-reaction of the gauge field on the geometry and so the
leading change in the relative entropy is quadratic in the corresponding
coefficient $q$.

\section{General analysis} \labell{general}

In this section, we would like to generalize the previous analysis to examine
the inequality \reef{123} for more general holographic states. As long as we
focus our attention on a spherical entangling surface, it is straightforward to
evaluate $\Delta\langle H\rangle$ using eq.~\reef{sphereH} since a standard
holographic prescription allows us to determine $\langle T_{\mu\nu}\rangle$
\cite{construct,nice}. In principle, the calculation of the entanglement
entropy using eq.~\reef{define} is more challenging because we must determine
the extremal surface in the bulk geometry describing the second state $\rho_1$.
However, as we saw above, if this state describes a `small' perturbation of the
initial vacuum state $\rho_0$, our calculations are restricted to considering
asymptotic perturbations of the AdS geometry. Hence, our analysis of the
holographic entanglement entropy was greatly simplified in this perturbative
context. It also suggests that it is natural to formulate these calculations in
the framework of the asymptotic Fefferman-Graham (FG) expansion \cite{feffer}
--- see also \cite{construct}. In particular, such an approach will allow us to consider a much broader
class of perturbed states without concerning ourselves with the details of the
bulk geometry in the far infrared.

Using the FG expansion, we consider three distinct calculations in the
following: We begin by considering states described by purely gravitational
excitations in the AdS bulk. That is, the stress tensor is the only operator
that has a nonvanishing expectation value in these states. Now let us introduce
a small perturbative parameter $\vep$ which controls the magnitude of $\langle
T_{\mu\nu}\rangle$. Our first result is to demonstrate that we always saturate
the inequality \reef{123}, \ie $\Delta\langle H\rangle=\Delta S$, when working
to linear order in $\vep$. We emphasize that this equality holds even when
$\langle T_{\mu\nu}\rangle$ varies on scales comparable to $R$, the size of the
spherical entangling surface. Secondly, we extend these calculations to second
order in $\vep$ in section \ref{quad}. There while $\Delta\langle H\rangle$ is
unchanged, we show that the additional contributions to the entanglement
entropy have a definite sign ensuring that $\Delta\langle H\rangle>\Delta S$.
The third case, which we consider in section \ref{matter}, involves states in
which additional matter fields are excited in the dual AdS spacetime and hence
additional operators acquire expectation values. As we saw in section
\ref{charge}, it is relatively easy to determine quadratic corrections to the
entanglement entropy coming from such perturbations. Below, we extend this
analysis to a much broader class of states and verify that the quadratic
contributions again ensure that $\Delta\langle H\rangle>\Delta S$.

As commented above, our general analysis will be formulated in the context of
the Fefferman-Graham expansion of the asymptotic bulk solutions
\cite{construct,feffer}. Hence we begin by considering a general bulk metric
written in FG coordinates
 \be
ds^2 =\frac{L^2}{z^2} \left( dz^2 + g_{\mu\nu}(z,x^\mu) dx^\mu dx^\nu\right)\,.
 \labell{bulkG}
 \ee
We are considering the asymptotic geometry where $z\simeq0$. We will always
choose the asymptotic metric (on which the boundary CFT is defined) to be flat
and so we may write
 \be
g_{\mu\nu}(z,x^\mu) = \eta_{\mu\nu} + \delta g_{\mu\nu}(z,x^\mu)
 \labell{expand}
 \ee
where $\delta g_{\mu\nu}$ begins with terms of order $z^d$. We are interested
in calculations of holographic entanglement entropy \reef{define} and so we
will want to evaluate the area of various extremal surfaces in the bulk. In
principle, for situations where the background geometry is not static, the
profile of these $(d-1)$-dimensional surfaces would be specified by giving both
the radial position and time in the bulk as functions of the remaining spatial
coordinates, \ie $z=z(x^i)$ and $t=t(x^i)$. However, our goal is to evaluate
the change in the entanglement entropy $\Delta S$ and discussed in section
\ref{boost}, it will suffice to consider bulk surfaces that live in a constant
time slice. Hence with a radial profile $z=z(x^i)$ alone, the induced metric
$h_{ij}$ on this surface is given by
 \be
h_{ij}dx^i dx^j = \frac{L^2}{z^2}\left(g_{ij} + \partial_i z\partial_j z
\right) dx^i dx^j
 \labell{induced}
 \ee
and the corresponding area is then
 \be
A= L^{d-1}\int d^{d-1}x  \sqrt{h} = L^{d-1}\int d^{d-1}x \sqrt{{\rm
det}g_{ij}}\,\sqrt{1+ g^{ij}\,\partial_i z\,\partial_j z}\,.
 \labell{action}
 \ee
In principle, eq.~\reef{action} can now be used as an effective action to
determine the extremal profile $z=z(x^i)$. However, as before, to determine the
leading change $\Delta S$, we will be evaluating the area in the new background
geometry with the original profile \reef{sphereb}.

\subsection{Linear corrections to relative entropy} \labell{line}

We begin by considering states $\rho_1$ whose small deviation of the vacuum
state $\rho_0$ is characterized by an expectation value of stress tensor
$T^0_{\mu\nu}$ in the boundary CFT.\footnote{For simplicity, we drop the angle
brackets in denoting this expectation value throughout our calculations here.}
We suppose the latter is `very small' and that the smallness is characterized
by a (dimensionless) parameter $\vep \ll 1$. As before, we will limit our
attention to a spherical entangling surface for which the (vacuum) \modu
Hamiltonian \reef{sphereH} is linear in the stress tensor and so $\Delta\langle
H\rangle$ is linear in $\vep$. However, in eq.~\reef{123}, the change in the
entanglement entropy will receive contributions at all orders in $\vep$. In the
present section, we will only evaluate $\Delta S$ to linear order in $\vep$.

In general, using the FG expansion, the deviation of the bulk metric from pure
AdS in eq.~\reef{expand} takes the form.
 \be
\delta g_{\mu\nu} = \frac{2}d\frac{\lp^{d-1}}{L^{d-1}} z^d\sum_{n=0} z^{2n}\,
T^{(n)}_{\mu\nu}\,.
 \labell{FGexp}
 \ee
The bulk Einstein equations will determine $T^{(n)}_{\mu\nu}$ for $n>0$ in
terms of expectation value $T^{(0)}_{\mu\nu}$. Following the above discussion,
our strategy will be to only solve for $T^{(n)}_{\mu\nu}$ to leading order in
$\vep$ (or to linear order in $T^{{(0)}}_{\mu\nu}$).

Before we solve the Einstein equations, we let us recall that the goal is to
evaluate the change in the holographic entanglement entropy in the perturbed
metric. Here, we may apply the same reasoning as in section \ref{simple}. In
particular, in the vacuum AdS, there is an analytic solution \reef{sphereb} for
the extremal bulk surface corresponding to a spherical entangling surface of
radius $R$ in the boundary
 \be
z_0^2 + r^2 = R^2, \qquad {\rm where}\ \ r^2 = \sum_{i=1}^{d-1} x_i^2\,.
\labell{wox}
 \ee
Now in the perturbed background, the bulk entangling surface can also be given
as an expansion in $\vep$, \ie $z(x^i)= z_0(x^i) + \vep z_1(x^i) + \cdots$.
However, as described in the previous section, since the profile $z_0$ is
extremal to leading order, the perturbation $z_1$ only contributes at order
$\vep^2$. Hence we can evaluate the linear change in the area by simply
evaluating the area \reef{action} with the original profile $z_0$ in the
perturbed background. Hence given (\ref{FGexp}), one finds to linear order in
$\vep$ that
 \be
\Delta S=2\pi\frac{\Delta A}{\lp^{d-1}} = \frac{2\pi R}{d} \int_{|{x}|\le
R}\!\!\!d^{d-1}x\ \sum_{n=0} z_0^{2n}\left(T^{(n)}{}_{i}{}^{i} -
T^{(n)}{}_{\,ij}\, \frac{x^i\,x^j}{R^2}\right)\,.
  \labell{linearA}
  \ee

Now we return to solving the Einstein equations, which can be written as
 \be
\hat{R}_{AB} - \frac{1}{2}G_{AB}\left(\hat{R} + \frac{d(d-1)}{L^2}\right)=0\,,
 \ee
where $\hat{R}_{AB}$ is the bulk Ricci tensor evaluated on the bulk metric
$G_{AB}$ given as in eq.~(\ref{bulkG}). Using the results from \cite{useful},
we can write to linear order in $\vep$,
\begin{eqnarray}
\hat{R}_{\rho\rho} &=&  -\frac{d}{4\rho^2}-\frac{1}{2} \partial^2_\rho
\delta g^{\mu}{}_{\mu} \,,\nonumber \\
\hat{R}_{\mu\rho}  &=& \frac{1}{2}\left(\partial_\rho
\partial_{\nu}\delta g^{\nu}{}_{\mu}-
\partial_{\mu}\partial_\rho\delta g^{\nu}{}_{\nu}\right) \nonumber \,,\\
\hat{R}_{\mu\nu} &=& R_{\mu\nu} -2\rho\partial^2_{\rho}\delta g_{\mu\nu}
+ (d-2)\partial_{\rho}\delta g_{\mu\nu}
+ \eta_{\mu\nu}\partial_{\rho}\delta g_{\gamma}{}^{\gamma} -
\frac{d}{\rho} (\eta_{\mu\nu} + \delta g_{\mu\nu}) \nonumber \,,\\
\hat{R} &=& -d(d+1) + \rho R + 2(d-1) \rho \partial_\rho \delta g_{\mu}{}^{\mu} -
4\rho^2 \partial^2_\rho\delta g_{\mu}{}^{\mu}\,,
\end{eqnarray}
where we have chosen a (dimensionless) radial coordinate $\rho = z^2/L^2$. Also,
$R_{\mu\nu}$ and $R$ are curvature tensors evaluated on $g_{\mu\nu}$ treating
$z$ (or $\rho$) as an external parameter. Explicitly, then the linear order in
$\vep$, we have
 \be
R_{\mu\nu} = \frac{1}{2}\left(\partial_{\nu}\partial_{\gamma}\delta
g^{\gamma}{}_{\mu} +
\partial_\mu\partial_\gamma\delta g^{\gamma}{}_{\nu} - \Box \delta g_{\mu\nu} -
\partial_\mu\partial_\nu\delta g^{\gamma}{}_{\gamma}\right)\,.
 \ee

Substituting eq.~(\ref{FGexp}) and the above expression into the Einstein
equations, we obtain the following equations for $T^{(n)}$ using the $\rho\rho$
and $\mu\rho$ components, respectively:
\begin{eqnarray}
&& \partial^\mu\partial^\nu T^{(n)}_{\mu\nu} - \Box T^{(n)\,\mu}{}_{\mu} +
(d-1)(d+2n+2)\,T^{(n+1)\,\mu}{}_{\mu} =0\,,
 \quad T^{(0)\,\mu}{}_{\mu}=0\,,
 \\
&& \partial_{\nu}T^{(n)}{}_\mu{}^{\nu} -\partial_\nu
T^{(n)\,\mu}{}_{\mu} =0 \,.
\end{eqnarray}
Together, these two equations imply that
 \be
T^{(n)\,\mu}{}_{\mu} =0\,,\qquad
\partial_{\nu}{T^{(n)}}_{\mu}{}^{\nu} =0\,,
 \labell{boat}
 \ee
for all $n$. Hence we note that Einstein equations automatically ensure that
$T^{(n)}$ is traceless and conserved for \emph{all} $n$. Finally, the $\mu\nu$
components of Einstein equations then reduce to
 \be
T^{(n)}_{\mu\nu} = -\frac{\Box T^{{(n-1)}}_{\mu\nu}}{2n(d+2n)}\,,
 \ee
which implies
 \be
T^{(n)}_{\mu\nu} =    \frac{ (-1)^n \Gamma[d/2+1]}{2^{2n} n! \Gamma[d/2+n+1]}
\,\Box^n T^{(0)}_{\mu\nu}\,.
 \labell{box}
 \ee
Of course, we can substitute these results back into eq.~(\ref{linearA}) to
express $\Delta S$ entirely in terms of $T^{(0)}_{\mu\nu}$.

For the following, it will be more convenient to express the stress tensor in a
Fourier expansion
 \be
T^{(0)}_{\mu\nu}(x) = \int d^{d}p \,\, \exp(-i p\cdot x)\
\widehat{T}_{\mu\nu}(p)\,.
 \labell{fourier}
 \ee
Using the previous results, the change in the entanglement entropy
\reef{linearA} then becomes
\begin{eqnarray}
&&\Delta S = \frac{2\pi R}{d}\int d^{d-1}x \, \int d^dp \, \exp(-i p\cdot x)\,\times
 \labell{main} \\
&&\frac{\Gamma[d/2+1]}{(z_0|p|/2)^{d/2}}
\sum_{n=0} \left[\frac{1}{n!\Gamma[d/2+n+1]}\left(\frac{|p|z_0}{2}\right)^{2n+
d/2}\right]\left(\widehat{T}_{i}{}^{i}(p)- \widehat{T}_{ij}(p)\frac{x^ix^j}{R^2}
\right)\,,
 \nonumber
\end{eqnarray}
where $|p|= |\sqrt{p_\mu p^\mu}|$. Now we may recognize that the sum in the
square brackets yields precisely
 \be
\sum_{n=0} \left[\frac{1}{n!\Gamma[d/2+n+1]}\left(\frac{|p|z_0}{2} \right)^{2n+
d/2}\right] = I_{d/2}(|p|z_0)\,.
 \labell{sumI}
 \ee
For time-like momenta $p$ in Lorentzian signature, it gives instead
$J_{d/2}(|p|z_0)$. That is, we recover an expression that is precisely
proportional to the Green's function of the graviton in AdS$_{d+1}$. However,
note that the asymptotic boundary condition is taken to be one where the
leading constant term is set to zero --- for example, see \cite{vijay0}. The
latter can be contrasted with the usual bulk-to-boundary Green's function which
is proportional to $K_{d/2}(|p|z_0)$, where the boundary condition is chosen
such that the leading term near the AdS boundary is a constant.

\subsection*{Saturating the inequality in eq.~\reef{123}:}

Turning to eq.~\reef{123}, we would like to establish that this inequality is
in fact saturated at linear order in $\vep$ for the general class of states
considered here. Given the \modu Hamiltonian \reef{sphereH} (for a spherical
entangling surface), we may write
\begin{equation}
\Delta\langle H\rangle=\frac{\pi}{R} \int_{|x|\le R} d^{d-1}x\, z_0^2\
T^{(0)}_{00} \,.
\labell{spheredH}
\end{equation}
where $z_0$ is the extremal profile in eq.~\reef{wox}. A priori, this
expression bares no resemblance to the expression for $\Delta S$ in
eq.~\reef{linearA}, even after we substitute in the results in eq.~\reef{box}.

To prove the inequality \reef{123} is saturated, we begin by examining
eq.~(\ref{main}) for a single momentum component with the corresponding
quantity in $\delta H$ above. We can set the spatial direction of momentum in
direction $x^1$, \ie
\begin{equation}
 T^{(0)}_{\mu\nu}(x)=\widehat{T}_{\mu\nu}\, e^{-i p\cdot x}\,.
 \labell{2four}
\end{equation}
We take the momentum to be time-like for definiteness. An analogous calculation
holds for space-like momentum.

Conservation and tracelessness of $T^{(0)}_{\mu\nu}$ imply
 \be
 \widehat{T}_i{}^i=\widehat{T}_{00}\,,\quad
\widehat{T}_{10}=-\frac{p^0}{p^1}\,\widehat{T}_{00}\quad{\rm and}\quad
\widehat{T}_{11}=\frac{(p^0)^2}{(p^1)^2}\, \widehat{T}_{00}\,.
 \labell{polar}
 \ee
Then we note that given the stress tensor chosen in eq.~\reef{2four}, the
integral of eq.~(\ref{main}) is symmetric under rotations leaving $x^1$ fixed.
This implies the integral containing the term $\widehat{T}_{ij}\, x^i x^j$ will
vanish for $i\neq j$. Also for $i=j=2,\cdots,(d-2)$, all the integrals are
equal. Then inside the integral, we can replace
\begin{eqnarray}
&& \widehat{T}_i{}^i-\widehat{T}_{ij} \frac{x^i x^j}{R^2}\rightarrow
\widehat{T}_{00}-\widehat{T}_{11}\, \frac{(x^1)^2}{R^2}-
\sum_{i=2}^{d-2}\widehat{T}_{ii}\, \frac{(x^i)^2}{R^2}\\
 && \rightarrow \widehat{T}_{00}-\widehat{T}_{11} \frac{(x^1)^2}{R^2}- \sum_{i=2}^{d-2}
 \widehat{T}_{ii}\frac{\sum_{j=2}^{d-2} (x^j)^2}{(d-2) R^2}
\rightarrow
 \widehat{T}_{00}-\widehat{T}_{11} \frac{(x^1)^2}{R^2}- \sum_{i=2}^{d-2}
 \widehat{T}_{ii} \frac{r^2-(x^1)^2}{(d-2) R^2}
 \nonumber \\
 &&\rightarrow \widehat{T}_{00}-\widehat{T}_{11} \frac{(x^1)^2}{R^2}-
 (\widehat{T}_i{}^i-\widehat{T}_{11})\,
 \frac{r^2-(x^1)^2}{(d-2) R^2}
\\&&\hspace{2cm}\rightarrow
 \widehat{T}_{00} \left(1-\frac{(p^0)^2}{(p^1)^2}\frac{(x^1)^2}{R^2}-
\frac{\left(1-\frac{(p^0)^2}{(p^1)^2}\right)
(r^2-(x^1)^2)}{(d-2) R^2}\right)\,.\nonumber
\end{eqnarray}
In the last transformation we have used (\ref{polar}). This final expression depends only on $\widehat{T}_{00}$, which is necessary
for the equality with $\Delta\langle H\rangle$.

Then $\Delta S$ reads in polar coordinates
\begin{eqnarray}
\Delta S&=& \frac{2^{(d+2)/2} \pi R }{d |p|^{d/2}}\,\Gamma[d/2+1]
\Omega_{d-3}\,\widehat{T}_{00}e^{ip^0 t}\,\int_0^R dr\, r^{d-2}\int_0^\pi d \theta\,
\sin^{d-3}\!\theta\, e^{-i p^1 r \cos(\theta)} \nonumber\\
 && \times   \frac{J_{d/2}(|p|\sqrt{R^2-r^2})}{(R^2-r^2)^{d/4}}
\left(1-\frac{(p^0)^2}{(p^1)^2}\frac{r^2 \cos^2\theta}{R^2}-
\frac{\left(1-\frac{(p^0)^2}{(p^1)^2}\right)r^2\sin^2\theta}{(d-2) R^2}\right)\,.\labell{dss}
 \end{eqnarray}
The integrals over $\theta$ can then be done explicitly using
\begin{equation}
\int_0^\pi d \theta\, \sin^q(\theta) e^{-i x \cos(\theta)}=2^{q/2}
\sqrt{\pi}\,\Gamma[(q+1)/2]\, \frac{J_{q/2}(| x|)}{|x |^{q/2}}\,.\labell{angle}
\end{equation}

Now for the variation of the \modu Hamiltonian, we substitute eq.~\reef{2four}
into the eq.~\reef{spheredH} which yields
\begin{eqnarray}
\Delta\langle H\rangle &=&  2\pi \Omega_{d-3}\,\widehat{T}_{00}e^{ip^0 t}\,
\int_0^R dr\,r^{d-2}
\int_0^\pi d\theta\, \sin^{d-3}\!\theta\, \frac{R^2-r^2}{2 R}
e^{-i p_1 r \cos(\theta)}\labell{modu2}\\
&=&
 2^{(d-1)/2}\pi^{3/2} \Omega_{d-3}\,
\Gamma[(d-2)/2]\,\widehat{T}_{00}e^{ip^0 t}\,
 \frac{R^{(d-1)/2}}{ |p^1|^{(d+1)/2}} J_{(d+1)/2}(|p^1| R)\,.\nonumber
\end{eqnarray}

Note the integral for $\Delta S$ in eq.~\reef{dss} depends on an additional
parameter $|p|$ which is not present in the integral in eq.~\reef{modu2}. Then
the equality between $\Delta S$ and $\Delta\langle H\rangle$ requires that the
expression in eq.~(\ref{dss}) is miraculously independent of $p$ for a given
fixed value of $p^1$. One can check this actually happens by making an
expansion in powers of $p$ and $p^1$  and also replacing $(p^0)^2=p^2+(p^1)^2$
in the integral in eq.~(\ref{dss}). Collecting the terms with the same powers
of $p$ and $p^1$, one arrives at expressions which are possible to integrate in
$\theta$ and $r$ analytically. The result is that the coefficient of $(p^1)^m
p^n $ in the expansion of $\Delta S$ is zero for any $n>0$. Hence, we may take
the limit of $p\rightarrow 0$ in the integrand to simplify the calculation
 and eq.~(\ref{dss}) becomes
\begin{eqnarray}
 \delta S &=& \frac{4\pi}{d}\Omega_{d-3}\,\widehat{T}_{00}e^{ip^0 t}\,
 \int_0^R
 dr\,r^{d-2}\int_0^\pi d\theta\, \sin^{d-3}\!\theta\,
\frac{R^2-r^2 \cos(\theta)^2}{2 R} e^{-i p_1 r \cos(\theta)} \labell{bango}\\
&=&  2^{(d-1)/2}\pi^{3/2} \Omega_{d-3}\,
\Gamma[(d-2)/2]\,\widehat{T}_{00}e^{ip^0 t}\,
 \frac{R^{(d-1)/2}}{ |p^1|^{(d+1)/2}} J_{(d+1)/2}(|p^1| R)\,.\nonumber
\end{eqnarray}
Now comparing eqs.~(\ref{modu2}) and \reef{bango}, we see
\begin{equation}
\Delta\langle H\rangle=\Delta S\,. \labell{equalz}
\end{equation}
While this analysis was done for a single plane wave \reef{2four}, since we are
considering linear perturbations, the same equality must hold for a general
Fourier expansion \reef{fourier}. Therefore, we conclude that eq.~\reef{equalz}
holds for any first order perturbation of the stress tensor. In particular,
this equality still applies even when $T^{(0)}_{\mu\nu}$ varies on scales
comparable to $R$, the size of the spherical entangling surface.

\subsection{Quadratic corrections to relative entropy} \labell{quad}

While it was technically difficult to establish, the equality in
eq.~\reef{equalz} should have been expected given the discussion preceding
eq.~\reef{eeex}. Similarly, if we extend the previous calculation of $\Delta S$
to second order in $\vep$, we should expect that the new contributions at this
order result in the required inequality \reef{123}. In this section, we verify
that this expectation is indeed correct. For simplicity, we will restrict our
attention to constant stress tensors.

To obtain the quadratic correction to the relative entropy, we proceed in three
steps: First, we expand the bulk metric to quadratic order in the stress
tensor. Then we expand the area functional \reef{action} to quadratic order in
the perturbation parameter $\vep$. In particular, we obtain the equations of
motion governing the deformation of the minimal surface to linear order in the
stress tensor. Then solving the equations of motion, we substitute the results
back into the area functional and collect the aggregate quadratic correction in
the relative entropy.

\subsubsection*{Step 1: Bulk metric}
In eq.~(\ref{FGexp}) the bulk metric is expanded to linear order. To quadratic
order, the expansion will take the from
 \be
\delta g_{\mu\nu} = \eta_{\mu\nu} + a\,z^d\, T_{\mu\nu} + a^2 \,z^{2d}  \left(
n_1\, T_{\mu\alpha}T^{\alpha}{}_{\nu} + n_2\, \eta_{\mu\nu}T_{\alpha\beta}
T^{\alpha\beta}\right)+\cdots\,,
 \labell{secbulk}
 \ee
where
 \be
a= \frac{2}{d}\, \frac{\lp^{d-1}}{L^{d-1}}\,.
 \labell{aaax}
 \ee
The term, which is quadratic in the stress tensor, has the most general form
allowed by Lorentz invariance, symmetry between $\mu$ and $\nu$, and that the
trace of $T_{\mu\nu}$ vanishes.\footnote{Recall that we are limiting our
attention to $T_{\mu\nu}$ being a constant and hence the derivative terms
\reef{box}, which appeared at linear order in $\vep$ above, vanish here.}
Further the power of $z^{2d}$ in this term is simply determined by dimensional
grounds. It remains to fix the coefficients $n_{1,2}$, which can be done by
comparing this expression to the black brane metric \reef{static} when the
latter is re-expressed in FG coordinates \reef{bulkG}. The latter requires
transforming to a new radial coordinate in the asymptotic AdS geometry
 \be
\tz = z\,\left(1+\frac{1}{2d}\frac{z^d}{\zh^d}+\frac{2+3d}{16d^2}
\frac{z^{2d}}{\zh^{2d}}+\cdots\right)\,.
 \labell{newz}
 \ee
This new coordinate is chosen to produce $G_{zz}=L^2/\tz^2$, as required in
eq.~\reef{bulkG}. With this radial coordinate, the remaining metric components
in the asymptotic expansion take the form:
 \bea
g_{00} &=& -1 + \frac{d-1}{d}\frac{\tz^d}{\zh^d}
-\frac{4d^2-9d+4}{8d^2}\frac{\tz^{2d}}{\zh^{2d}}
+\cdots \nonumber\\
g_{ij}&=&\delta_{ij}\left(1 +
\frac{1}{d}\frac{\tz^d}{\zh^d}-\frac{d-4}{8d^2}\frac{\tz^{2d}}{\zh^{2d}}+\cdots
\right)\,.
 \labell{asmetric}
 \eea
Recall that the stress tensor takes the form given in eqs.~\reef{enerden1a} and
\reef{enerden1}, as can be read off from the leading terms above. Then
comparing eqs.~\reef{secbulk} and \reef{asmetric}, we can read off $n_1$ and
$n_2$ as
 \be
n_1 = \frac{1}{2}\qquad{\rm and}\qquad n_2 = -\frac{1}{8(d-1)}\,.
 \labell{coeff3}
 \ee

\subsubsection*{Step 2: Expansion of area functional and equations of motion}

The profile of the extremal surface receives corrections since the bulk is
altered. Recall from the previous section that the minimal surface in static
gauge can be described by $z(x^i)$, \ie the bulk radial coordinate is specified
as a function of the spatial coordinates $x^i$. In the present perturbative
construction, we can expand
 \be z(x^i) = z_0(x_i) + \vep\, z_1(x_i) +
\vep^2 z_2(x_i) + \cdots\,,
 \labell{profler}
 \ee
where $z_0$ is given in eq.~\reef{wox}. Note that since we are only interested
in quadratic corrections to the entanglement entropy, $z_2$ is not needed since
it would appear linearly in the area functional and hence would vanish by
virtue of equations of motion.

The order $\vep^2$ correction to the area functional \reef{action} can be
written as
 \be
A_{(2)} = A_{2,0} + A_{2,1} + A_{2,2}\,,
 \ee
where we are separating the contributions into three terms, according to the
power of $z_1$ appearing in the expressions, which is denoted by the second
index. Only $A_{2,1}$ and $A_{2,2}$ contribute to the linearized equations of
motion for $z_1$.

Carefully expanding, we find
\begin{eqnarray}
A_{2,0} &&=L^{d-1} a^2 \int d^{d-1}x\,  R z_0^{d} \bigg(-\frac{1}{16}
\left(1-\frac{r^2}{(d-1)R^2}\right)
(T_{00}^2 + T_{ij}T^{ij})
\nonumber \\
&&
+\frac{T_{i0}T^{i0}}{8}\left(1+ \frac{r^2}{(d-1)R^2}\right)+
\frac{x^i x^k}{4R^2}T_{i\alpha}T^{\alpha}{}_{k} +
\frac{1}{8}(T^2 - T_x^2 - 2T T_x)\bigg)\,,
\end{eqnarray}
where
 \be
T\equiv T_i{}^i\qquad{\rm and}\qquad T_x \equiv T_{ij}\frac{x^ix^j}{R^2}\,.
 \ee
Note that we have made use of $z_0^2 = R^2- r^2$ to simplify the above
expression.\footnote{We emphasize that the Greek indices $\mu,\nu,\cdots$ run
through all the indices corresponding to the flat boundary directions, whereas
Latin indices $i,j,\cdots$ are restricted to the spatial directions.} Further,
we find:
\begin{eqnarray}
A_{2,1} &=&L^{d-1} a \int d^{d-1}x\,\frac{R}{2z_0}\bigg(
T\big(z_1 - \frac{z_0^2}{R^2} x^i\partial_i z_1\big)  \nonumber \\
&&\qquad
+T_{ij}\left(\frac{2z_0^2 x^i \partial^j z_1}{R^2} -\frac{z_1 x^ix^j}{R^2}
 - \frac{z_0^2 x^ix^j x^k\partial_k z_1}{R^4}\right)\bigg)\,,
\end{eqnarray}
and
\begin{eqnarray}
A_{2,2}&=& L^{d-1}\int d^{d-1}x\, \frac{R}{z_0^d}\bigg( \frac{d(d-1)z_1^2}{2z_0^2} +
\frac{z_0^2(\partial z_1)^2}{2R^2}\nonumber\\
&&\qquad -\frac{z_0^2(x^i\partial_i z_1)^2}{2R^4} +
\frac{(d-1)}{2}\frac{x^i\partial_i z_1^2}{R^2}
\bigg)\,.
\end{eqnarray}
Note that in $A_{2,1}$, we have already dropped terms that vanish upon
evaluating them on the minimal surface $z_0$. We also remind the reader that
the boundary terms do not contribute. Now the equations of motion for $z_1$ are
derived by varying $A_{2,1} + A_{2,2}$ and can be written as
 \be
 \frac{1}{z_0^{d-1}R}\left(\partial^2 (z_0\,z_1) -
\frac{x^ix^j}{R^2}\partial_i\partial_j(z_0\,z_1)\right) = \frac{z_0}{2R}
\left((d-2) T + (d+2) T_x \right)\,. \labell{bingo}
 \ee
The perturbation $z_1$ is expected to take the form $T f_1(r) + T_{ij}x^i x^j
f_2(r)$. After some trial an error to solve for $f_2$, and setting the
appropriate boundary conditions by adding suitable choice of solutions to the
homogeneous equation, we arrive at the following very simple solution in
general $d$:
 \be
 z_1 = -\frac{a R^2z_0^{d-1}}{2(d+1)}(T + T_x)\,.
 \labell{solz1}
 \ee

\subsubsection*{Step 3: Substitution into the area functional}

With all the ingredients in place, we are ready to substitute everything back
into the area functional. This amounts to some more tedious algebra resulting
in seven tensor structures:
\begin{eqnarray}
A_{(2)} &=& L^{d-1}a^2\int d^{d-1}x \,\,\bigg(c_1 T^2 + c_2 T_x^2 + c_3 T_{ij}^2 + \nonumber \\
&& \hspace{2cm} c_4 T_{i0}T^{i0} + c_5 \frac{x^i T_{ij}T^{j}{}_{k}x^k}{R^2}
+ c_6  \frac{x^i T_{i0}T^{0}{}_{j}x^j}{R^2} + c_7 T T_x\bigg)\,.
\end{eqnarray}

The coefficients are given by
\begin{eqnarray}
c_1 &=& \frac{z_0^{d-4}}{16(d+1)^2(d-1)R} \bigg((d+1)^2 r^6 + (3 + d (3d^2+d-15)) r^4 R^2
 \nonumber \\
&&\qquad+(d^2(13 - 8 d)+ 2d) -3) r^2 R^4 +(3 d^3- 7 d^2+ d +3) R^6 \bigg)\,,\\
c_2 &=& \frac{ z_0^{d-4}}{8(d+1)^2} \bigg((1 - 5 d^2) r^2 R^3 + ( d(4d+3)-3) R^5\bigg)\,,  \\
c_3 &=& \frac{(\frac{r^2}{d-1}- R^2) z_0^d}{16 R}\,,\\
c_4 &=& \frac{(\frac{r^2}{d-1}+ R^2) z_0^d}{8 R}\,,\\
c_5 &=& R z_0^{d}\frac{d(d-2)-1}{4(d+1)^2}\,,\\
c_6 &=& \frac{R}{4} z_0^d \,, \\
c_7 &=& z_0^{d-4}\frac{R^3(d-1)}{4(d+1)^2}\bigg((1 - 3 d) r^2 + (2d+1) R^2\bigg)\,.
\end{eqnarray}

Proceeding with the remaining integrals, it is useful to note that by symmetry,
whenever an integral has the form $\int d^{d-1}x\,\, (x^i x^j x^k x^l \cdots)
f(r)$, \ie there are $n$ pairs of $x^i$'s in the integrand, we can simply
replace them by
 \be
N(\delta_{ij}\delta_{kl}\cdots + \textrm{permutations})\int d^{d-1}\!x\
r^{2n}\, f(r)\,,
 \labell{gummy}
 \ee
with some appropriate normalization constant $N$. Using this fact, we are left
with a final result of the form
 \be
A_{(2)} =a^2 L^{d-1} \Omega_{d-2}\left(C_1 T^2 + C_2 T_{ij}^2 + C_3 T_{0i}^2\right)\,,
 \labell{final}
 \ee
where
\begin{eqnarray}
C_1 &&= -\frac{d\sqrt{\pi}R^{2d}\Gamma[d+1]}{2^{d+4}(d+1)\Gamma[d+\frac{3}{2}]}
\,,\nonumber\\
C_2 &&= C_1\,, \labell{corel}\\
C_3 &&= -\frac{(d+2)\sqrt{\pi}R^{2d}\Gamma[d+1]}{2^{d+3}(d-1) \Gamma[d+\frac{3}{2}]}\,.
 \nonumber
\end{eqnarray}
Note that in the above expression, $T_{0i}^2 \equiv T_{0i}
T_{0j}\delta^{ij}\ge0$. Therefore given that the three coefficients are
negative, we are assured that the second order perturbation to the area is
negative and hence the second order contribution to the holographic
entanglement entropy ensures that the inequality \reef{123} is satisfied. Since
at second order, we have $\Delta\langle H\rangle\ne\Delta S$, it is nontrivial
to check the monotonicity property in eq.~\reef{include}. However, from the
above result, we find that $S(\rho_1|\rho_0)\propto R^{2d}$ and hence this
inequality is simply satisfied, \ie $\partial_RS(\rho_1|\rho_0)>0$.

As an example, we might apply these general results to the static thermal gas
described by the planar AdS black hole. The corresponding stress tensor is
given by eqs.~\reef{enerden1a} and \reef{enerden1},\footnote{In keeping with
the above analysis, we might introduce an explicit expansion parameter $\vep$
to these expressions. However, we adopt the simpler approach of formally
setting $\vep=1$ in the above expansion. From our previous examination of the
thermal bath, as well as the results here, we can infer that $\Delta S$ appears
as an expansion in the small parameter $aR^d\veps$.} \ie we have $T_{00} =
\veps$ and $T_{ij} = \delta_{ij} \veps /(d-1)$. The solution of
eq.~\reef{bingo} can be written as
 \be
z_1(r) = \frac{k_1}{\sqrt{R^2-r^2}} + a\veps \left( \frac{ ((d-1 ) R^{d+2} -
(R^2-r^2)^{d/2} (r^2 + (d-1) R^2))}{
 2 (d^2-1 ) \sqrt{R^2-r^2 }}\right)\,,
 \ee
where $k_1$ is an undetermined integration constant. To ensure that $r\to R$ as
$z\to 0$, which is already satisfied by $z_0$, we must choose
 \be k_1 = -\frac{a \veps\, R^{d+2}}{2 (d+1)}\,. \ee
This choice yields precisely the solution for $z_1$ given in eq.~(\ref{solz1}).
Substituting this solution into the area functional, we find
 \be \Delta S_{(2)}
= - \frac{\pi^{3/2}d\,\Omega_{d-2}\,
   \Gamma[d-1]}{2^{d+1}(d+1)\, \Gamma[d+\frac32]}\,\frac{L^{d-1}}{\lp^{d-1}}
   \,R^{2 d}\veps^2\,.
\ee which as required is a negative contribution. One should appreciate the
fact that the integrand involves a complicated collection of polynomials in
$d$. However, the final result reduces to the above simple form.

\subsection{Corrections from additional operators} \labell{matter}

To this point, we have only considered a special class of states that give rise
to a nontrivial expectation value for the stress tensor. For generic
perturbations away from the vacuum, we would expect that other operators will
acquire nontrivial expectation values. Hence in this section, we consider
states in which certain operators beyond the stress tensor acquire an
expectation value. The dual description will involve bulk gravity solutions in
which additional matter fields are excited. As we saw with the charge black
brane in section \ref{charge}, it is relatively easy to determine quadratic
corrections to the entanglement entropy coming from such matter field
perturbations. Below, we evaluate the analogous contributions to $\Delta S$ for
two types of states: The first will involve a scalar operator acquiring an
expectation value. The dual description involves adding a massive scalar field
to the gravitational theory. The second class will involve perturbations by a
conserved current in the boundary theory or a gauge field in the bulk. Hence
analyzing these latter configurations is a simple generalization of that for
the charged black brane. For both families of states, we find that the
quadratic contributions again ensure that $\Delta\langle H\rangle>\Delta S$.

\subsubsection*{Perturbing with a scalar condensate}

In our first class of states, a scalar operator $\mathcal{O}$ of dimension
$\Delta$ acquires a non-trivial expectation value (in the absence of any
sources). The corresponding dual description is that a scalar field has been
turned on and subsequently back reacts on the geometry to change the
entanglement entropy. We will limit ourselves here to calculate only the
leading contribution of this back reaction. The bulk action, which we are
considering here, is given by
 \be
I = \frac{1}{2\lp^{d-1}}\int d^{d+1}x\, \sqrt{G} \left[R -\frac{1}{2}(\partial
\phi)^2 - V(\phi) \right]\,.
 \ee
Since we are only solving perturbatively in $\phi$, we need only to keep up to
quadratic terms in the scalar, and thus the potential can be taken simply as
 \be V(\phi)=-\frac{d(d-1)}{L^2} + \frac{1}{2}m^2 \phi^2\,, \ee
where the first term provides the negative cosmological constant.

A standard result \cite{revue} in the AdS/CFT correspondence is that to leading
order in the condensate, the scalar field $\phi$ of mass $m = \Delta
(d-\Delta)$ behaves asymptotically as
 \be
\phi = \gamma\, \mathcal{O}\, z^\Delta + \cdots\,,
 \ee
with some normalization constant $\gamma$. This can be substituted into the
Einstein equation which, in the presence of the scalar, can be written as
 \be
\hat{R}_{AB} = \frac{1}{2}\partial_A \phi
\partial_B \phi + \frac{1}{d-1} G_{AB} V(\phi)\,. \labell{albert}
 \ee
In the presence of the scalar field, we expect that the boundary expansion of
the metric is altered \cite{construct}. However, since we are only interested
in the leading contribution of the perturbation, cross terms between the
boundary stress tensor and the scalar condensates  need not be included
here. To linear order in the boundary stress tensor and quadratic order in the
operator, the expansion of the metric $\delta g_{\mu\nu}$ in eq.~\reef{expand}
takes the form
 \be
\delta g_{\mu\nu} = a z^d \sum_{n=0}z^{2n} T_{\mu\nu}^{(n)}+z^{2\Delta}
\sum_{n=0} z^{2n} \, \sigma^{(n)}_{\mu\nu}+ \cdots\,,\labell{poww}
 \ee
where, of course, terms in the first sum were analyzed in section \reef{line}.
In both sums, the superscript $(n)$ indicates that the corresponding operator
contains a total of $2n$ derivatives, \eg see eq.~\reef{box}. Hence, for $n=0$,
the only possible contribution of the scalar is $\sigma^{(0)}_{\mu\nu}=
\alpha_0\,\eta_{\mu\nu} \mathcal{O}^2$ where $\alpha_0$ is some constant. The
latter is easily determined by substituting the expansion of the metric and
also that of the scalar field into the Einstein equations \reef{albert}, which
yields
 \be \sigma^{(0)}_{\mu\nu}
= -\frac{\gamma^2}{4(d-1)}\,\eta_{\mu\nu}\,\mathcal{O}^2\,.
 \labell{sigma0}
 \ee
Note that the coefficient here is negative definite, which will be crucial in
evaluating the change in the entanglement entropy below.

For the interested reader, we also consider the next term
$\sigma^{(1)}_{\mu\nu}$, which carries two derivatives acting on the condensate
$\mathcal{O}$. Demanding Lorentz invariance and symmetry in $\mu,\nu$, lets one
to write the general form
 \be
\sigma^{(1)}_{\mu\nu}=\alpha_1
\partial_\mu\mathcal{O}\partial_\nu\mathcal{O} +\alpha_2
\mathcal{O}\partial_{\mu}\partial_{\nu}\mathcal{O} +\alpha_3
\eta_{\mu\nu}\mathcal{O}\Box \mathcal{O}+ \alpha_4 \eta_{\mu\nu}(\partial
\mathcal{O})^2\,,
 \ee
with some undetermined coefficients $\alpha_i$. Again using the equations of
motion \reef{albert}, we arrive at:
\begin{eqnarray}
\sigma^{(1)}_{\mu\nu}&&= \frac{\gamma^2}{4(d-1)(\Delta+1)(2\Delta+2-d)}\bigg(
\big((d-2)\,\mathcal{O}\partial_{\mu}\partial_{\nu}\mathcal{O} + \Delta \,
\eta_{\mu\nu}\mathcal{O}\Box\mathcal{O}\big)\nonumber \\
&&\qquad-\big(d\,\partial_\mu\mathcal{O}\partial_\nu\mathcal{O}
- \eta_{\mu\nu} (\partial\mathcal{O})^2\big)\bigg)\,.
\end{eqnarray}

For general $\mathcal{O}(x)$, we would have to consider the sums in
eq.~\reef{poww} to all orders in derivatives. However, if $\mathcal{O}$ is
slowly varying on the scale of $R$, $\sigma^{(0)}$ provides the leading
contribution to the change in the entanglement entropy and we focus on this
scenario here. As in our previous calculations, we determine this leading
contribution by evaluating the area functional \reef{action} with the perturbed
metric but the leading order profile \reef{wox} for the extremal surface. The
resulting change in the entanglement entropy is given simply by
\begin{eqnarray}
\Delta S(\mathcal{O}) &&= \frac{\pi L^{d-1 }R}{\lp^{d-1}}
\int \frac{d^{d-1}x}{z_0^{d-2\Delta}} (\sigma^{(0)\,i}{}_{i} -
 \sigma^{0}_{ij}\frac{x^ix^j}{R^2}) \nonumber \\
&&= -\frac{\pi\gamma^2 L^{d-1}R}{4\lp^{d-1}}\, \mathcal{O}^2\,
\int \frac{d^{d-1}x}{z_0^{d-2\Delta}}
 \left(1-\frac{r^2}{(d-1)R^2}\right)
 \nonumber\\
&&= -\frac{\gamma^2 L^{d-1}}{\lp^{d-1}}\, \frac{\pi^{3/2} \(
\Delta-\frac{(d-2)^2}{2(d-1)}   \) \Gamma[\Delta - \frac{d}{2} + 1]}{
8 \Gamma[\Delta - \frac{d}{2} + \frac52]}\,\Omega_{d-2}\,R^{2\Delta}
\mathcal{O}^2\,.
 \labell{apple}
\end{eqnarray}
Note that the unitarity bound $\Delta>\frac{d}2-1$ ensures that the numerical
prefactor in the last line is positive and hence the overall result for $\Delta
S$ is negative. We note that this overall minus sign descends directly from
eq.~(\ref{sigma0}). Hence it is interesting that at the level of the FG
expansion, the metric appears to know already about the positivity of the
relative entropy!

It is interesting to compare the above contribution of the scalar condensate
$\mathcal O$ with the leading order contribution coming from the stress tensor.
In particular, one might consider a scenario where the expectation value of
both operators is set by a single scale $\mu$ (\eg the temperature), in which
case, we would have $\mathcal{O}\sim \mu^\Delta$ and $T_{\mu\nu}\sim \mu^d$.
Then the corresponding contributions to the entropy would scale like $\Delta
S(\mathcal{O})\sim (R\mu)^{2\Delta}$ and $\Delta S(T_{\mu\nu})\sim (R\mu)^{d}$
where our calculations would hold in a regime where $R\mu\ll1$. Hence if
$\mathcal O$ is sufficiently relevant, \ie $\frac{d}2-1<\Delta<\frac{d}2$, then
its contribution would be the dominant contribution. Of course, with
$\frac{d}2<\Delta<d$, the stress energy would produce the dominant contribution
while for the special case $\Delta=\frac{d}2$, the scaling of both
contributions would be the same. In a more general situation where there are
several scales in the problem, the scale of $T_{\mu\nu}$ would necessarily be
related to that of $\mathcal O$ and then there would be no obvious way to
compare their respective contributions to $\Delta S$.

It follows from the above expression \reef{apple} that relative entropy is
proportional to $R^{2\Delta}$ and hence it also satisfies the monotonicity
inequality \reef{include}.

\subsubsection*{Perturbing with a current} \labell{river}

Here we provide a brief description of the extension of the analysis in section
\ref{charge} to a state with a general boundary current $J_\mu$. Recall first
we wish to construct a metric in the FG form, as given in eqs.~\reef{bulkG} and
\reef{expand}. For simplicity, we will assume that the expectation value of the
current is constant and then to leading order the metric perturbation takes the
form
\begin{equation}
\delta g_{\mu\nu}=a \, z^d \, T^{(0)}_{\mu\nu}+
z^{2d-2}\,(b \, J_\mu J_\nu+c\, \eta_{\mu\nu} J^2)\,,\labell{ff}
\end{equation}
where the constants, $a$, $b$ and $c$, are all dimensionless. Since we are
working to linear order in the metric perturbation, we can consider the
contribution of each of the two terms in eq.~\reef{ff} independently, as above
for the scalar operator. We know that the $T^{(0)}_{\mu\nu}$ contribution
saturates the inequality \reef{123} and hence the current perturbations must
produce a negative contribution to the change in the entanglement entropy.

Recall that $a$ is given in eq.~\reef{aaax}. To determine the remaining
constants, we compare to the charged black brane metric \reef{chargemet}. It is
convenient to write the metric function in eq.~\reef{mimi} as simply
\begin{equation}
h=1-\gamma \tilde{z}^d+\beta \tilde{z}^{2d-2}\,,
\end{equation}
with $\gamma$ and $\beta$ being positive constants. We have to change the
radial coordinate $z$ in order to put the metric in the desired FG form
\reef{bulkG}. After this is done, we find to leading order that the remaining
metric components take the form
\begin{eqnarray}
g_{00}&=&-\left(1-\gamma \left(1-\frac{1}{d}\right)z^d+\beta
\left(1-\frac{1}{2d-2}\right)z^{2d-2}\right)\,,
\nonumber\\
g_{ij}&=&\delta_{ij}\left(1+\gamma \frac{z^d}{d}-\beta \frac{z^{2d-2}}{2d-2}\right)\,.
\labell{popper}
\end{eqnarray}
Setting $J_i = 0$ in eq.~\reef{ff}, we may compare the resulting expression
with the above and find:
\begin{eqnarray}
b = - 2(d-1)\,c \,,\qquad
c =\frac{\beta}{2(d - 1)J_0^2} \,.\labell{si}
\end{eqnarray}
Further identifying $J_0\equiv\lim_{z\to0}z^{d-3}\partial_zA_0$ in the charged
black brane solution we find that $c$ as a positive constant independent of the
current, \ie
\begin{equation}
c =\frac{1}{4(d-1)^2(d-2)}\,.
\end{equation}
 Now the relevant part of the metric
perturbation becomes
\begin{equation}
\delta g_{\mu\nu}= c\, z^{2d-2}(-2(d - 1)J_{\mu}J_\nu + \eta_{\mu\nu}J^2) \,.
\end{equation}
Inserting this expression into the area functional \reef{action} yields
\begin{equation}
\Delta S =\frac{\pi R L^{d-1}}{\lp^{d-1}}
\int d^{d-1}x \, \frac{1}{z_0^d}\left(\delta g_i{}^i -\delta g_{ij}
 \frac{x^i x^j}{R^2}\right)\,.
\end{equation}
 For a constant current, we then find
that the integral yields
\begin{equation}
\Delta S= -\frac{\pi^{3/2} (d-3)!\, \Omega_{d-2} }{2^{d+1} \Gamma[d+\frac12]}
\,\frac{L^{d-1} R^{2d-2}}{\lp^{d-1}}\,( \vec{J}^2 + (J^0)^2) \,.\labell{contri}
\end{equation}
Then from eq.~(\ref{contri}), it follows that relative entropy $\Delta \langle
H\rangle-\Delta S$ is positive, and it also increasing as $R^{2d-2}$,
satisfying the monotonicity inequality \reef{include}.

\subsection{Corrections for general entangling surfaces} \labell{generic}

In this section, we consider extending our analysis to entangling surfaces,
which are not simply spheres. Let us begin by considering the area functional
\reef{action} with a generic entangling surface in the boundary and a
perturbation of the vacuum state in which the stress tensor is excited. At
linear order, the perturbation of the bulk geometry still takes the form
presented in eq.~\reef{FGexp} where the coefficients $T^{(n)}_{\mu\nu}$ are
given by eq.~\reef{box}. As in our previous examples, the holographic
calculation of the entanglement entropy in the AdS vacuum will yield some
extremal profile $z_0(x^i)$ depending on the geometry of the entangling
surface. Now while this profile is perturbed in the excited state, the
perturbation will only contributes to the change in the area at second order.
Hence we can evaluate the linear change of the area by simply evaluating the
area \reef{action} with the original profile $z_0$ in the perturbed background.
Therefore with a generic entangling surface, the linear perturbation of the
entanglement entropy becomes
 \be
\Delta S=2\pi\frac{\Delta A}{\lp^{d-1}} = \frac{2\pi}{d} \int d^{d-1}x\,
\sqrt{1+(\partial z_0)^2}\,
 \sum_{n=0} z_0^{2n+1}\left(T^{(n)}{}_{i}{}^{i} -
T^{(n)}{}_{\,ij}\, \frac{\partial^iz_0\,\partial^jz_0}{1+(\partial
z_0)^2}\right)\,,
  \labell{linearA2}
  \ee
where $(\partial z_0)^2 = \delta^{ij}\partial_i z_0\partial_j z_0$ and
implicitly, the boundary geometry is simply flat space. Previously we concluded
in eq.~\reef{boat} that all of the tensors $T^{(n)}_{\mu\nu}$ are traceless and
hence we can replace $T^{(n)}{}_{i}{}^{i}=T^{(n)}_{00}$, which in turn are all
related to the local energy density $T^{(0)}_{00}$ by eq.~\reef{box}. Hence the
first term above is controlled entirely by the energy density. However, there
is no clear connection to the energy density in the second term. In section
\ref{line}, the rotational symmetry of the spherical entangling surface and the
corresponding bulk profile \reef{wox} was essential in reducing this expression
to a contribution which again was controlled by $T^{(0)}_{00}$. Hence our
observation here is simply that we should expect other components of the stress
tensor to contribute to $\Delta S$, even at linear order, for entangling
surfaces with a less symmetric geometry.

To explicitly illustrate this behavior, we consider the well-studied case of a
`slab' geometry where the entangling surface is comprised of two flat planes at
$x=\pm\ell/2$ \cite{rt1}. The extremal surface in the AdS vacuum has a profile
$z(x)$ and the area becomes
 \be
A=L^{d-1} B^{d-2}\int_{-\ell/2}^{\ell/2} \frac{dx}{z^{d-1}}\sqrt{1+z'^2}\,,
 \labell{actions}
 \ee
where $B$ is an IR length scale that regulates the size of the two planes, \ie
$B^{d-2}$ is the area of one plane. Further regarding this area as an action
for $z(x)$, the profile is constrained by a conserved quantity \cite{rt1}
 \be
 z^{d-1} \,\sqrt{1+z'^2} = z_*^{d-1}\,.
 \labell{conserve}
 \ee
Here $z_*$ is the maximum value of $z$ which the extremal surface reaches in
the bulk at $x=0$,
 \be
z_* = \frac{ \Gamma[\frac{1}{2 (d-1)}]}{2 \sqrt{\pi} \, \Gamma[\frac{d}{2 (d-1
)}]}\ \ell\,.
 \labell{zmax}
 \ee
The change in the entropy \reef{linearA2} then becomes
 \be
\Delta S = \frac{2\pi}{d} B^{d-2} z_*^{d-1} \int_{-\ell/2}^{\ell/2} dx
\sum_{n=0} z^{2n+2-d}\left[T^{(n)}_{\,00} - T^{(n)}_{\,xx}\,
\left(1-\frac{z^{2(d-1)}}{z_*^{2(d-1)}}\right)\right]\,.
 \labell{change8}
 \ee
Hence we see that both the energy density and the pressure along the $x$-axis
are contributing in this result. To produce a more explicit result, we can
simplify the calculation by assuming that the expectation value of the stress
tensor is uniform, \ie $T^{(n)}{}_{\mu\nu}=0$ for $n\ge1$. Then
eq.~\reef{change8} becomes
 \bea
\Delta S&=& \frac{2\pi}{d} B^{d-2} z_*^{d-1} \int_{-\ell/2}^{\ell/2}
\frac{dx}{z^{d-2}}\left[T_{00} - T_{xx}\,
\left(1-\frac{z^{2(d-1)}}{z_*^{2(d-1)}}\right)\right]
 \nonumber\\
 &=&  \frac{\pi^{1/2}
  \Gamma[\frac{d}{d-1}] \Gamma[\frac{1}{2 (d-1)}]^2}{8d \Gamma[
\frac{3d-1 }{2 (d-1)}] \Gamma[\frac{d}{2 (d-1)}]^2}\, B^{d-2} \ell^{2}\,
\left[\left(\frac{d+1}{d-1}\right)\,T_{00} - T_{xx}\right] \,,
  \labell{linearA3}
  \eea
where we have used eqs.~\reef{conserve} and \reef{zmax} to evaluate the final
expression above. Here again, we see that the result contains a term
proportional to $T_{xx}$.

Then we observe that with the first order calculations described here, we
expect that the inequality \reef{123} must be saturated, \ie $\Delta\langle
H\rangle=\Delta S$. Therefore from this result, we can also infer that the
\modu Hamiltonian for the slab geometry also contains terms which are linear in
the operator $T_{xx}$. Hence from these calculations, we can begin to see the
appearance of new operators, \ie other components of the stress tensor beyond
$T_{00}$, appearing in the \modu Hamiltonian for regions with general
entangling surfaces.

Let us add a few more observations about $\Delta S$ for general entangling
surfaces. First, we note that if we make a Fourier transform of the stress
tensor, as in eq.~\reef{fourier}, then eq.~\reef{linearA2} can be rewritten
using eq.~\reef{sumI} as
 \bea
\Delta S &=& \pi\,\Gamma[d/2] \int d^{d-1}x \, \int d^dp \, \exp(-i p\cdot x)
\, \sqrt{1+(\partial z_0)^2}
 \labell{linearA2x}\\
&&\qquad\qquad\qquad\qquad
\frac{I_{d/2}(|p|z_0)}{(z_0|p|/2)^{d/2}}\,\left(\widehat{T}_{00}(p) -
\widehat{T}_{ij}(p)\, \frac{\partial^iz_0\,\partial^jz_0}{1+(\partial
z_0)^2}\right)\,,
  \nonumber
  \eea
where $|p|= |\sqrt{p_\mu p^\mu}|$. Hence the same Green's function
$I_{d/2}(|p|z_0)$ appears in evaluating this leading contribution to $\Delta S$
for general entangling surfaces. Unfortunately, without the symmetry of a
spherical entangling surface, this expression does not simplify in any obvious
way.

In fact, eq.~\reef{linearA2} makes an important assumption about the extremal
surface in the bulk. Namely, that it is single-valued as a function of the
boundary coordinates $x^i$ or alternatively, that the extremal surface does not
extend to values of $x^i$ beyond the region $V$. Unfortunately, this assumption
can be shown not to apply in many cases. For example, a standard FG-like
expansion of the extremal surface describes the bulk surface as $X^\mu(y^a,z)$
where $y^a$ are coordinates along the entangling surface and $z$ is the usual
radial coordinate in the bulk \cite{adam,calc}. Then near the AdS boundary, one
finds
 \be
X^i = X^{i}_0(y^a) - \frac{1}{2(d-2)}K^i(y^a)\, z^2 +\cdots
 \labell{expansion}
 \ee
where $X^{i}_0(y^a)$ describes the position of the entangling surface in the
boundary and $K^i$ is the trace of the extrinsic curvature for the spatial
normal to the entangling surface. Our conventions are such that $X^i <
X^{i}_0(y^a)$ corresponds to the region inside the entangling surface and
$K^i=+(d-2)X^i/R^2$ for a sphere of radius $R$, centered at $X^i=0$. Hence for a spherical entangling
surface, the above expression shows how the extremal surface begins towards the
interior of $V$ as it extends into the bulk geometry. However, if the geometry
is such that $K^i<0$ on some portion of the entangling surface, then the
extremal surface actually extends to $X^i > X^{i}_0(y^a)$. Clearly,
eq.~\reef{linearA2} does not accommodate this situation where the integration
would include contributions from outside of the region $V$ -- see section
\ref{puzzle} for further discussion.

We can also use the above expansion \reef{expansion} to make an interesting
observation about the contributions to $\Delta S$ from near the entangling
surface. Let us assume that $K^i$ is positive everywhere and then use
eq.~\reef{expansion} to evaluate $\partial_i z$ to leading order in small $z$,
or equivalently to leading order in $X^i- X^i_0(y^a)$,

\be
\partial_i z=-\frac{d-2}{z}\left(\frac{1}{K^i(y^a)}-\frac{ \frac{\partial X^{i}_0}{\partial y^b}\frac{\partial y^b}{\partial X^i}}{K^i(y^a) }\right) +\cdots\,.
 \ee
We can choose coordinates $y^a$ to coincide with $d-2$ of the coordinates $X^i$ at linear order in the vicinity of a point in the boundary, and we call $r$ the remaining $X$ coordinate, orthogonal to the boundary. Substituting into eq.~(\ref{linearA2}), we find to leading order
 \be
\Delta S = 2\pi\frac{d-2}{d}  \int d^{d-1}x \,K^{-1}\(T_{00} -
T_{rr}\) + \cdots\,,
  \label{generalK}
  \ee
  where $K=\sqrt{\sum(K^i)^2}$.
We have dropped the higher derivative contributions with $T^{(n)}_{\mu\nu}$ in
the above expression. Further note the integrand is only well approximated
above in the vicinity of the entangling surface. Now as we noted above, in the
special case of a sphere of radius $R$, we have $K^r=+(d-2)/R$. Then
the general expression (\ref{generalK}) reduces to
 \be
\Delta S = \frac{2\pi R}{d}  \int d^{d-1}x \,\bigg(T_{00} - T_{rr}\bigg) +
\cdots\,,
 \labell{roundS}
 \ee
which agrees with expanding \reef{linearA} to leading order in $(R-r)$.
However, we note that this does not appear a good approximation of $\Delta H$
as given in eq.~\reef{spheredH}, even for small $(R-r)$. This suggests that the
infinite derivative expansion in \reef{linearA} is crucial to the ultimate
agreement between $\Delta S$ and $\Delta \langle H\rangle$, if we want to introduce
 localized sources which test the vicinity of the region boundary.

As explained in section \reef{discuss}, one expects quite generally that if
$T_{\mu\nu}$ was localized sufficiently close to the entangling surface, then
$\Delta \langle H\rangle$ should reduce to that of the Rindler modular
Hamiltonian \reef{siete}. Further then, in the regime where $\Delta S=\Delta
\langle H\rangle$, one must expect this form to be reflected in the result for
$\Delta S$. However, as demonstrated above, this agreement cannot be obtained
in our holographic calculations purely by expanding to leading order in $z$
near the boundary, no matter how close and sharply localized near the
entangling surface $T_{\mu\nu}$ is. In fact, the more localized $T_{\mu\nu}$
becomes, the more important the higher derivative terms will be, which leads to
a significant correction to the leading $z$ term. As concluded above therefore,
knowledge of the infrared completion of the bulk minimal surface is always
important.

\section{Two-dimensional boundary theories} \labell{two}

For a two-dimensional boundary theory, we can describe a thermal state with the
BTZ black hole \cite{btz}. However, in this case, the bulk geometry is still
locally AdS$_3$ space. Further, in calculations of holographic entanglement
entropy, the extremal surfaces are simply geodesics. Combining these two
observations, we are able to determine the extremal surfaces analytically and
hence we can extend our previous analysis beyond perturbation theory. That is,
in contrast with the results in section \ref{simple}, in the following we can
evaluate $\Delta \langle H\rangle$ and $\Delta S$ for arbitrary values of $R
T$. The present analysis also allows us to see the effect of compactifying the
AdS boundary and also to check the validity of the inequality \reef{123} in a
situation where the extremal surface exhibits a `phase transition.'

Eq.~\reef{static} already describes the appropriate three-dimensional black
hole. However, since we wish to consider the spatial direction as compact, we
write the (Euclidean) BTZ metric \cite{btz} in more familiar coordinates as
\begin{equation}
ds^2_\mt{E}=\frac{r^2-r_+^2}{R^2} d\tau^2+ \frac{L^2\,dr^2}{r^2-r_+^2} +r^2\,
d\phi^2\,,
 \labell{btzmet}
\end{equation}
where, as usual, $L$ is the AdS radius and the period of $\phi$ is $2\pi$. The
above geometry is smooth as long as $\tau$ is chosen with period $\beta=2\pi
LR/r_+$ and so the temperature is given by simply $T = 1/\beta= r_+/(2\pi LR)$.
The coordinates in eq.~\reef{btzmet} are normalized so that the boundary metric
is
 \be
ds^2_{boundary} = d\tau^2 + R^2\, d\phi^2\,.
 \labell{bound9}
 \ee
Hence the periodicity of the spatial direction is $2\pi R$ and the boundary is
a cylinder with a total area $2\pi R\beta$. We should note that because the
spatial direction is compact, there is a Hawking-Page phase transition
\cite{HP}. The above black hole geometry is the dominant saddle-point in the
gravity path integral for $T>1/(2\pi R)$, while for $T<1/(2\pi R)$, the
dominant saddle-point is simply the thermal AdS$_3$ geometry. We may write the
metric for the latter as
\begin{equation}
ds^2_\mt{E}=\frac{r^2+L^2}{R^2} d\tau^2+ \frac{L^2\,dr^2}{r^2+L^2} +r^2\,
d\phi^2\,.
 \labell{thermalmet}
\end{equation}
Implicitly, $\tau$ and $\phi$ are chosen with the same periodicity as in the
previous case and the boundary metric is again given by eq.~\reef{bound9}.

Let us begin with the high temperature phase for which eq.~\reef{btzmet}
describes the correct bulk geometry. It is relatively straightforward to
evaluate the entanglement entropy of an interval with an angular width
$\Delta\phi$ (and on a constant $\tau$ surface). Of course following
eq.~\reef{define}, it is given by the length of the geodesic connecting the
endpoints of the interval $V$ on the boundary \cite{rt1},
\begin{equation}
S(V)=\frac{c}{3}\log\left[\frac{\beta}{\pi \epsilon}\sinh\(\frac{\pi R
\Delta \phi}{\beta}\)\right]\,,
 \labell{finiteT}
\end{equation}
where $c= 
12\pi L/\lp$ is the central charge of the boundary CFT and $\epsilon$ is the
short-distance cut-off in the CFT.\footnote{The latter appears in the
holographic calculation by terminating the geodesic at a UV regulator surface
positioned at $r=r_\mt{UV}=LR/\epsilon$ in the bulk geometry.}

This expression precisely matches the known result previously derived for
two-dimensional CFT's at finite temperature \cite{korepin,cardy0}. However, we should
note that this previous result was derived for the case where the spatial
direction was noncompact. That is, this same expression \reef{finiteT} was
derived for any two-dimensional CFT but only in the limit $R\to\infty$ while
holding $\Delta x=R\Delta\phi$ fixed. Hence, we see here that in a holographic
$d=2$ CFT, compactifying the spatial direction does not affect this finite
temperature entanglement entropy \reef{finiteT}. Of course, this statement
holds when the bulk physics is accurately described by classical Einstein
gravity and hence eq.~\reef{finiteT} only represents the leading contribution
in an expansion for large $c$.

\begin{figure}
\centering
\leavevmode
\epsfysize=6cm
 \begin{tabular}{ccc}
\includegraphics[width=0.4\textwidth]{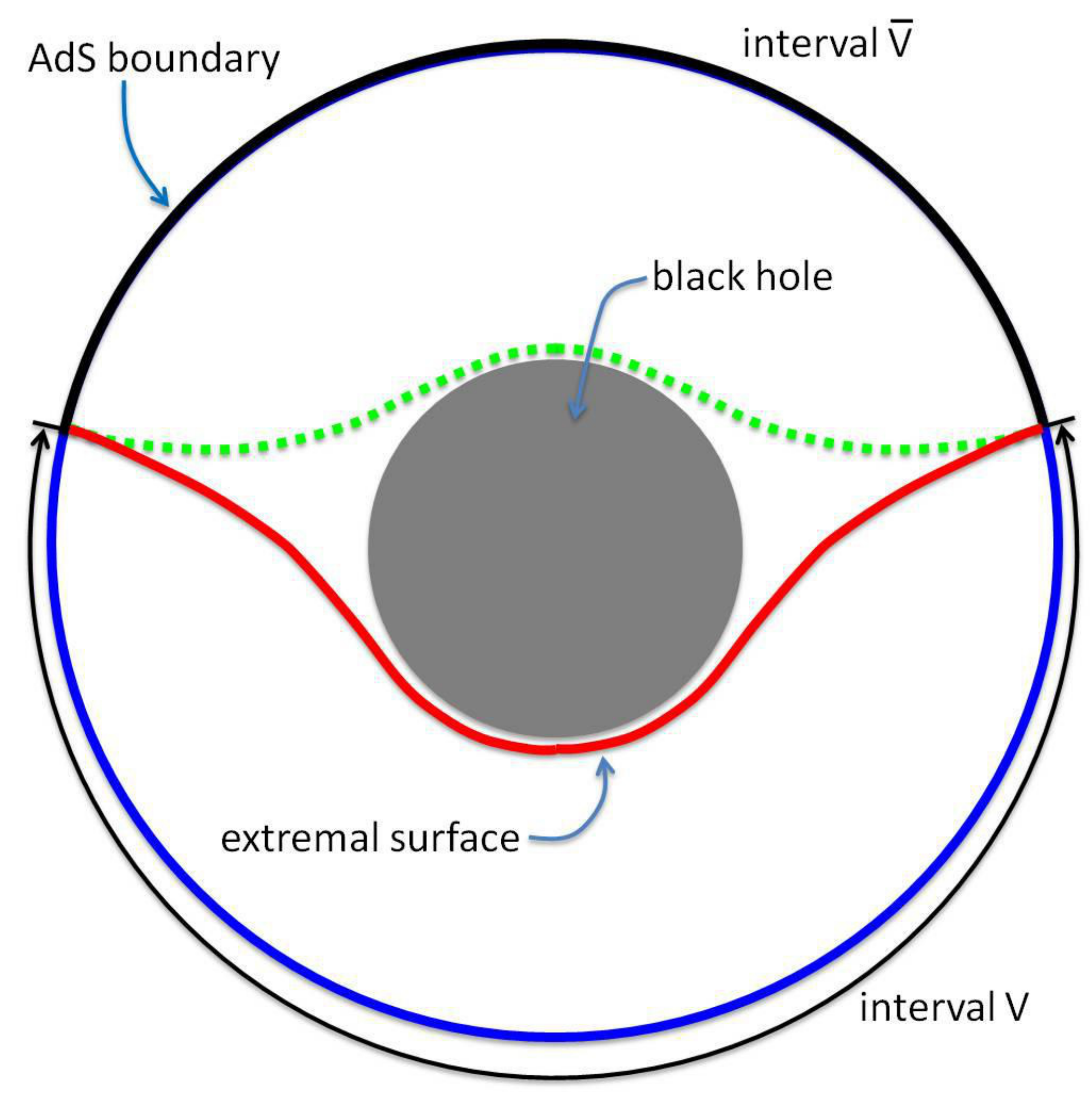}
&$\quad$&
\includegraphics[width=0.4\textwidth]{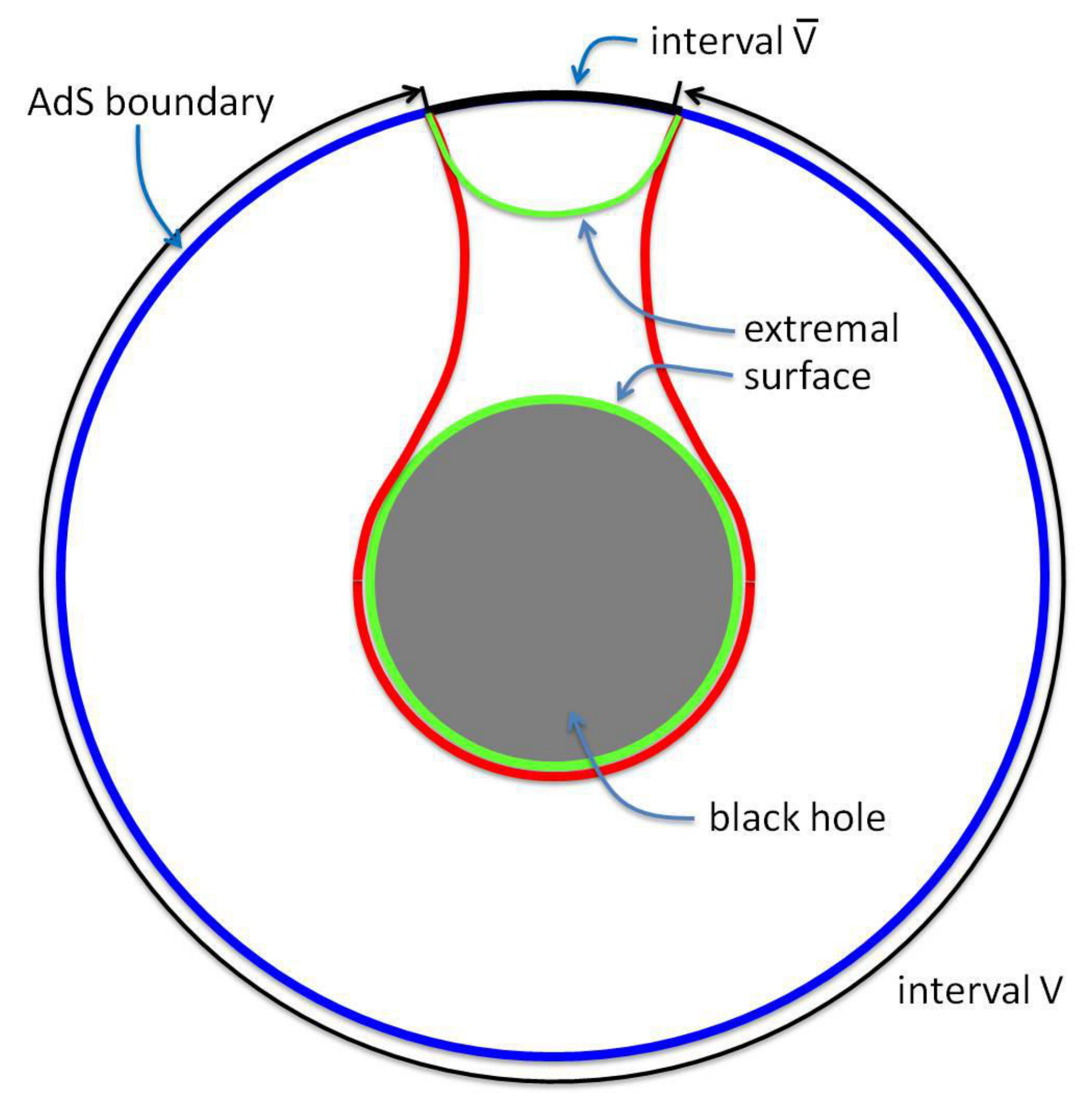}
\\
\ \ (a) & &\ (b)
\end{tabular}
\caption{(Colour Online) Extremal surfaces in the high temperature phase.
The figures show a cross-section of the AdS$_3$ black hole at constant $t$. (a)
For sufficiently small $\Delta\phi$, the holographic entanglement entropy
\reef{define} is evaluated with the red geodesic. The dashed green geodesic
passing on the other side of the black hole is not homologous to the interval
$V$, however, it would yield the entanglement entropy for the complementary
interval $\bar V$. (b) For large $\Delta\phi$, the dominant saddle-point (in
green) has two disconnected components, \ie the geodesic homologous to $\bar V$
and the geodesic wrapping around the horizon.}
\labell{disconnect}
\end{figure}
Implicitly, the above result also assumes that $\Delta\phi$ is sufficiently
small. In this high temperature phase, one finds for large enough $\Delta
\phi$, that the holographic entanglement entropy experiences a `phase
transition,' as described in figure \ref{disconnect}. For any value of
$\Delta\phi$, there are two geodesics connecting the endpoints of the interval
on the boundary, which pass on either side of the black hole, as shown in
figure \ref{disconnect}a. However, only one of these (the green geodesic in the
figure) is homologous to the boundary interval $V$ and hence this one must be
chosen to evaluate the holographic entanglement entropy. The other (the dashed
red geodesic) can be used to evaluate the entanglement entropy for the
complementary region $\bar V$, with the result
\begin{equation}
S(\bar{V})=\frac{c}{3}\log\left[\frac{\beta}{\pi \epsilon}\sinh\(\frac{\pi R
(2\pi-\Delta \phi)}{\beta}\)\right]\,.
 \labell{finiteT2}
\end{equation}
Of course, for $\Delta\phi>\pi$, the latter expression is smaller than $S(V)$
in eq.~\reef{finiteT}. While this geodesic by itself is not homologous to the
region of interest, it can be used to construct another extremal surface with
two disconnected components, as shown in figure \ref{disconnect}b, which is
homologous to $V$. The second component consists of a closed (spatial) geodesic
which wraps around (the bifurcation surface of) the black hole horizon. The
latter contributes the standard horizon entropy, \ie
 \be
S_\mt{BH}=\frac{2\pi}{\lp}\, A(r_+)=\frac{2\pi^2
r_+}{\lp}=\frac{2\pi^2c}{3}\,\frac{R}{\beta}\,.
 \labell{finiteT3}
 \ee
Hence combining these results, the entropy for a general interval is given by
\begin{equation}
S=\frac{c}{3}\min\left[\,\log\left(\frac{\beta}{\pi \epsilon}\sinh\left(\frac{\pi R
\Delta \phi}{\beta}\right)\right)\, ,\,\log\left(\frac{\beta}{\pi \epsilon}\sinh\left(
\frac{\pi R(2\pi-\Delta \phi)}{\beta}\right)\right)+\frac{2 \pi^2 R}{\beta} \,\right]\,.
\labell{finiteT4}
\end{equation}
For general values of $R/\beta$, it would require a numerical evaluation to
determine the precise value of $\Delta\phi$ at which there is a phase
transition between the two saddle-points occurs. However, in the high
temperature limit with $R/\beta\gg1$, it is straightforward to show that the
phase transition occurs at\footnote{We thank Ian Morrison and Matt Roberts for
pointing out an error in the result given here in our original manuscript.}
 \be
\Delta\phi\simeq 2\pi - \log 2\,\frac{\beta}{2\pi R} + \cdots\,,
 \labell{ouch}
 \ee
where the $\cdots$ denotes corrections that are exponentially suppressed by
$e^{-2\pi^2 R/\beta}$.

Recall in the low temperature phase with $R/\beta<1/(2\pi)$, the bulk geometry
is simply the thermal AdS$_3$ geometry \reef{thermalmet}. In this case, there
is always a single geodesic joining the endpoints of the boundary interval and
we have
\begin{equation}
S=\frac{c}{3}\log\left(\frac{2R}{\epsilon}\sin(\Delta\phi/2)\right)\,.
\labell{lowT}
\end{equation}
Again this expression precisely matches a known result derived for general
two-dimensional CFT's \cite{wilcCFT,cardy0}. In this case, this expression \reef{lowT}
holds for any two-dimensional CFT but only in the limit $T=0$. Hence, we see
here that in a holographic $d=2$ CFT, turning on a small temperature does not
affect the entanglement entropy \reef{finiteT} to the leading order in the
large-$c$ expansion.

Hence comparing the entropy of a low temperature state to that of the vacuum
(\ie $T=0$) for a fixed interval, we find $\Delta S=0$. Rather the order $c$
contributions cancel and hence $\Delta S$ is only a quantity of order one. If
instead, we compare the entropy of a state in the high temperature phase to
that of the vacuum, we find
 \bea
\Delta S&=& \frac{c}3\,\log\(\frac{1}{2\pi R T} \frac{\sinh\(\pi
RT\Delta\phi\)}{\sin\(\Delta\phi/2\)} \)
 \labell{deltaST1}\\
 &=& \frac{\pi^2}{18}\,c\,\(R^2T^2+\frac{1}{4\pi^2}\)\Delta\phi^2+O\(\Delta\phi^4\)\,.
 \eea
In the first line, we have assumed that $\Delta\phi$ is small enough that the
finite temperature entropy is given by eq.~\reef{finiteT}. In the second line,
we are expanding the result for $\Delta\phi\ll1$. Note that the two expressions
in eqs.~\reef{finiteT} and \reef{lowT} are approximately equal in this limit
$\Delta \phi\ll 1$ where the effects of compactification and finite temperature
can be neglected.

The \modu Hamiltonian of a $d$-dimensional CFT for the vacuum on the
cylindrical geometry $R\times S^{d-1}$ can be obtained by conformally
transforming the result \reef{sphereH} for the sphere in Minkowski space
\cite{casini9}. Applying this transformation in the present case with $d=2$, we
have
\begin{equation}
H=2 \pi R^2\int_{-\Delta \phi/2}^{\Delta\phi/2}d\phi\,
\frac{\cos(\phi)-\cos(\Delta \phi/2)}{\sin(\Delta \phi/2)}\, T_{00}   \,.
\labell{cylinH}
\end{equation}
In the vacuum, on the cylinder, the energy density is given by $T_{00}=-\frac{
c}{24\pi R^2}$ \cite{difra}. In general at finite temperature, the
expression for the energy density will be quite complicated but to leading order in the central charge
the energy density does not change until the temperature reaches the high temperature phase $RT>(2 \pi)^{-1}$ \cite{bh3d}. In this high temperature phase we have $T_{00}=\frac{\pi}{6}\,c\,  T^2$, which is the standard result for any CFT in the high temperature limit (or in decompactified space) \cite{difra}.

 Combining these results
gives
 \bea
\Delta \langle H\rangle&=&\frac{2 \pi^2c}{3} \
\left[1-\frac{\Delta\phi/2}{\tan(\Delta\phi/2)} \right]\,
\(R^2T^2+\frac1{4\pi^2}\)
 \labell{dhtx8}\\
&=& \frac{\pi^2}{18}\,c\,
\(R^2T^2+\frac{1}{4\pi^2}\)\Delta\phi^2+O\(\Delta\phi^4\)\,.
 \nonumber
 \eea
The second line gives an expansion of the result for $\Delta\phi\ll1$.
Comparing with the expansion in eq.~\reef{deltaST1}, we see the leading term in
both cases agrees and so we saturate the inequality \reef{123} for small
$\Delta\phi$.

Our results above apply for any value of $\Delta\phi$ and so we may also
examine the inequality \reef{123} for finite values.  Figure \ref{paralelas1}a
shows the difference $\Delta \langle H\rangle-\Delta S$ as a function of
$\Delta\phi$ for the high temperature phase. There we see that this difference
is positive and increasing for all angles. Hence the inequalities in both
eqs.~\reef{123} and \reef{include} are satisfied throughout the full range of
$\Delta\phi$.  Note the phase transition at large angular sizes, which was
discussed above, contributes very little to the slope of the curves. Figure
\ref{paralelas1}b shows the ratio $\Delta S/ \Delta \langle H\rangle$. This
ratio decreases with size and the figure clearly shows that $\Delta S\simeq
\Delta \langle H\rangle$ for intervals of small size, as noted above.
\begin{figure}
\centering
\leavevmode
\epsfysize=4.8cm
 \begin{tabular}{ccc}
\includegraphics[width=0.45\textwidth]{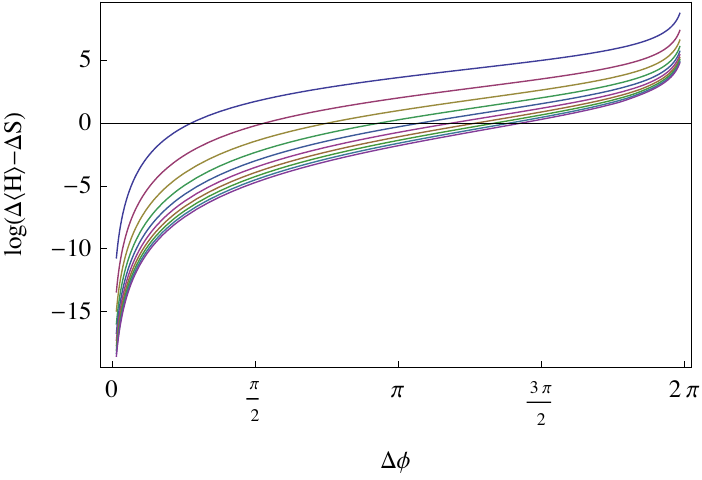}
&$\quad$&
\includegraphics[width=0.44\textwidth]{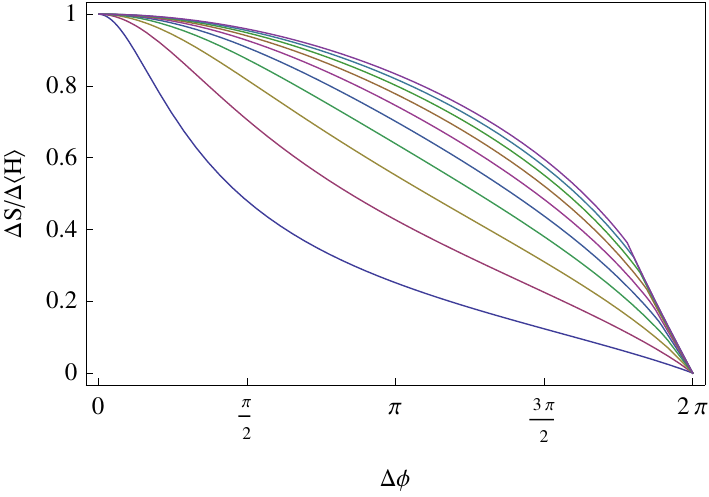}
\\
\ \ (a) & &\ (b)
\end{tabular}
\caption{Comparing $\Delta \langle H\rangle$ and $\Delta S$ in the high temperature
phase. Panel (a) shows the log of the relative entropy and panel (b), the ratio
$\Delta S/\Delta \langle H\rangle$, both as functions of angular size
$\Delta\phi\in (0,2\pi)$. The different curves are for $\beta/R=2\pi
\frac{i}{10}$ with $i=1,...,10$. Curves corresponding to higher temperature
(smaller $\beta$) have greater relative entropy in (a) and lower ratios $\Delta
S/\Delta \langle H\rangle$ in (b).} \labell{paralelas1}
\end{figure}

\subsection{Thermal Rindler space} \labell{RindlerT2}

In this section, we consider a two-dimensional CFT in a thermal state in the
Rindler wedge. The \modu Hamiltonian for this case is given in
eq.~\reef{2dcftRinT}. We will use this to compute the relative entropy between
states at different temperatures, \ie both $\rho_0$ and $\rho_1$ will describe
thermal states with temperatures, $T_0$ and $T_1$, respectively. The
expectation value of the stress tensor for both of these states is $T_{00}(x)=
\frac{\pi}6\, c T_i^2$, where $T_i$ corresponds to the appropriate temperature.
Since Rindler space has infinite volume, we need to introduce a  long-distance
infrared cut-off $\Lambda$, \ie we integrate only over $0\le x\le \Lambda$.
Given eq.~\reef{2dcftRinT}, we fix the \modu Hamiltonian to be $H_0=H(T=T_0)$
corresponding to $\rho_0$. Then the change of the expectation value of modular
Hamiltonian between $\rho_1$ and $\rho_0$ given by
\begin{equation}
\Delta \langle H \rangle=\Tr\(\rho_1 H_0\)-\Tr\(\rho_0 H_0\)=
 \frac{\pi }{6}\,c\Lambda\left(\frac{T_1^2}{T_0}
-T_0\right)-\frac{c}{12}\left(\frac{T_1^2}{T_0^2}-1\right)\,.\label{dooce}
\end{equation}
Here, we have dropped terms proportional to $\exp(-2\pi T_0\Lambda)$. The first
term on the right hand side is the purely thermal and extensive ($\propto
\Lambda$) contribution, which comes from the large part of the Rindler wedge
which is at distances larger than $T^{-1}$ from $x=0$. One can regard the
second term as the contribution of the entanglement across the entangling
surface $x=0$.

Turning to the holographic calculation of the entanglement entropy, we use
the original metric (\ref{static}) with $d=2$ to describe the black hole
geometry. The appropriate extremal surface with which to evaluate
eq.~\reef{define} is the geodesic which begins at $x=0$ on the AdS boundary
($z=0$) and extends out along the event horizon ($z=\zh$) at large positive
$x$. This geodesic is given by
\begin{eqnarray}
x(s)&=&\frac{1}{2} \zh \log\left(4 e^{2 s/L}+1\right)\,,
\labell{geodesic9}\\
z(s)&=&\frac{\zh}{\left(\frac{1}{4}e^{-2 s/L}+1\right)^{1/2}}\,,
\nonumber
 \end{eqnarray}
where $s$ is the affine parameter along the geodesic. Note that the geodesic
approaches the AdS boundary as $s\to-\infty$ and extends out along the horizon
as $s\to+\infty$. With $d=2$, eq.~\reef{temper1} yields $T=1/(2\pi \zh)$, and
we recall that $c=12\pi L/\lp$. Imposing an ultraviolet cut-off $z=\epsilon$
and an infrared cut-off at $x=\Lambda$, the entropy at a generic temperature
$T$ becomes
\begin{equation}
S(T)=\frac{2\pi}{\lp}\(s(x=\Lambda)-s(z=\epsilon)\)=\frac{c}{12}\log
\left(\frac{e^{4 \pi T \Lambda}-1}{4 \pi^2 \epsilon^2 T^2}\right)\,.
 \labell{entropy87x}
\end{equation}
Given this expression, it follows that
\begin{equation}
\Delta S=S(T_1)-S(T_0)=\frac{\pi}{3}\,c\Lambda  (T_1-T_0)
-\frac{c}{6}\log\left(\frac{T_1}{T_0}\right)\,,
 \labell{entropy87y}
\end{equation}
where again we are dropping terms that are exponentially small in $\Lambda$.

Combining eqs.~\reef{dooce} and \reef{entropy87y}, the relative entropy is
\begin{equation}
S(\rho_1|\rho_0)=\Delta \langle H \rangle-\Delta S=\frac{\pi }{6}\,c\Lambda T_0
  \(\frac{T_1}{T_0}-1\)^2+\frac{c}{12}\left(1
+2\log\left(\frac{T_1}{T_0}\right)-\frac{T_1^2}{T_0^2}\right)\,.
 \labell{relent98}
\end{equation}
For generic $T_1$, this result is always
positive because it is dominated by the first term since $\Lambda T_{0,1}\gg
1$. Of course, one must treat the region $T_1\sim T_0$ more carefully. With
$T_1=T_0$, both $S(\rho_1|\rho_0)$ and the first derivative
$\partial_{T_1}S(\rho_1|\rho_0)$ vanish. The second derivative yields
 \be
\partial^2_{T_1}S(\rho_1|\rho_0)=\frac{c}{6}\(2\pi\frac{\Lambda}{T_0}
-\frac1{T_0^2} -\frac1{T_1^2}\)\,.
 \labell{secondX}
 \ee
This quantity is again positive given $\Lambda
T_{0,1}\gg 1$ and so the relative entropy is positive in the vicinity of $T_1=
T_0$. Because of the vanishing first derivative, we also have the equality for
small deviations $\delta T=T_1-T_0$
\begin{equation}
\Delta S=\Delta \langle H\rangle=c\,\left(\frac{\pi }{3}\ \Lambda
-\frac{1}{6 T_0}\right)\, \delta T\,.
\end{equation}

In previous calculations, we compared the vacuum state and a thermal state. To
compare the thermal state with the vacuum on Rindler space, we can set $T_1=0$
and $\Delta \langle H \rangle$ follows from eq.~(\ref{dooce}) as
\begin{equation}
\Delta \langle H \rangle=-\frac{ \pi }{6}\,c  \Lambda T_0+\frac{c}{12}\,.
 \labell{fireX}
\end{equation}
The vacuum in Rindler space has logarithmic entropy $S\sim c/6 \log
(\Lambda/\epsilon)$. Hence the difference in entropies is
\begin{equation}
\Delta S=-\frac{ \pi }{3}\,c  \Lambda T_0 +\frac{c}{6}\log(\Lambda T_0)
 + {\cal O}(\Lambda^0)\,.
\end{equation}
Thus the inequality $\Delta \langle H\rangle>\Delta S$ is always valid. Note
that there is no meaningful way to say that the relative entropy\footnote{Note
that in evaluating $S(\rho_1|\rho_0)=\Delta \langle H\rangle-\Delta S$,
$\rho_1$ corresponds to the vacuum while $\rho_0$ is the thermal state. In our
previous calculations, these roles were reversed.} approaches zero, \ie $\Delta
\langle H\rangle\to\Delta S$, for small temperatures since we must keep
$\Lambda T_0 \gg 1$. In fact, the vacuum in Rindler space always remains at an
infinite statistical distance from a thermal state since far enough from the
origin, \ie $x\gg 1/T_0$, the thermally excited modes are in presence of a
nearly zero Unruh temperature vacuum. This does not happen in comparing the
vacuum and a thermal state over a finite interval of size $\ell$. At
sufficiently small temperatures, \ie $T_0\lesssim 1/\ell$, the change in the
\modu Hamiltonian will essentially match the change in the entanglement
entropy. In particular, in the previous section, we saw that $\Delta \langle
H\rangle$ and $\Delta S$ were always nearly identical for sufficiently small
$\Delta\phi$, irrespective of the temperature.


\section{Puzzles about localization} \labell{puzzle}

Most of our previous calculations only probed the asymptotic region in the bulk
geometry and in particular, the analysis in section \ref{general} relied
heavily on the asymptotic FG expansion. With the latter approach, one can
construct the asymptotic geometry for states with an essentially arbitrary
expectation value for the stress tensor and other operators. However, one
should be aware that in many cases, these expectation values will not
correspond to a physical state. In other words, if one really goes beyond the
asymptotic expansion to construct the full nonlinear gravity solution, one
would find that in many cases, the solution has a naked singularity somewhere
in the infrared region. Of course, string theory may be able to resolve some
such singularities \cite{sing}, however, one should expect that most of these
singular solutions are simply unphysical. Certainly, our previous analysis does
not consider such issues which might arise in defining a global state from
imposing a `smoothness' boundary condition in the infrared. In this section, we
consider some apparent paradoxes (and their resolution) which appear from
localizing the expectation values which contribute to $\Delta \langle H\rangle$
and $\Delta S$. From this perspective, the relative entropy provides
interesting probe of the AdS/CFT correspondence, which reveals constraints on
the properties of physical states which would not be easily seen by other
means.

\subsection{Complementary regions in a pure state} \labell{compliment}

Our general arguments from the previous sections indicate that under the
conditions of a small linear perturbation $\delta T_{\mu\nu}$, the inequality
in eq.~\reef{123} is saturated with $\Delta S=\Delta \langle H\rangle $, for a
spherical entangling surface. Further it is clear from the holographic
calculations that if the perturbation $\delta T_{\mu\nu}$ was completely
localized outside of the sphere, it would not change the entanglement entropy,
\ie $\Delta S=0$. Further given the form of the modular Hamiltonian
\reef{sphereH}, it is also clear that for this situation that we also have
$\Delta \langle H\rangle=0$. Of course, this is as it must be, since $H_V$
 is a operator in the algebra generated by local operators in
the region $V$, \ie the interior of the sphere.\footnote{The full generator of modular flow is $H_V-H_{-V}$,
while $H_V$ is the generator for the modular flow inside $V$.} The latter would then also extend to more general
regions $V$, for which we also expect $\Delta \langle H_V\rangle$ to be given
by contributions from the expectation values of operators inside $V$.

Now our first apparent paradox arise from considering instead the case where
$\delta T_{\mu\nu}$ is entirely localized inside the sphere. Again we suppose
that eq.~\reef{123} is saturated with
\begin{equation}
\Delta S_V=\Delta \langle H_V\rangle\,.\labell{un}
\end{equation}
The modular Hamiltonian of the vacuum state in the region $\bV$ outside the
sphere is given by
\begin{equation}
H_{\bV}=2\pi\int_{|x|>R} d^{d-1}x\, \frac{r^2-R^2}{2R}\, T_{00}(x)\,.
\labell{sphereHx}
\end{equation}
Our assumption is that
the stress tensor vanishes in this complementary region. Hence since $\langle
T_{00}(x)\rangle=0$ in $\bV$, we have
\begin{equation}
\Delta \langle H_{\bV}\rangle=0\,.\labell{do}
\end{equation}

However, if the perturbed state is pure, the entanglement entropy for the two
complementary regions, the interior and the exterior of the sphere, must be
equal. Holographically, $\Delta S_V$ came from the changes in the corresponding
extremal surface in the bulk. However, assuming there are no additional
horizons in the bulk, as should be the case for a pure state, the same two
extremal surfaces (\ie the one for the vacuum and the one for the perturbed
state) also determine $S_{\bV}$. Thus, in this case we have
\begin{equation}
\Delta S_V=\Delta S_{\bV}\,.\labell{tre}
\end{equation}

Now combining eqs.~(\ref{un}), (\ref{do}) and (\ref{tre}), we see $\Delta
\langle H_{\bV}\rangle\neq \Delta S_{\bV}$. In particular then, the equality
can not be achieved for $\bV$ no matter how small $\delta T_{\mu\nu}$ is. In
fact, assuming we have injected a small positive energy inside the sphere, \ie
$\delta T_{00}>0$, then $\Delta \langle H_{V}\rangle=\Delta S_{V}=\Delta
S_{\bV}>0$. Then we have arrived at a clear contradiction with the positivity
of relative entropy since $\Delta \langle H_{\bV}\rangle-\Delta S_{\bV}<0$. Of
course, the resolution to this apparent paradox is that it is not possible to
choose to inject (positive) energy only in $V$ and not in $\bV$ for a pure
state near the vacuum. There must be enough energy in both $V$ and $\bV$ to
ensure the equality of the expectation values of the modular Hamiltonians for
the two complementary regions. In the context of the AdS/CFT correspondence,
this is a constraint that would not be visible with the FG expansion but that one
can imagine arises from global issues in defining a smooth bulk geometry.

We can also make a field theory argument to directly demonstrate this
conclusion that the energy of the perturbed state cannot be strictly localized.
To see this, we construct the combination
\begin{equation}
 H=H_V - H_{\bV}\,.
 \labell{combo}
\end{equation}
This operator generates the conformal transformations which keep the sphere
fixed \cite{casini9}. It annihilates the global vacuum state
\begin{equation}
H|0\rangle=(H_V-H_{\bV})|0\rangle=0\,.\labell{gigi}
\end{equation}
Now we can write an arbitrary pure state which approaches the vacuum as
\begin{equation}
 \left|\psi\right> =| 0\rangle+\epsilon |\phi\rangle \labell{doa}
\end{equation}
with small $\epsilon$. Then using eq.~(\ref{gigi}), we have
\begin{eqnarray}
\Delta\langle  H_V \rangle=
\left< \psi\right|H_V\left|\psi\right>&\simeq& \epsilon
(\left< \phi\right| H_V\left|0\right>+ \left< 0\right| H_V\left|\phi\right>)\nonumber \\
&=&
\epsilon (\left< \phi\right| H_{\bV}\left|0\right>+
\left< 0\right| H_{\bV}\left|\phi\right>)\nonumber\\
&=&\Delta \langle H_{\bV} \rangle\,.\labell{nuni}
\end{eqnarray}
For example then, $\left|\phi\right>$ might be generated by creation operators
associated to wave packets concentrated inside the sphere. However, the above
equality indicates that there is also some energy density built outside the
sphere, to linear order in $\epsilon$.

Moreover, the relation (\ref{gigi}) is completely general, valid for the
modular Hamiltonian of any region. To see this note the vacuum state is a pure
state belonging to the Hilbert space ${\cal H}_V\otimes{\cal H}_{\bV}$, and
hence can be written in a Schmidt decomposition \cite{nielsen}
\begin{equation}
|0\rangle=\sum_i \sqrt{\lambda_i} \,\,|\psi^V_i\rangle\otimes |\psi^{\bV}_i\rangle \,.
\end{equation}
One readily checks doing the partial traces of this state that the
$|\psi^V_i\rangle$ are the eigenvectors of $\rho_V$ and $|\psi^{\bV}_i\rangle$
are those for $\rho_{\bV}$, while $\lambda_i$ are the common eigenvalues of
both density matrices. Then a simple calculation shows
\begin{equation}
(\rho_V)^{i \tau}\otimes (\rho_{\bV})^{-i \tau}|0\rangle=|0\rangle\,.
\end{equation}
These unitary operators leave the vacuum invariant for any
$\tau$.\footnote{These unitary operators implement  an evolution for an
internal time $\tau$. This time flow is called the modular flow \cite{haag}.}
Expanding for small $\tau$, and taking into account that $\rho_V\sim e^{-H_V}$,
$\rho_{\bV}\sim e^{-H_{\bV}}$, we obtain (\ref{gigi}) for any region. In the
limit which we are considering, where a perturbed state is approaching the
vacuum, there is no way to make a pure state with localized modular energy.
This guarantees once $\Delta\langle H_V\rangle=\Delta S_V$ we also have $\Delta
\langle H_{\bV}\rangle=\Delta S_{\bV}$ for any region and any pure state in
this approximation.

Of course, the latter also represents a restriction that applies for
holographic pure states in the AdS/CFT correspondence. However, this
observation has a limited utility in general because, as we noted before, the
precise form of the modular Hamiltonian is not know except in certain special
cases and in general, it is not even local (though it is generated by local fields inside $V$).
However, in section
\ref{discuss}, we will argue that for holographic CFT's dual to Einstein
gravity, expectation values of the  \modu Hamiltonian for any region, to first order in
 pure state deviations from the vacuum state, are in fact given by expressions linear in the
expectation value of the stress tensor. Hence in this case, the above observation becomes a constraint
on the localization of the stress energy in pure states for such a holographic
theory.

\subsection{An inequality for $\Delta \langle H\rangle$}

We can use the previous result to relate boundary data in the FG expansion with
the formation of horizons or singularities in the infrared region. Suppose as
before that we have a global state for which the stress energy inside of a
given sphere is small enough that $\Delta S_V=\Delta \langle H_V\rangle$.
Further, if this global state is pure, it is $S_V=S_{\bV}$. However, even if
the density matrix describing $\bV$ is not near the vacuum and there is no
equality between $\Delta \langle H_{\bV}\rangle$ and $\Delta S_{\bV}$, we still
have from relative entropy in $\bV$ that for a pure state
\begin{equation}
\Delta \langle H_{\bV}\rangle\ge \Delta \langle H_V\rangle\,,\label{aque}
\end{equation}
for any sphere $V$ with small stress tensor $\langle T_{\mu\nu}(x)\rangle$. If
this inequality is not maintained then either the state is impure or the
boundary data does not describe a consistent physical state. In particular, as
described above, we expect that the boundary data yields a full gravity
solution containing a naked singularity in the infrared region.

Let us consider further the case of an impure state, in which case we expect
that the bulk develops a horizon. The discussion in section \ref{two} provides
an explicit example of the following considerations. A state near the vacuum
inside the sphere $V$ gives again $\Delta S_V=\Delta \langle H_V\rangle$. In
this situation, we also know that generally $\Delta S_V\neq \Delta S_{\bV}$,
because for a global impure state the entropies of complementary regions do not
coincide. We may still ask what is the possible value of $\Delta S_{\bV}$.
In the vacuum, the minimal surfaces determining the entanglement entropy for
$V$ and $\bV$ coincide, yielding $S_{V}=S_{\bV}$. In the perturbed state,
$\Delta S_V$ comes from a small variation of this minimal surface. It seems
reasonable to expect that the minimal surfaces determining $S_{\bV}$ will
contain as one component the same (perturbed) minimal surface. Then this
surface would contribute the quantity $\Delta S_V$ to $\Delta S_{\bV}$.
However, there may also be a horizon contributing positively some $S_H$ to
$S_{\bV}$, which of course is not present in the vacuum entropy. In this case,
we have $\Delta S_{\bV}=\Delta S_{V}+S_H>\Delta S_{V}$. Thus, from the
positivity of the relative entropy applied to $\bV$, we again find that the
inequality (\ref{aque}) is satisfied. It is a logical possibility that the
extremal surface determining $S_{\bV}$ does not contain the minimal surface
yielding $S_V$. In this case, we would expect that there is again a horizon in
the interior between the two extremal surfaces preventing one from collapsing
to the other. We think this possibility is not probable if we are in the regime
where $\Delta S_V=\Delta \langle H_V\rangle$. The area of any such putative
minimal surface would be very large compared to the two component surface
comprised of the horizon and the surface in the asymptotic region near $V$.
However, even in this situation, the area of the minimal surface determining
$S_{\bV}$ would be much larger than that for $S_V$ and eq.~(\ref{aque}) would
still hold.

In conclusion, it seems that inequality (\ref{aque}) cannot be violated even
for impure states. Hence violations of this inequality should signal that the
boundary data appearing in the FG expansion does not correspond to a physical
state. Recall that the modular Hamiltonian for the interior and exterior of the
sphere are explicitly given in eqs.~\reef{sphereH} and \reef{sphereHx}. Hence
it is straightforward to explicitly evaluate $\Delta \langle H\rangle$ on both
sides of eq.~\reef{aque} and test this inequality.

It would be interesting to have a purely QFT understanding on why $\langle
T_{\mu\nu}\rangle$ not satisfying this inequality is unphysical. Returning the
QFT discussion above, eq.~(\ref{nuni}) need not apply in general because the
order $\epsilon^2$ terms are important in $\bV$. Instead one would have $\Delta
\langle H_{\bV} \rangle= \Delta\langle H_V \rangle+\epsilon^2 \langle \phi|
H_{\bV}|\phi\rangle$ and hence eq.~(\ref{aque}) demands that $\langle \phi|
H_{\bV}|\phi\rangle\ge0$. Examining $H_{\bV}$ in eq.~\reef{sphereHx}, this
inequality seems to indicate that CFT states in Minkowski space cannot support
a negative energy density over large regions, which certainly seems an
intuitive conclusion.

\subsection{Annular regions}

\begin{figure}
\centering
\leavevmode
\epsfxsize=12cm
\includegraphics[width=0.8\textwidth]{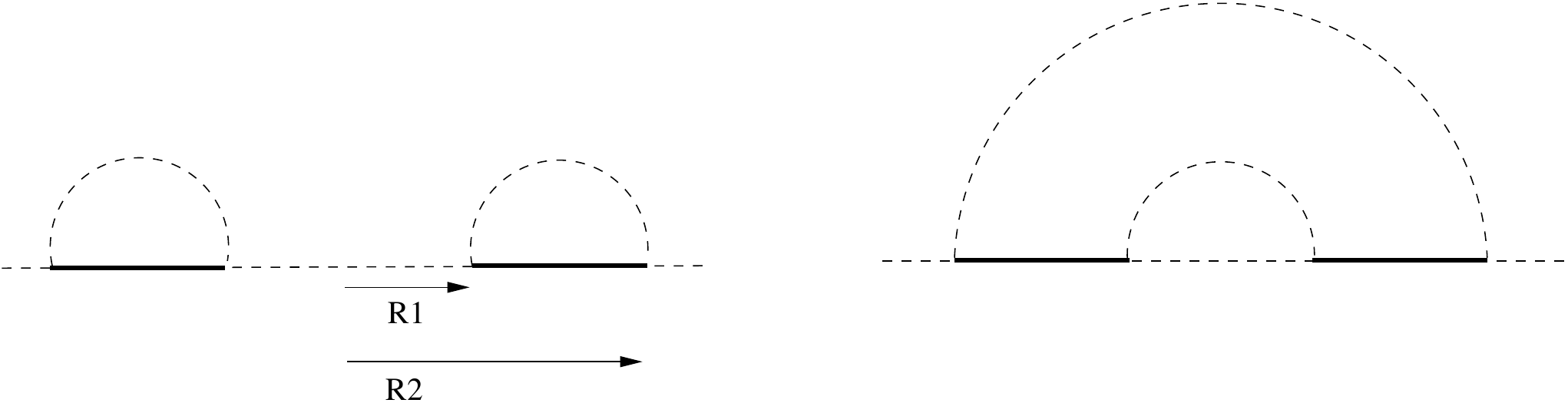}
\bigskip
\caption{The annular region on the AdS boundary is shown with the two solid lines. When
the radius $R_1$ and $R_2$ of the annulus approach  each other the minimal
surface has the shape of a half torus connecting the two spheres (left panel).
When $R_2/R_1$ is greater than a certain value the minimal surface is formed by
the two spherical caps ending at the spheres of radius $R_1$ and $R_2$ at the
boundary (right pannel).  } \labell{concentric}
\end{figure}
Consider now an annular region $A$ bounded by two concentric spheres with radii
$R_1<R_2$. We denote the regions within the two spheres as $V_1$ and $V_2$. In
the holographic context, depending on the ratio $R_2/R_1$ and the dimension
$d$, the minimal surface can have two different topologies. In one regime where
$R_2 \sim R_1$, the minimal surface has the shape of a half torus connecting
the two spheres in the asymptotic boundary. In the opposite regime where $R_2
\gg R_1$, the surface is formed by two separate spherical caps, each one
attached to one of the spheres on the boundary\footnote{Note that the inner
sphere of radius $R_1$ provides an example where a portion of the entangling
surface has $K^i<0$ and so, as discussed below eq.~\reef{expansion}, the
extremal surface in the bulk bends away from the interior of $A$. In fact, this
behaviour also persists in the regime where $R_2 \sim R_1$ and the minimal
surface has the topology of a half torus, as observed in \cite{prep}.} --- see
figure \ref{concentric}. We focus on the latter regime in the following.

Now if we turn on a small expectation value for $\langle T_{\mu\nu}\rangle$
using the FG expansion, we obtain a variation $\Delta S_A$ for the annulus
which is linear in $\langle T_{\mu\nu}\rangle$, and according to the general
arguments above, this variation will equal $\Delta \langle H_A\rangle$.
However, we note that $\Delta \langle H\rangle$ is the expectation value of an
operator with support entirely inside the annular region.

For the phase where the minimal surface has two disconnected components, one
attached to each spherical boundary of the annular region, we know the
contribution to $\Delta \langle H\rangle$ for any linear $\langle
T_{\mu\nu}\rangle$ exactly. The contribution from each cap can be evaluated
independently, and as shown in section \ref{line}, each of these contributions
satisfies $\Delta \langle H\rangle=\Delta S$. Hence we find
\begin{equation}
\Delta\langle H_A\rangle=\int_{|x|<R_1} d^{d-1}x\, \frac{R_1^2-r^2}{2 R_1} \,T_{00}(x)
+\int_{|x|<R_2} d^{d-1}x\, \frac{R_2^2-r^2}{2 R_2}\, T_{00}(x)\,.\labell{tyty}
\end{equation}
Clearly the expression on the right-hand side includes contributions of
$T_{\mu\nu}(x)$ from outside of $A$, \ie from $x<R_1$, inside the smaller
sphere. While we do not have a precise expression for $H_A$ at this point, we
re-iterate that it only has support inside of $A$. As such, eq.~\reef{tyty}
becomes a nonlocal constraint on small stress tensor perturbations of physical
states in the holographic framework.

We might also consider the exterior of the annulus, $\overline{A}=V_1\cup
\bV_2$. Working again in the regime with $R_2 \gg R_1$, holographic
entanglement entropy is again determined by the area of the two spherical caps
in the bulk. In this case, we would have
\begin{equation}
\Delta\langle H_{\bar{A}}\rangle=\int_{|x|<R_1} d^{d-1}x\, \frac{R_1^2-r^2}{2 R_1} \,T_{00}(x)
+\int_{|x|>R_2} d^{d-1}x\, \frac{r^2-R_2^2}{2 R_2}\, T_{00}(x)\,,\labell{tyty2}
\end{equation}
where the contributions come entirely from the region external to the annulus
or alternatively from within $\overline{A}$. Hence one might guess that
$H_{\overline{A}}=H_{V_1}+H_{\bV_2}$. In particular, this structure would yield
a density matrix with the product form $\rho_{\overline{A}}=\rho_{V_1}\otimes
\rho_{\bV_2}$.


\section{Discussion} \labell{discuss}

In this paper, we have examined relative entropy for some particular states and
entangling surfaces in the context of the AdS/CFT correspondence using the
standard prescription for holographic entanglement entropy \reef{define}. Our
results here constitute a strong test of this holographic entropy formula. A
notable case is the sphere for which we have shown in section \ref{line} by
direct calculation that holographic entanglement entropy yields the correct
entropy for any perturbation of the vacuum, to linear order.

It is remarkable the inequality \reef{123} expressing the positivity of the
relative entropy, is in fact saturated at leading order and so this equality
provides an equation that any first order deviations of holographic entropy
must satisfy. The equality $\Delta S=\Delta\langle H\rangle$ then becomes an
interesting tool. In fact, we can think of reversing the logic of our tests and
trying to obtain information about the modular Hamiltonian, or equivalently the
reduced density matrix, from the holographic entanglement entropy. In this
sense, the entanglement entropy has the potential to provide a full `vacuum
state tomography.' Let us recall that {\it any} pure perturbation of the vacuum can
be written as
\begin{equation}
 \left|\psi\right> =\left| 0\right>+\epsilon \left|\phi\right> \label{doa1}
\end{equation}
with some small $\epsilon$. Then the expected change in the entropy and modular
energy are
 \begin{equation}
 \Delta S=\Delta\langle H_V\rangle=\epsilon (\langle 0 | H_V | \phi \rangle+
 \langle \phi | H_V | 0 \rangle)\,.\label{72}
 \end{equation}
The knowledge of $\Delta S$ for any perturbation gives us the expectation
values on the right hand side. The knowledge of these expectation values for
any $|\phi\rangle$ and the fact that $H_V$ is an operator localized in the
region  $V$ imply  we can in principle reconstruct the full density matrix from
the entropy functional. To see this let us recall the expression for the
Schmidt decomposition of the vacuum: $|0\rangle=\sum_i \sqrt{\lambda_i}
\,\,|\psi^V_i\rangle\otimes |\psi^{\bV}_i\rangle$. In this basis, an arbitrary
global pure state writes
\begin{equation}
|\phi\rangle=\sum_{i,j} \beta_{ij} |\psi^V_i\rangle\otimes |\psi^{\bV}_j\rangle\,.
\end{equation}
Writing $\rho^0_V=e^{-H_V}$, with the particular normalization $\Tr\[
e^{-H_V}\]=1$, the modular Hamiltonian is simply
\begin{equation}
H_V=\sum_i -\log(\lambda_i )|\psi^V_i\rangle\langle \psi^V_i|\,.
\end{equation}
Then, after a little algebra, eq. (\ref{72}) gives
\begin{equation}
\Delta S=\epsilon\, \Tr \left( (\beta + \beta^\dagger) H_V \, e^{-H_V/2}\right)\,.
\end{equation}
If $\Delta S$ is known for any $\phi$, represented by the arbitrary matrix
$\beta$ in this equation, we can obtain the matrix $H_Ve^{-H_V/2}$ as a
solution of a set of linear equations.\footnote{Here all the eigenvalues $\lambda_i$
 are assumed to be different from zero, or equivalently the density matrix $\rho_V$ has inverse.
  Otherwise the modular Hamiltonian has infinite coefficient for $| \psi_i^V\rangle\langle\psi_i^V |$, and hence is undefined under finite additions of this projector. This ambiguity for $H_V$ is seen clearly from \ref{72} since in this case $\langle \phi_i^V|0\rangle=0$ and then additions of $| \psi_i^V\rangle\langle\psi_i^V |$ in $H_V$ do not change $\Delta S$. In QFT these local excitations in $V$ completely orthogonal to the vaccum are not allowed as a consequence of the Reeh-Slieder theorem \cite{Reeh}.}
  In other words, there is a unique
operator in $V$ such that all linear entropy perturbations for pure state
deformations coincide with the value of $\Delta \langle H_V\rangle$.

In principle, this idea allows us to reconstruct the full density matrix of a
region based only in the entanglement entropy functional. In particular then,
in the context of the AdS/CFT correspondence, it seems that the latter is
readily accessible using the standard holographic prescription \reef{define}.
For example, based on our results in section \ref{general}, we can reconstruct
the full modular Hamiltonian operator for the vacuum reduced to the region
within a sphere and the result coincides precisely with the standard expression
\reef{sphereH} for a CFT. In order to show this, we note we are doing an
experiment devised to produce pure deviations of the vacuum as in
eq.~(\ref{doa1}) in the boundary theory. In the AdS/CFT context, this
excitation is translated to the bulk language by the effect which it has on
expectation values of operators. We can say the excitation will be defined by a
series of expectation values for certain operators on the boundary which are
turned on linearly in $\epsilon$,
  \begin{equation}
\Delta \langle {\cal O}\rangle\simeq \langle \psi| {\cal O}|\psi\rangle-\langle 0|
{\cal O}|0\rangle=\epsilon \left(\langle \phi|{\cal O}|0\rangle+\langle 0|{\cal O}|\phi\rangle\right)\,.
  \end{equation}
Hence, the reasoning which leads to the vacuum state tomography for the sphere
is given by the following steps:
\begin{enumerate}
 \item Since the holographic prescription \reef{define} yields the
     entanglement entropy purely in terms of the geometry, linear (order
     $\epsilon$) perturbations in the entropy should depend only on linear
     (order $\epsilon$) perturbations of the bulk metric.
 \item  Linear perturbations in the bulk metric can be separated as having
     two different origins. The first one is due to order $\epsilon$ terms
     in the boundary data giving the expectation value of the stress
     tensor, which modifies the boundary conditions. The second one is due
     to perturbations of the bulk stress tensor which are also linear in
     $\epsilon$. The latter modify the source of Einstein equations.
 \item  The bulk stress tensor gets corrections from deviations of bulk
     matter fields and these are in turn related to the boundary data for
     expectation values of dual operators. However, the bulk stress tensor
     is quadratic in the matter fields and so does not yield corrections
     linear in $\epsilon$ in the vacuum.\footnote{We stress that this is
     only for variations around the vacuum, where the bulk stress tensor is
     given by the cosmological constant alone. For other states, it is
     clear there will be linear terms in expectation values of other
     operators, \ie. charge density operator for a state with non zero
     chemical potential. The argument fails in this case because the change
     of order $\epsilon$ in the stress tensor is infinitesimal with respect
     to the stress tensor for $\epsilon =0$, which is non zero for non
     vacuum states. Hence we can change the sign of $\epsilon$ without
     implying a failure of null energy condition.} Alternatively, one can
     argue the absence of order $\epsilon$ corrections because if they did
     exist, changing the sign of $\epsilon$ would lead to an unphysical
     bulk stress tensor, \ie not satisfying the null energy conditions.
 \item  Hence, correction to $S$ linear in $\epsilon$ can only depend on
     the linear perturbations of the boundary stress tensor  for the
     minimal surface of {\sl any region}.
 \item  For the case of the sphere, we have shown the linear terms on
     $\Delta S$ coincide with the ones in the expectation value of the
     operator $H=2 \pi\int dx^{d-1}\, \frac{R^2-r^2}{2 R} T_{00}(x)$.
 \item  This operator $H$ is localized inside the sphere (\ie belongs to
     the algebra of operators generated by local fields on the sphere).
     Hence, it is the unique operator in the sphere which does the job of
     satisfying the equation $\Delta \langle H\rangle=\Delta S$ for any
     pure deviation of the vacuum to linear order in $\epsilon$. Hence it
     is the modular Hamiltonian of the sphere.
\end{enumerate}

It is interesting to see what obstacles arise to reconstructing the modular
Hamiltonian for other regions. As above, we have that for a general region
$\Delta S$ is linear in $T_{\mu\nu}(x)$ to first order in $\epsilon$. However,
here we find two related problems. First consider the case of a minimal surface
corresponding to a region $V$ which in terms of the FG coordinate $z$ is
single-valued, that is, for any $x\in V$ we have a unique $z(x)$ describing the
surface (\ie the one corresponding to an ellipsoidal region with $K^i>0$
everywhere along the entangling surface). Using the FG expansion, the
contribution to $\Delta S$ in eq.~(\ref{linearA2}) involves time and spatial
derivatives of $T_{\mu\nu}(x)$ of arbitrarily high order. Even if the spatial
derivatives can be eliminated by integration by parts, the time derivatives
remain. In section \ref{line}, it was a surprising result that for the sphere
these time derivatives finally disappear from the final expression. However,
the result for a general surface cannot be considered as the expectation value of
an operator localized in $V$, because even if it depends only of
$T_{\mu\nu}(x)$ for $x\in V$, it depends on arbitrary derivatives of the stress
tensor. For a minimal surface extending outside $V$, such as the annulus
discussed in section \ref{puzzle}, this nonlocality of the contribution is seen
more directly since $\Delta S$ involves $T_{\mu\nu}$ outside $V$. In fact,
these two types of nonlocality can be put on the same basis by writing the
variation of the metric tensor in terms of the boundary-to-bulk Green's
function in coordinate representation. That is, we can write eqs.~(\ref{FGexp})
and (\ref{box}) as
\begin{equation}
\delta g_{\mu\nu}(y,z)= \int d^{d} x\,\, G(x-y,z)\, T_{\mu\nu}(x)\,,
 \labell{fof}
\end{equation}
where $G(x-y,z)$ is proportional to the Fourier transform of the Green's
function of section \ref{line} in momentum space,\footnote{The integration is
over time-like momentum since only time-like momentum appears in the
contributions $\langle 0|T_{\mu\nu}(x)|\psi\rangle+\langle
\psi|T_{\mu\nu}(x)|0\rangle$ because physical states have momentum inside the
light cone.}
\begin{equation}
G(x-y,z)=\frac{l_p^{d-1}}{d L^{d-1} 2^{d/2-1}\Gamma[d/2+1]}\int
\frac{d^dp}{(2\pi)^d}\,\, \theta(-p^2)\, \frac{z^{d/2}}{p^{d/2}}\,
 J_{d/2}(|p| z) e^{-i p x}\,.\labell{greengauge}
\end{equation}
Hence, generically, because  this Green's function is not of compact support,
the contribution of $T_{\mu \nu}(x)$ for any spacetime point $x$ will not
vanish in the variation of the area for a given minimal surface. This is so
unless some conspiracy between the particular minimal surface and the
tracelessness and conservation of $T_{\mu\nu}$ occurs. This is the case of the
sphere, where the contribution is localized inside $V$, but we do not expect
the latter property to extend to the case of general surfaces.

The question is how is this possible. The expectation value of the modular
Hamiltonian should be localized in $V$. The answer to this apparent
contradictions has to reside in the fact that the expression of the
result for $\Delta S$ in terms of operator expectation values suffer from
 two different types of ambiguities.
First, we do not have full control on which
perturbations for $T_{\mu\nu}$ are generated by genuine pure deviations from
the vacuum, as discussed in section \ref{puzzle}. Some of the constraints we
know, for example, the expectation value of $T_{\mu\nu}$ has to satisfy $\Delta
\langle H_V\rangle=\Delta \langle H_{\overline V}\rangle$ for any sphere. The
same equality holds for the (unknown) modular Hamiltonian for any general
region. The second source of ambiguities is due to the fact that since operators
obey time evolution laws the expectation values at different times could in principle
 be rewritten as expectation values for other operators at a single time.
Hence, is its natural to suppose that the expression of $\Delta S$ for
a general surface, given in terms of $T_{\mu\nu}$ in all spacetime, could be
converted into one of some other operator inside $V$ once these constraints are
fully understood.

For the case of two (or more) well separated spheres $A$ and $B$, where the
minimal surface consists of the separate minimal surfaces for the spheres, it
is evident $\Delta S$ is also the sum of $\Delta S$ for both spheres
separately. In this case, one has that the modular Hamiltonian can be
reconstructed again, and coincides with the sum of the those for separated
spheres, because $\Delta S$ depends on $T_{\mu\nu}$ inside the region $A\cup B$
only. This is consistent with the mutual information $I(A,B)=S(A)+S(B)-S(A\cup
B)$ being zero in this case. Mutual information is an upper bound to connected
correlators, and if it is zero it means correlators of operators in $A$ and $B$
factor out, to leading order in large $N$ (or large central
charge).\footnote{It is worthwhile to note that our discussion of the
reconstruction of the \modu Hamiltonian for holographic theories relies on the
geometric prescription \reef{define} to calculate the entanglement entropy. Of
course, this formula is only expected to yield the leading contribution in an
expansion of large central charge. In principle, this limitation represents
another obstacle to recovering the full \modu Hamiltonian.}

It is intriguing that Renyi entropies \cite{renyi0,karol} for these same
configurations do not separate into the contributions of $A$ and $B$ to this
same order in $N$, for $d=2$ \cite{head,twod} or higher dimensions
\cite{twistop}. However, recall that Renyi entropies are given by
$S_n=(1-n)^{-1}\log\[\tr\rho^n\]$ and so depend on powers of the density
matrix. These powers are very different from the density matrix itself, as well
as, very different from any finite energy density state in the region. In
contrast, the entanglement entropy is the limit $n\rightarrow 1$ and hence only
feels states near to $\rho$. Let us look at a simpler case which, while it is
quite different to the specific situation we are considering above, still
exemplifies the relevant ideas. Hence we think of a global thermal state with
$\rho=e^{-\cH/T}/Z$ where $\cH$ is the standard Hamiltonian. Then Renyi
entropies are quantities related to states at different temperatures. Now it is
always possible to have a phase transition at some critical temperature where
the $N$ dependence of various physical quantities changes, \eg section
\ref{two} described an example where the energy density suffers a phase
transition with different dependence on $N$.\footnote{In the case of two
decoupled regions in $d=2$, Renyi entropies for integer $n>1$ do not decouple.
This corresponds to lower ``internal'' temperature for the region. Hence, the
phase transition is better described as a screening phase transition, where the
entropy, corresponding to a state of higher internal temperature (the vacuum
state in $A\cup B$ here), does not see correlations, while Renyi entropies
detect these correlations at lower internal temperatures. We owe this
observation to Hong Liu.} It follows then that the corresponding Renyi
entropies exhibit the same phase transition, since changing the Renyi order $n$
and the temperature $T$ are the same thing in this case. However, we can also
expect similar behaviour for the Renyi entropies in a more general context. In
particular, our discussion above indicates that the effective modular
Hamiltonian, which is the one relevant for computation of correlation
functions, decouples for the two well separated spheres. It does not include
all information on higher order $N$ corrections, which must still play an
important role in determining the Renyi entropies.

While in general the modular Hamiltonian for a generic region is not a local
operator, one expects that very close to the entangling surface of any region
$V$, $H_V$ will approach the simple local form given in eq.~\reef{siete} for
Rindler space. We readily see this behaviour in eq.~\reef{sphereH} as we
approach $r\sim R$ for the spherical entangling surface. However, the local
Rindler expression should be the leading contribution in the \modu Hamiltonian
independent of the shape of the surface. One approach \cite{rocket} to
understanding this general result is this Rindler term provides short distance
part of $\rho_V$ that encodes the correlators in the vicinity of the boundary
of the causal domain defined by the entangling surface and in the UV, these
correlators have the same structure as in flat space, \ie in the vicinity of a
Rindler horizon. Alternatively, as alluded to in various points in our
discussion, one can think of the Rindler Hamiltonian as defining a thermal
density matrix with a local effective temperature of $T_{\textrm{eff}}=1/(2\pi
x)$ where $x$ is the (orthogonal) distance to the entangling surface. Hence as
we approach this boundary, the effective temperature diverges and this Rindler
term overwhelms any other fixed contributions to the density matrix. Hence
along the lines of our discussion of vacuum state tomography, we might attempt
to verify the appearance of a Rindler-like contribution in the present
holographic setting. In particular, we are thinking here of evaluating $\Delta
S$ for perturbations localized near the boundary of the region. This
independence of shape for the contribution of these localized sources should
then be associated to the surface being {\sl minimal} --- remember in this
calculation the contribution would depend on a localized $\delta g_{\mu\nu}$
near the boundary and the dependence on the variation of the surface shape far
from the source would not contribute precisely because it is minimal. However,
the results of section \ref{generic} suggest that to have a sufficiently
localized $\delta g_{\mu\nu}$ we might need to choose a gauge for the
boundary-to-bulk Green's function which is different from the one in
eq.~(\ref{greengauge}). This reasoning would then lead us closer to a purely
thermodynamic understanding of the standard prescription \reef{define} for
holographic entanglement entropy.

In our previous discussion, we argued that the expectation values of operators
other than $T_{\mu\nu}$ only appear quadratically in $\Delta S$ and one form of
this argument relies on the null energy condition of the bulk stress tensor.
Further, these quadratic order contributions must be negative in order to
preserve the positivity of the relative entropy. It seems natural the sign of
these contributions could be directly related to the null energy condition. In
fact, for the sake of the argument, we can think directly in terms of the
change of entropy due to the bulk stress tensor perturbation. This encodes all
the information from the expectation value of fields at the boundary which is
relevant for the calculation of $\Delta S$. The variation of the metric due to
perturbations on the bulk stress tensor can be written with an expression
similar to eq.~(\ref{fof}), but where now the integration is over bulk
spacetime, the boundary stress tensor is replaced by the perturbation of the
bulk stress tensor, and the Green's function is the bulk-to-bulk Green's
function \cite{bulkbulk}. It would be interesting to find out if null energy
condition alone can ensure a definite sign for this contribution of $\delta
g_{\mu\nu}$ to the change in the area of the minimal surface in a general
situation. In other words, the area of the minimal surface of any boundary
region $V$ in presence of a bulk $T_{\mu\nu}$ satisfying the null energy
condition has to be smaller than the one corresponding to $V$ in pure AdS
spacetime.

Bekenstein argued that all systems must satisfy an inequality of the form
\cite{beke0}
\begin{equation}
S\le 2\pi R\, E \,,\labell{bek9}
\end{equation}
where $S$ and $E$ are the entropy and energy of a system confined to a region
of size $R$ --- see appendix \ref{bound}. While this bound was originally
derived with a thought experiment involving dropping an object into a black
hole, eq.~\reef{bek9} does not involve Newton's constant and so it should be
possible to understand the bound entirely in terms of flat space physics.
Unfortunately, as presented, all of the physical quantities in eq.~\reef{bek9}
are ambiguous. However, these ambiguities can be eliminated by re-interpreting
the bound in terms of the inequality \reef{123} expressing the positivity of
the relative entropy \cite{beke1,beke2} \ie $\Delta S\le \Delta \langle
H\rangle$. As described in the appendix, one can apply eq.~\reef{123} in
Bekenstein's thought experiment where the region of interest is Rindler space
and the result is precisely the inequality in eq.~\reef{bek9}. Relating
eq.~\reef{bek9} to relative entropy makes clear that the physics behind the
Bekenstein bound is simply quantum mechanics and special relativity.

Of course, to make progress with this
approach, we must know the modular Hamiltonian for a given situation.
Therefore, let us turn to the example of a spherical entangling surface for a
CFT in which case the modular Hamiltonian is given by eq.~\reef{sphereH}. One
observation is that if the bound is expressed in terms of the total energy
enclosed, as in eq.~\reef{bek9}, then the precise bound depends very much on
how the energy is deposited within the sphere. For example, for a smooth
distribution of energy, analogous to those considered in section \reef{simple},
one finds $\Delta S\le \frac{2 \pi}{d+1} R\,E$ while if the energy is localized
near the center of the ball enclosed by the sphere $\Delta S\le \pi R\,E$. Both
of these inequalities have the same form as that in eq.~\reef{bek9} and only
the overall numerical factor changes on the right-hand side. A more dramatic
change arises if the energy is deposited in a spherical shell of roughly radius
$R$ and width $w$ with $w\ll R$. In this case, eq.~\reef{123} becomes $\Delta
S\le 2\pi w\,E$ and so the relevant length scale that emerges here is, in fact,
the width of the shell. This behaviour is reminiscent of the result in
\cite{boussoa}, where it was argued that the Bekenstein bound is controlled by
the shortest dimension (rather than the largest) for matter confined to an
elongated region. Of course, the discussion there relied on considerations of how
the weakly gravitating matter focussed light rays passing through the region. A
similar result can be inferred from our holographic calculations for the strip
geometry in section \ref{generic}. To linear order where eq.~\reef{123} is
saturated, we find for a smooth energy distribution that $\Delta\langle
H\rangle \propto \ell\,E$ where $\ell$ is the width of the strip. Hence again
it appears that the shortest distance sets the geometric scale for the
Bekenstein bound.

Of course, the example of the strip reminds us that in general the `modular
energy' in eq.~\reef{123} can be quite dissimilar to the energy appearing in
eq.~\reef{bek9}. Our holographic result for $\Delta S$ in eq.~\reef{linearA3}
shows that the pressure $T_{xx}$ appears on a more or less equal footing with
the energy density $T_{00}$. Hence using the saturation of eq.~\reef{123}, we
expect for homogenous (CFT) matter distributions that the bound will be set by
\begin{equation}
\Delta \langle H\rangle \simeq  \ell\,V\,
\left[\frac{d+1}{d-1}\,T_{00} - T_{xx}\right]\,,\label{quant}
 \end{equation}
where $V=B^{d-2}\ell$ is the volume of the strip. The resulting bound is
qualitatively different from the Bekenstein bound in eq.~\reef{bek9} since we
can not expect the quantity in eq.~\reef{quant} to be proportional to the
energy in the strip. In principle, for quantum matter, $T_{00}$ and $T_{xx}$ do
not need to satisfy any relation and are not even constrained by classical
energy conditions. So the bound set by eq.~(\ref{quant}) can be much more (or
less) constraining than a bound set by $T_{00}$ alone. In particular, if one
could realize $T_{xx}\simeq \frac{d+1}{d-1} T_{00}$, we would have the
interesting conclusion that $\Delta S\le 0$, \ie the entropy in the perturbed
state has to be smaller than that in the vacuum state. Of course, these results
are symptomatic of the fact that in general, the modular Hamiltonian will
contain contributions involving operators other than the energy density and in
fact, operators unrelated to the stress-energy tensor. Hence the bounds set by
eq.~\reef{123} will generically be far more complicated than the simple
expression appearing in eq.~\reef{bek9}. Further we must add that although the
interpretation of the Bekenstein bound in terms of eq.~\reef{123} gives a
general prescription which is free of ambiguities, unfortunately, without a
clear understanding of the modular Hamiltonian for a given situation, this
interpretation is left somewhat lacking.

Some recent references \cite{tak1,bianchi} also consider relations similar to
the first law of thermodynamics, \ie $dE=T\, dS$, for entanglement entropy ---
see also \cite{tak31,newest}. In particular, the discussion in \cite{tak1}
centers on the proportionality between the energy within a small region and the
entanglement entropy of the same region, which is seen in several examples. In
the present paper, we have seen the origin of this proportionality is the
equation $\Delta \langle H\rangle=\Delta S$. However, we must again remark that
it is in general a different `type' of energy, the modular `energy,' that
enters into a proper definition of the equation. For example, if this is not
taken into account, the proportionality factor between energy and entropy for a
spherical entangling surface depends on the distribution of the energy inside
it --- as was already observed in \cite{tak1}. For more general (\ie
non-spherical) geometries, $\Delta S$ is simply not proportional to the energy,
but rather other operators will appear in the \modu Hamiltonian and in the
expression for $\Delta \langle H\rangle$.

\vskip 1cm

\section*{Acknowledgements:} We would like to thank Eugenio Bianchi, Patrick
Hayden, Marina Huerta, Andreas Karch, Hong Liu, Masahiro Nozaki and Rafael
Sorkin for discussions. Research at Perimeter Institute is supported by the
Government of Canada through Industry Canada and by the Province of Ontario
through the Ministry of Research \& Innovation. RCM also acknowledges support
from an NSERC Discovery grant and funding from the Canadian Institute for
Advanced Research. Work by DDB and HC was supported by CONICET, Universidad
Nacional de Cuyo, and CNEA, Argentina. LYH is supported by the Croucher
Foundation.

\appendix

\section{Comments on Relative Entropy} \labell{review}

Relative entropy provides a precise measure of the statistical distance between
two states. Given a state $\rho_1$,  the probability of confounding it with
$\rho_0$ after $n$ trials of some  measurement is asymptotically exponentially
decreasing for large $n$ as
\begin{equation}
e^{-n S(\rho_1|\rho_0)}\,.\labell{sam}
\end{equation}
In this sense. relative entropy is commonly thought as a measure of the
distinguishability between states \cite{vedral}.

As mentioned, relative entropy is positive and increasing with system size,
\begin{eqnarray}
&& S(\rho^V_1|\rho^V_0)\ge 0\,,\\
&& S(\rho_1^V|\rho_0^V)\le S(\rho_1^W|\rho_0^W)\,,\hspace{2cm} V\subseteq W\,.\labell{mon}
\end{eqnarray}
The monotonicity property (\ref{mon}) is a particular case of monotonicity
under general completely positive trace preserving maps (CPTP). These are
linear maps of density matrices in one space into density matrices in another
one, which are physical in the sense they are combinations of operations such
as unitary evolution, partial tracing and enlarging the system with a new
subsystem. The general expression of a CPTP map is \cite{nielsen}
\begin{equation}
\rho^\prime=\sum_i M_i \rho M_i^\dagger\,,\hspace{2cm} \sum_i M_i^\dagger M_i=I\,,
\end{equation}
for matrices $M_i$ with arbitrary dimension, \ie not necessarily square
matrices. Then, more generally the relative entropy satisfies
\begin{equation}
S(\rho_1|\rho_0)\ge S(\rho_1^\prime|\rho_0^\prime)\,.
\end{equation}
The partial trace over a subsystem as in (\ref{mon}) is one example of CPTP
map. Such CPTP maps then generally entail the loss of distinguishability
between states.

\subsection{$\Delta S=\Delta \langle H\rangle$ for first order perturbations}\labell{equal}

Recall that the relative entropy only vanishes for identical states. Here we
expand on the discussion around eq.~\reef{eeex} to see what to expect for the
relative entropy of nearby states. Keeping our reference state $\rho_0$ fixed,
we move through a family of states $\rho_1(\lambda)$ with a parameter $\lambda$
such that $\rho_1(\lambda=0)=\rho_0$, \ie the states coincide for $\lambda=0$.
Hence we have that $S(\rho_1(0)|\rho_2)=0$ but $S(\rho_1(\lambda)|\rho_2)>0$
for both $\lambda>0$ and $\lambda<0$. Therefore assuming that
$S(\rho_1(\lambda)|\rho_2)$ describes a smooth curve, it must have zero first
derivative at $\lambda=0$. This then implies
\begin{equation}
\Delta S=\Delta \langle H\rangle\labell{eee}
\end{equation}
to first order in $\lambda$ at $\lambda=0$. For nearby thermal equilibrium
states, this relation is just the well known thermodynamic equation $\Delta S =
\Delta E/T$.

Another way to see the above equality is to evaluate the first order
perturbation of $S(\rho)$ and $H(\rho)$ for a density matrix
 \be
\rho=\frac{e^{-(H+\delta H)}}{\textrm{tr}(e^{-(H+\delta H)})}\,.
 \ee
Then to linear order in $\delta H$, we have that both coincide with
\begin{equation}
\Delta S=\Delta \langle H\rangle=\frac{\textrm{tr}(e^{-H} H )\textrm{tr}(e^{-H}
\delta H )}{(\textrm{tr}(e^{-H} ))^2}-\frac{\textrm{tr}(e^{-H} H \delta H )}{
\textrm{tr}(e^{-H} )}=\langle H\rangle\langle\delta H\rangle-\langle H \delta
H\rangle\,,\labell{deviation}
\end{equation}
where in the last expression the expectation values are computed with the
unperturbed density matrix. In deriving eq.~(\ref{deviation}), we have treated
$\delta H$ as a numerical perturbation rather than as an operator. This
approach is justified here because we are manipulating the operators under the
trace and taking only terms which are functions of $H$ with only a single
operator $\delta H$. Hence it is not necessary to keep track of the ordering of
operators.

However, this formula assumes the perturbation of $\rho$ is small with respect
to $\rho$. At this point, we have to be careful in QFT because density matrices
have an infinite number of eigenvalues, which have to suffer small deviations.
For example, inserting a pure particle excitation, which is well localized
inside the bulk of a large region $A$, should not change very much the entropy
with respect to the vacuum state. In particular, as the particle is far from
the boundary $\partial A$, where most of the entanglement is produced, the
entropy should be approximately the same as in the vacuum state. However
$\Delta \langle H\rangle$ will measure the energy of the particle wave packet.
Of course, the reason for the discrepancy between $\Delta S$ and $\Delta
\langle H\rangle$ in this case is that the particle state never approaches the
vacuum state while the distance $R$ between the wave packet and the boundary of
the region is greater than the wavelength $\lambda$ of the wave packet. In
fact, the global state with the particle excitation is always orthogonal to the
global vacuum and we expect the relative entropy to increase to infinity in the
limit of large $R/\lambda$, corresponding to perfect distinguishability.
Further, due to the uncertainty relations the energy of the particle scales as
$1/\lambda$ and $\Delta \langle H\rangle \sim R/\lambda$.

We can formulate the following intuitive picture as to when the equality
(\ref{eee}) is applicable. Near the boundary of a region, the density matrices
will have a Rindler form \reef{siete}, which suggests a thermal interpretation
in the sense of Unruh \cite{Unruh}. In particular, there is a high temperature
near the boundary and the temperature decreases with $1/x$ as we move into the
bulk of the region, where $x$ is the distance to the boundary. For a finite
region of size $R$ then, there is a minimal temperature $T\sim 1/R$
\cite{rovelli}. Now we want to change the state by adding some perturbation.
Suppose then that we have a thermal state and mix it with a state $|E\rangle
\langle E|$ of energy $E$ with some small probability $p$. In order that the
change in eigenvalues is small, we must take $p\ll e^{-\beta E}/Z=p_E$, \ie $p$
must be smaller than the probability with which the same state appears in the
thermal ensemble. The latter is always be achieved if the change in energy is
smaller than the typical average energy for the same state in the thermal bath.
Hence in our original problem, we require that the energy density deposited at
a location, where the local temperature is roughly $T(x)$, must be much smaller
that $T^{d}$. Then the change in the entropy satisfies $\Delta S\sim \Delta
E/T\ll 1$ and we are perturbing the thermal bath by our thermodynamical analogy.
Otherwise, the injection of excitations in the region produces a
far-from-equilibrium state.

The conclusion is that we can probe the equality (\ref{eee}) for compact
regions with any state in the limit of small stress tensor expectation value.
We can have small energy density perturbations inside $A$ by taking an
admixture (pure or impure) of the wave packet with the vacuum. For example
$\vert 0 \rangle+\epsilon \vert \phi\rangle$ for small $\epsilon$. In this case
we can make the energy density of the state as low as we want without requiring
the state to be of large wavelength.

\subsection{Strong subadditivity}

Mutual information $I(A,B)=S(A)+S(B)-S(AB)$ between two subsystems $A$ and $B$ is a measure
of the information shared by them. It can be written as a
particular relative entropy,
 \be I(A,B)=S(\rho_{AB}|\rho_A\otimes \rho_B)\,.\ee
Mutual information is positive and increasing with size as a consequence of the
positivity and monotonicity of relative entropy.  The monotonicity of the
mutual information gives
 \be I(A,BC)-I(A,B)=S(AB)+S(BC)-S(ABC)-S(B)\ge 0\,. \ee
Then, strong subadditivity, which is the last inequality, is implied by
monotonicity of relative entropy. Note that using other properties of the
entropy, one can also prove the monotonicity of relative entropy starting from
strong subadditivity \cite{wehrl}. However, the monotonicity of relative
entropy that we discuss in this paper does not reduce to strong subadditivity
of entropy for different regions in space.\footnote{The latter has been
discussed previously in the AdS/CFT context by \cite{headrick}.} Instead, if
written in terms of strong subadditivity, it would involve a different kind of
partition of the global Hilbert space, combined with the property that the
entropy is concave.

\subsection{Second law of thermodynamics} \labell{secondlaw}

The oldest physical interpretation of the positivity of the relative entropy
$S(\rho_1|\rho_0)$ is in terms of thermodynamics. As we described in the
introduction, if $\rho_0$ is the equilibrium state at temperature $T$, then the
relative entropy takes the form $S(\rho_1|\rho_0)=(F(\rho_1)-F(\rho_0)/T)$,
where $F(\rho)=\textrm{tr} (\rho E )- T S(\rho)$ is the free energy evaluated
for a general state $\rho$ but at a fixed temperature $T$. Hence, the
positivity of relative entropy has the meaning that the free energy is minimal
for the equilibrium state.

The thermodynamical version of this inequality is now a consequence of the
second law. In general, for a system held in contact with a thermal bath at
temperature $T$, the second law implies that the following the inequality holds
in any process:
\begin{equation}
\delta F\le \delta W\,,
\end{equation}
where $\delta W$ is the work done on the system. Hence it must be that for a
spontaneous transformation, in which no work is done, one must have $\delta
F\le 0$. That is, the free energy must decrease as the system evolves towards
equilibrium.

The second law can be proved using  properties of the relative entropy under
certain assumptions for the quantum time evolution \cite{vedral,paper}. We also
note that relative entropy inequalities have been applied  to prove the
generalized second law in the context of black hole evaporation
\cite{sorkin1,wall}.

In these proofs the second law is related to a generalized monotonicity
property: The relative entropy always decreases under completely positive trace
preserving (CPTP) maps between states. The CPTP maps are thought as very
general class physical quantum evolutions of states \cite{nielsen}. For
example, the evolution a subsystem which is initially decoupled from the rest,
and where the global system undergoes unitary evolution, is CPTP.

The second law states that the entropy of an isolated system cannot decrease.
Of course a completely isolated system in quantum mechanics evolves unitarily
and the entropy does not change. We have to soften the condition of being
completely isolated in order to allow for some interchange of information with
the ambient space. As a model for this evolution consider the case of a quantum
system with state $\rho(t)$ evolving under  CPTP maps. Assume, in accordance
with the idea of an ``isolated'' system, that the total energy $E$ is
conserved. Also assume that time evolution preserves the thermal equilibrium
state\footnote{In fact it is only necessary that there is a state such that its
entropy and energies are preserved.} $\rho_T=e^{-H/T}/\textrm{tr}(e^{-H/T})$ at
some temperature $T$, which corresponds to the conserved energy $E$,
$\textrm{tr}(\rho_T H)=E$.

Then the relative entropy $S(\rho(t),\rho_T)$ is decreased by the CPTP
evolution, and we have for $t_1<t_2$,
\begin{equation}
F(\rho(t_2))-F(\rho_T)<F(\rho(t_1))-F(\rho_T)\,,
\end{equation}
where we used that the thermal state is invariant under time evolution.
Expressing this relation in terms of entropy and energy, and considering all
the involved energies are the same by assumption, we have
\begin{equation}
S(t_2)>S(t_1)\,,
\end{equation}
as required by the second law of thermodynamics.  Note that the difference in
free energies between the state and the thermal state is positive and decreases
in time. As a consequence, the state approaches the thermal equilibrium state
during evolution. Eventually, if thermal equilibrium is reached, this free
energy difference goes to zero.

Another case where the relative entropy allows one to prove the second law is
when the totally random state $\rho_0=I/n$, where $n$ is the dimension of the
Hilbert space, is preserved under a CPTP evolution. This state can be regarded
as the microcanonical distribution. The second law follows from the fact that
the relative entropy is in this case
\begin{equation}
S(\rho(t)|\rho_0)=\log(n)-S(\rho(t))\,.
\end{equation}
The increase in entropy then follows again by the decrease of relative entropy.

\subsection{Bekenstein bound} \labell{bound}

The Bekenstein bound \cite{beke0} is a proposal that all systems in nature
should satisfy an inequality of the form
\begin{equation}
S\le 2\pi R\, E \,,\labell{bek}
\end{equation}
where $S$ and $E$ are the entropy and energy of a system confined to a region
of size $R$. This proposed bound follows from considerations of thought
experiments involving black holes. However eq.~\reef{bek} does not involve
Newton's constant and thus it should express a general property that even
applies outside of the context of gravity. In particular, it should be possible
to understand eq.~\reef{bek} purely in terms of flat space physics. While this
inequality appears to have a rather simple form, discussions of its possible
validity, \eg \cite{beke2,boussoa,bent,donrad}, revealed a variety of
subtleties in interpreting the various quantities appearing in eq.~\reef{bek}.
Eventually, it was realized that a well defined version of this bound in QFT is
given by the positivity of the relative entropy between two states reduced to a
given region \cite{beke1,beke2}. The connection between relative entropy and
the Bekenstein bound is essentially established by eq.~\reef{123}.

To better understand this connection between relative entropy and the
Bekenstein bound, let us re-visit Bekenstein's original thought experiment
\cite{beke0}. Imagine that a small probe is released to fall into a large black
hole, from a short distance $R$ above the horizon.\footnote{Implicitly, we
assume that the size $\lambda$ of the probe is smaller than the original
distance above the horizon, \ie $R\gtrsim \lambda$.} The object then disappears
behind the horizon carrying entropy $S$ and energy $E$, as measured by a local
observer at the point from which it was released. The energy swallowed by the
black hole as measured asymptotically is red-shifted to $E\,
T_\mt{BH}/T_\mt{rel}\simeq 2 \pi R\, E\, T_\mt{BH}$, where $T_\mt{BH}$ and
$T_\mt{rel}$ are the Hawking temperature measured at infinity and the local
temperature measure at the release radius, respectively. Hence the variation of
the black hole's mass is $\delta M=2 \pi R\,E\, T_{BH}$ and the corresponding
variation in the horizon entropy is given by $\delta S_\mt{BH}=\delta
M/T_\mt{BH}=2 \pi R\,E$. Finally the generalized second law demands that the
increase in the horizon entropy must at least compensate for the loss of
entropy in the exterior region, \ie $\delta S_\mt{BH}\ge S$, and hence we have
arrived at the bound (\ref{bek}).

A drawback of the expression (\ref{bek}) is that the entropy (and the energy)
of a finite region are not well defined quantities. In order to eliminate the
ambiguities in the definition of the entropy, it was argued in \cite{beke2}
that the relevant quantity for Bekenstein's thought experiment was the
difference of entropies between the state in the relevant region $V$ and the
vacuum entropy in the same region $\Delta S=S(\rho_V)-S(\rho_V^0)$. In
Bekenstein's thought experiment, $V$ is the near horizon region just outside of
the black hole. In fact then, there is a large entanglement entropy, which can
be seen as the entropy in the thermal atmosphere around the black hole, both
for the object localized to the region outside the event horizon and for the
vacuum state localized in the same region. These are the initial and final
states of the process and so only the change in entropy, \ie the difference
between the two entanglement entropies, enters into the inequality. That is, we
should interpret $S$ appearing on the left-hand side of eq.~\reef{bek} as
$\Delta S$, the same difference which appears on the right-hand side of
eq.~\reef{123}.

Further, the quantity $2\pi R\,E$ appearing on the right-hand side of
eq.~(\ref{bek}) suffers from similar ambiguities. However, this product can
also be given a precise meaning as $\Delta \langle H\rangle$, the difference in
expectation values of the \modu Hamiltonian (for $\rho_V^0$) between the two
states \cite{beke1}. To make this connection precise, we first note that in
Bekenstein's thought experiment, the relevant physics for the near horizon
region of a large black hole is very nearly the same as that for Rindler space.
Hence, recall the \modu Hamiltonian in Rindler space is given by
eq.~(\ref{siete}). Hence evaluating $\Delta \langle H\rangle$ between the state
with Bekenstein's probe near the horizon and the vacuum state, we find
 \be
\Delta \langle H\rangle =2\pi \int_{x>0} d^{d-1}x \ x \ \langle
T_{00}(x)\rangle_{\rho_V} \simeq 2 \pi R\,E  \,.
 \ee
Hence $\Delta \langle H\rangle$ reproduces the expression appearing on the
right-hand side of eq.~(\ref{bek}) in Bekenstein's thought experiment and the
inequality \reef{bek} found there is nothing but the inequality \reef{123}
expressing the positivity of the relative entropy \reef{123}. Of course,
$\Delta \langle H\rangle$ also provides an unambiguous definition for the
product of energy and size when applying the Bekenstein bound to more general
systems and more general regions.

This discussion shows a well-defined version of the Bekenstein bound in QFT is
given by the positivity of the relative entropy $\Delta S\le \Delta \langle
H\rangle$ between an arbitrary state and the vacuum state, both reduced to some
finite region $V$ \cite{beke1}. This relative entropy bound holds
automatically, implying, despite the use of black holes in Bekenstein's thought
experiment, that the physics behind the Bekenstein bound is simply quantum
mechanics and special relativity. It also generalizes the Bekenstein bound to
arbitrary regions, since the original derivation by Bekenstein is limited to
Rindler space.

\subsection*{Relative entropy kills the species problem}

Interestingly, the version of the Bekenstein bound arising from relative
entropy, \ie eq.~\reef{123} does not suffer from the species problem
\cite{beke2}. That is, considering theories with a large number of species or
different quantum fields will not lead to violations of eq.~\reef{123}. This is
because as the number of degrees of freedom is increased, the entropy of a
localized excitation can be made bigger for the same energy, but the entropy
already present in the vacuum entanglement also gets larger. Since $\Delta S$
is bounded by $\Delta \langle H\rangle$, the difference in the entropies must
converge to a fixed value as the number of species becomes arbitrarily large.
In terms of the relative entropy, adding more species makes the
distinguishability between the localized object and the localized vacuum
poorer, reducing the relative entropy. However, the distinguishability is
always positive, only becoming zero for the identical states. That is, an
increased number of species may mean that we will be closer to saturating the
bound but it can never produce violations of the inequality \reef{123}. The
role of Hawking radiation and black hole thermal atmosphere in preserving the bound in Bekenstein's thought
experiment is then information theoretical and not mechanical, in the sense
that radiation pressure on the infalling object does not play a decisive role,
as is sometimes considered, \eg \cite{bent}.

To see how the species problem is solved in more detail, we start by describing
the way it was originally posed, \ie let us look at a canonical case with many
species. In particular, let us consider a theory consisting $\N$ decoupled
copies of some QFT. For a moment, let us set aside the idea of bounded regions
and consider global states. Let $\hat{\rho}_0=|0\rangle \langle 0|$ be the
global vacuum for a single species, and $\hat{\rho}_1=|\psi \rangle \langle
\psi|$ is any other orthogonal pure state (\eg a one-particle state). We start
with the global vacuum $|\Omega\rangle=|0\rangle\otimes \cdots\otimes
|0\rangle$ and the corresponding density matrix
 \be
\rho_0=|\Omega\rangle\langle \Omega| =\hr_{0} \otimes
\cdots\otimes \hr_{0}\,.
 \labell{density}
 \ee
Now we replace the vacuum by the excited state $|\psi \rangle$ in the $i$'th
copy of the field theory, \ie $|\Psi_i\rangle=|0\rangle\otimes
\cdots\otimes|\psi\rangle \otimes \cdots\otimes|0\rangle$. Then the
corresponding density matrix becomes
 \be
\rho_{i}=|\Psi_i\rangle\langle \Psi_i| =\hr_{0} \otimes
\cdots\otimes \hr_{1} \otimes \cdots\otimes\hr_{0}
 \labell{den3s}
 \ee

So the states $\rho_i$ are pure and we also have they correspond to orthogonal
vectors, $\langle \Psi_i|\Psi_j \rangle=0$ if $i\neq j$. Hence, the mixed
density matrix obtained by combining these particle excitations for the
different species as
 \begin{equation}
\rho_{\textrm{mix}}=\frac{1}{{\cal N}}
\sum \rho_i=\frac{1}{{\cal N}} \sum |\Psi_i\rangle \langle\Psi_i|
\labell{upon}
 \end{equation}
is already diagonalized in the basis of the $|\Psi_i\rangle$. It has ${\cal N}$
non-zero eigenvalues with value $1/{\cal N}$.  Hence it follows that
$S(\rho_{\textrm{mix}})=\log({\cal N})$ and
 \begin{equation}
 \Delta S_{\textrm{tot}}=S(\rho_{\textrm{mix}})-S(\rho_0)=\log({\cal N})\,.
 \labell{oncex}
 \end{equation}
Here $\Delta S$ increases without bound as ${\cal N}$ grows, while the energy
in $\rho_{\textrm{mix}}$ is independent of ${\cal N}$.

Considerations of a similar nature have been used to produce contradictions
with Bekenstein bound \eg \cite{donrad}. However, note that as we are
considering global states here, it is natural to assume that $R\to\infty$ and
so our discussion leads to no contradiction with eq.~\reef{bek}. The discussion
is slightly different using the interpretation in terms of relative entropy. In
this case, one may note that the global orthogonal pure states $\hat{\rho}_0$
and $\hat{\rho_1}$ within a single copy are perfectly distinguishable and hence
their relative entropy is infinite. The same holds for the states in
eqs.~\reef{den3s} and \reef{upon}, \eg $S(\rho_{\textrm{mix}}|\rho_0)=\infty$.
What allows $\Delta S_{\textrm{tot}}$ in eq.~\reef{oncex} to increase without
bound is the fact that $\Delta \langle H\rangle$ is already divergent.
Formally, this divergence can be seen as arising in writing $|0 \rangle \langle
0|$ as $\sim e^{-H}$, we introduced an infinite coefficient for the orthogonal
projector $|\psi \rangle \langle \psi|$ in in the \modu Hamiltonian $H$. For a
more intuitive insight, let us instead consider a thermal ensemble $\rho_T\sim
e^{-{\cal H}/T}$ with $H={\cal H}/T$, where $\cal H$ is the usual Hamiltonian.
Now, the vacuum density matrix $\rho_0$ can be seen as the zero temperature
limit and hence given that $|\psi \rangle$ has a fixed finite energy, one finds
$\Delta H\to\infty$ as $T\to0$.

However, these are global states, and a finite size $R$ is necessary to
formulate the Bekenstein bound in a sensible way. For simplicity then, let us
consider the case of reduced states inside a ball $V$ of radius $R$. The
reduced state of the vacuum becomes
 \be
\rho_0=\Tr_{\bar V}[|\Omega\rangle\langle \Omega|] =\hr_{0} \otimes
\cdots\otimes \hr_{0}
 \labell{density1}
 \ee
where now $\hr_{0}= \tr_{\bar V}[|0\rangle\langle 0|]$ is the `vacuum' density
matrix in each individual copy of the field theory. Note that we are
introducing $\Tr$ to denote tracing in the full Hilbert space, \ie over all
copies of the field theory, and $\tr$ to denote a trace in a single copy of the
field theory. Now constructing the analogous density matrices for the excited
states \reef{upon} yields
 \be
\rho_{i}=\Tr_{\bar V}[|\Psi_i\rangle\langle \Psi_i|] =\hr_{0} \otimes
\cdots\otimes \hr_{1} \otimes \cdots\otimes\hr_{0}
 \labell{den3s1}
 \ee
where $\hr_{1}= \tr_{\bar V}[|\psi\rangle\langle \psi|]$. Further the
corresponding mixed state is
 \begin{equation}
 \rho_{\textrm{mix}}=\frac{1}{{\cal N}} \sum \rho_i \labell{mix}\,.
 \end{equation}
Now as the different copies are all decoupled, the \modu Hamiltonian takes the
form $H_{\textrm{tot}}=\sum H_i$ where
 \be
H_i = \mathbf{1}_1 \otimes \mathbf{1}_2 \otimes \cdots \otimes H \otimes
\cdots\otimes \mathbf{1}_\N\,.
 \labell{Hi}
 \ee
In this expression, the $H$ appearing as the $i$'th entry in the direct product
is precisely the \modu Hamiltonian for a single copy of the QFT.

Now let consider a situation analogous to the one above, where we have a pure
excitation which is as different as possible from the vacuum. For global
states, distinguishability of vacuum and particle states is infinite. However,
inside the sphere, this must be bounded. In order for the excited state to be
as different as possible from the vacuum in the sphere, we should construct a
wave packet with a very short wavelength $\lambda$ far from the spherical
boundary (well inside where the effective temperature is low). Now if we
specialize to the case where the QFT's under consideration are conformal field
theories, the \modu Hamiltonian $H$ is given by eq.~\reef{sphereH} and we can
make a precise statement. In particular, placing the wave packet at the
center of the sphere, we find
\begin{equation}
\Delta H= \pi \frac{R}{\lambda}\gg 1\,.
\end{equation}
Certainly this result can be very large and when $\frac{R}{\lambda}\gg 1$, we
approach the that situation the excited state is maximally distinguishable from
$\rho_0$. Note however, that while it can be large, $\Delta H$ will never be
divergent in the bounded region. Further, in this regime, the entropy
calculation is approximately same as described for the global states above and
we have
\begin{equation}
\Delta \langle H\rangle-\Delta S=\pi \frac{R}{\lambda}-\log ({\cal N})\,.\label{asa}
\end{equation}
As ${\cal N}$ increases the relative entropy decreases (the bound becomes
tighter) as expected, since relative entropy always decreases under mixing
\cite{wehrl}
\begin{equation}
S(\sum p_i \rho^{(1)}_i|\sum p_i \rho^{(2)}_i)\le \sum p_i S( \rho^{(1)}_i|
 \rho^{(2)}_i)\,,
\end{equation}
for $p_i>0$ and $\sum p_i=1$. However, since $\Delta \langle H\rangle $ is
independent of ${\cal N}$ and relative entropy is always positive, the
$\log({\cal N})$ behavior of $\Delta S$ can not subsist for a very large number
of species ${\cal N}\gtrsim e^{R/\lambda}$. Finally eq.~\reef{123} must be
saturated with $\Delta S=\Delta\langle H\rangle$. Clearly there must be a
change in the behavior $\Delta S$ away from the simple logarithmic growth found
in eq.~\reef{oncex} in the regime where ${\cal N}\gtrsim e^{R/\lambda}$.
Intuitively, the probability of finding an excited wave packet from the $i$'th
copy of the CFT in the vacuum density matrix (which has an effective
temperature of roughly $1/R$ at the wave packet location) is $e^{-R/\lambda}/Z$
{\sl independently} of ${\cal N}$. For the excited state in
$\rho_{\textrm{mix}}$, this probability becomes $\frac1{\cal
N}+\frac{e^{-R/\lambda}}{Z}$. Hence when ${\cal N}\gtrsim e^{R/\lambda}$, the
vacuum and the mixed state are no longer very different and we are actually in
a regime where $\Delta S\simeq \Delta \langle H\rangle$.

Hence, we see the importance both of expressing the original product $2 \pi R\,
E$ on the right-hand side of the bound \reef{bek} as the change in the modular
`energy' $\Delta\langle H\rangle$, and of considering the entropy difference
$\Delta S$, rather than simply the entropy $S$. This last step ensures that
$\Delta S$ saturates the bound in the case of large number of species. When the
number of species is sufficiently large, the particle excitation whose
probability is distributed amongst the various copies in the mixed state is
hidden behind the cloud of excitations produced simply localizing the vacuum to
a finite region. Hence $\rho_{\textrm{mix}}$ and $\rho_0$ are no longer easily
distinguished.

In general, the transition from the form in eq.~(\ref{asa}) to zero for large
enough ${\cal N}$ will be some complicated function.  However, let us determine
the first nontrivial corrections for the case of small deviations from vacuum
state, \ie the opposite regime to that just analyzed above.\footnote{Note that
the following analysis would apply for any finite region and for a tensor
product of $\cal N$ copies of any QFT.} Let us begin by considering the pure
states \reef{den3s1}. Within any individual copy of the QFT, if $\hr_{1}$ is a
small perturbation of the vacuum density matrix $\hr_{0}$, then we will find as
usual
 \be
\Delta\langle H\rangle = \Delta S\quad \,.
 \labell{express1}
 \ee
Of course, for the copies containing no excitations, we find simply
$\Delta\langle H\rangle = \Delta S=0$. Hence for these pure states, we find
$\Delta\langle H_{\textrm{tot}}\rangle|_{\rho_i}=\Delta
S_{\textrm{tot}}|_{\rho_i}$, as expected.

Now for the mixed state (\ref{mix}), we find
 \bea
\Delta\langle H_{\tot}\rangle|_{\rho_{\textrm{mix}}}&=&\frac{1}{\N}\,\sum
\(\Tr[\rho_i\,H_{\tot}] -\Tr[\rho_0\,H_{\tot}]\)
 \nonumber\\
&=&\frac{1}{\N}\,\sum \(\tr[\hr_1\,H] -\tr[\hr_0\,H]\)_i
 \nonumber\\
&=&\frac{1}{\N}\,\sum \Delta\langle H\rangle_i= \Delta\langle H\rangle
 \labell{newener}
 \eea
where the subscript $i$ in the second and third sums indicates that the
corresponding expression is evaluated only in the $i$'th copy of the QFT.  The
final $\Delta\langle H\rangle$ can be evaluated in any single copy of the field
theory and so the shift in the expectation value of the \modu Hamiltonian is
unchanged that would be found for any of the pure states $\rho_i$. Similarly,
following our standard reasoning, one also finds $\Delta
S_{\tot}|_{\rho_{\textrm{mix}}}=\Delta\langle H\rangle$, as usual for small
deviations from the vacuum. That is, the new mixed state saturates the
inequality \reef{123} with precisely the same values as the individual pure
states $\rho_i$, {\it to first order}. That is, these first order calculations
do not distinguish the pure and mixed states.

 However,
the mixed state should have more entropy than the pure states and so we must go
to higher orders, we should see this difference. As in the holographic
calculations in section \ref{general}, going to higher orders means evaluating
the change in entropy to higher orders since the linear calculations of
$\Delta\langle H_{\textrm{tot}}\rangle$ are complete. To begin let us write the
excited state within a single copy of the field theory as
 \be
\hr_1 = \hr_0 + \delta\hr = \hr_0\ (\mathbf{1} + \hr_0^{-1}\delta\hr)\,.
 \labell{changehr}
 \ee
Further note that since $\tr[\hr_1]=1=\tr[\hr_0]$, we must have
$\tr[\delta\hr]=0$. To introduce some more notation, let us write the $i$'th
pure state as
 \be
\rho_i 
\equiv \rho_0\ \[\mathbf{1} + \tp_i\] \equiv e^{-H_{\tot}}\,e^{-\tH_i}
 \labell{changerho}
 \ee
where
 \be
 \tp_i\equiv\mathbf{1}_1 \otimes  \cdots \otimes
\hr_0^{-1}\delta\hr\otimes \cdots\otimes \mathbf{1}_\N
 \labell{changerho1}
 \ee
with $\hr_0^{-1}\delta\hr$ appearing in the $i$'th factor of the tensor
product. The `effective' shift in the modular Hamiltonian $\tH_i$ defined by
eq.~\reef{changerho} is related to $\tp_i$ by
 \be
\tH_i = -\log\left(\mathbf{1}+\tp_i\right) = -\tp_i + \frac12 \tp_i^2 - \frac13
\tp_i^3 +\cdots.
 \labell{changerho2}
 \ee
Note that the definition of $\tH_i$ involves the product of two separate
exponentials. So in general, it does not precisely match the shift $\delta H_i$
appearing in the conventional definition: $\rho_i \equiv \exp\[-H_{\tot}-\delta
H_i\]$ because $\delta H_i$ does not commute with $H_{\tot}$. That is,
$\tH_i=\delta H_i$ requires $[H_{\tot},\delta H_i]=0$.

Having established this notation, we would like to compare the shift in the
entanglement entropy for the perturbed pure states \reef{den3s1} with that for
the perturbed mixed state \reef{mix}. Towards that end, it is convenient to use
the Baker-Campbell-Hausdorff formula to expand the logarithm appearing in the
entanglement entropy. For example, we encounter
 \bea
\log\rho_i &=& \log\[ e^{-H_{\tot}}\,e^{-\tH_i} \] \labell{gumbo}\\
&=&   -H_{\tot}-\tH_i + \frac{1}{2}[H_{\tot},\tH_i]
-\frac{1}{12}[H_{\tot},[H_{\tot},\tH_i]] + \frac{1}{12}[\tH_i,[H_{\tot},\tH_i]]
 \nonumber\\
&&\qquad-\frac{1}{24}[\tH_i,[H_{\tot},[H_{\tot},\tH_i]]]
+\cdots\,,
 \nonumber
 \eea
where the terms denoted by the ellipsis will involve four and more commutators
of $H_{\tot}$ and $\tH_i$. Note that in the present calculation, we will only
concern ourselves with the terms with two or fewer $\tH_i$'s, however, there
are an infinite number of such contributions. However, we will only need to
understand the general form of these terms for the present comparison.

Applying the above definitions, we find for the pure states
 \bea
\Delta S_{\tot}|_{\rho_i} &=& -Tr[\rho_i\,\log\rho_i] +Tr[\rho_0\,\log\rho_0]
 \labell{kapow}\\
&=& \Tr\[\rho_0 \( \tH_i - \frac{1}{2}[H_{\tot},\tH_i]
+\frac{1}{12}[H_{\tot},[H_{\tot},\tH_i]] -
\frac{1}{12}[\tH_i,[H_{\tot},\tH_i]]+\cdots\)\]
 \nonumber\\
 &&\quad + \Tr\[\rho_0 \, \tp_i \,\(H_{\tot}+\tH_i - \frac{1}{2}[H_{\tot},\tH_i]
+\frac{1}{12}[H_{\tot},[H_{\tot},\tH_i]]-\cdots\)\] \,.
 \nonumber
 \eea
Again, there is an infinite number of terms for each order in $\tH_i$ (or
$\tp_i$) in the above expression. However, with the trace above, there is an
enormous simplification with
 \be
 \Tr\[\rho_0 [H_{\tot},Z]\]= \Tr\[\rho_0\, H_{\tot}\, Z\]-\Tr\[H_{\tot}\,\rho_0 \,Z\]=0
 \ee
for any matrix $Z$ since $H_{\tot}=-\log \rho_0$ commutes with $\rho_0$. Taking
this simplification into account, there are only two potential contributions at
linear order,
 \bea
\Delta S_{\tot}|_{\rho_i,\textrm{linear}} &=& -\Tr\[\rho_0 \, \tp_i\] +
\Tr\[\rho_0 \,
\tp_i \,H_{\tot}\] \labell{gumbo2}\\
&=&-\tr\[\delta\hr\] + \tr\[\delta\hr \,H\] =\Delta\langle H\rangle\,,
 \nonumber
 \eea
where the reduction between the first and second lines relies on the tensor
product structure of the various matrices and $\tr\[\delta\hr\]=0$. Of course,
this shift in the entropy at linear order agrees with $\Delta\langle
H_{\tot}\rangle|_{\rho_i}=\Delta\langle H\rangle$, as in our previous
discussion above. Now the quadratic contributions take the form
 \bea
\Delta S_{\tot}|_{\rho_i,\textrm{quad}} &=& \Tr\[\rho_0 \( \frac12 \tp_i^2 -
\frac{1}{12}[\tp_i,[H_{\tot},\tp_i]]+\frac{1}{24}[\tp_i,[H_{\tot},[H_{\tot},\tp_i]]]
+\cdots\)\]\nonumber \\
&&\quad - \Tr\[\rho_0 \, \tp_i \,\(\tp_i - \frac{1}{2}[H_{\tot},\tp_i]
+\frac{1}{12}[H_{\tot},[H_{\tot},\tp_i]]-\cdots\)\] \,.
 \labell{jumbo1}
 \eea
Again there is an infinite number of terms in both lines above. We will not
attempt to simplify eq.~\reef{jumbo1} for the states $\rho_i$ further. Rather we
now turn to considering the mixed state \reef{mix}.

Hence, for the mixed state \reef{mix}, we can define
 \be
\rho_{\textrm{mix}} = \rho_0\ \[\mathbf{1} + \frac{1}{{\cal N}} \sum \tp_i\]
\equiv e^{-H_{\tot}}\,e^{-\tH_{\textrm{mix}}}
 \labell{changerho4}
 \ee
where $\tp_i$ is defined in eq.~\reef{changerho1}. Further, the effective shift
in the \modu Hamiltonian $\tH_{\textrm{mix}}$ defined above can be written as
 \be
\tH_{\textrm{mix}} = -\log\left(\mathbf{1}+\frac{1}{{\cal N}} \sum\tp_i\right)
= -\frac{1}{{\cal N}} \sum \tp_i + \frac12 \frac{1}{{\cal N}^2}
\sum_{i,j}\tp_i\tp_j 
+\cdots.
 \labell{changerho5}
 \ee
Note that $\tH_{\textrm{mix}}\ne \frac{1}{{\cal N}} \sum \tH_i$ since the
latter sum would not contain all of the cross-terms appearing in
eq.~\reef{changerho5}.

Now it is a straightforward exercise to verify using the above expressions that
to linear order, we have: $\Delta S_{\tot}|_{\rho_\textrm{mix},
\textrm{linear}} =\Delta\langle H\rangle=\langle\Delta
H_{\tot}\rangle|_{\rho_\textrm{mix}}$. Turning then to the quadratic
contributions, we have
 \bea
\Delta S_{\tot}|_{\rho_\textrm{mix},\textrm{quad}} &=& \frac{1}{{\cal N}^2}
\sum_{i,j}\Tr\[\rho_0 \( \frac12 \tp_i\tp_j -
\frac{1}{12}[\tp_i,[H_{\tot},\tp_j]]+\frac{1}{24}[\tp_i,[H_{\tot},[H_{\tot},\tp_j]]]
+\cdots\)\]\nonumber \\
&&\quad - \frac{1}{{\cal N}^2} \sum_{i,j}\Tr\[\rho_0 \, \tp_i \,\(\tp_j -
\frac{1}{2}[H_{\tot},\tp_j] +\frac{1}{12}[H_{\tot},[H_{\tot},\tp_j]]-\cdots\)\]
 \labell{jumbo2}\\
&=& \frac{1}{{\cal N}^2} \sum_{i,j}\Tr\[\rho_0 \( \frac12 \tp_i\tp_j -
\frac{1}{12}[\tp_i,[H_j,\tp_j]]+\frac{1}{24}[\tp_i,[H_j,[H_j,\tp_j]]]
+\cdots\)\]\nonumber \\
&&\quad - \frac{1}{{\cal N}^2} \sum_{i,j}\Tr\[\rho_0 \, \tp_i \,\(\tp_j -
\frac{1}{2}[H_j,\tp_j] +\frac{1}{12}[H_j,[H_j,\tp_j]]-\cdots\)\] \,.
 \nonumber
 \eea
In the second equality, we have emphasized that because of the tensor product
structure of $\tp_i$ given in eq.~\reef{changerho1}, only the corresponding
terms of $H_{\tot}=\sum H_i$ contribute in the commutators. Further combining
this structure with $\tr[\hr]=0$, we have that all of the terms with $i\ne j$
above will vanish. Hence all of the double sums can be reduced as follows, \eg
 \bea
\frac{1}{\N^2}\sum_{i,j}\Tr\[\rho_0\tp_i  [H_j,\cdots [H_j, \tp_j]]\] &=&
\frac{1}{\N^2}\sum_i\Tr\[\rho_0 \tp_i [H_i,\cdots [H_i,\tp_i]]\]
 \nonumber\\
&=&\frac{1}{\N}\Tr\[\rho_0 \tp_1 [H_1,\cdots [H_1,\tp_1]]\] \,,
 \labell{zoom9}
 \eea
where we have eliminated the sum in the last expression and chosen $i=1$ as a
representative value, by using the fact that all of the terms in the previous
diagonal sum are identical. Hence the quadratic shift in the entropy simplifies
to
 \bea
\Delta S_{\tot}|_{\rho_\textrm{mix},\textrm{quad}}&=& \frac{1}{{\cal N}}
\Tr\[\rho_0 \( \frac12 \tp_1^2 -
\frac{1}{12}[\tp_1,[H_{\tot},\tp_1]]+\frac{1}{24}[\tp_1,[H_{\tot},[H_{\tot},\tp_1]]]
+\cdots\)\]\nonumber \\
&&\quad - \frac{1}{{\cal N}} \Tr\[\rho_0 \, \tp_1 \,\(\tp_1 -
\frac{1}{2}[H_{\tot},\tp_1] +\frac{1}{12}[H_{\tot},[H_{\tot},\tp_1]]-\cdots\)\]
\,.
 \labell{jumbo3}
 \eea
Here again, we have an infinite number of contributions above but comparing
this result with eq.~\reef{jumbo1}, it is clear that we have $\Delta
S_{\tot}|_{\rho_\textrm{mix},\textrm{quad}}= \frac1{\cal N} \Delta
S_{\tot}|_{\rho_i,\textrm{quad}}$. That is, at quadratic order, we have
 \be
\(\Delta\langle H_{\textrm{tot}}\rangle-\Delta
S_{\textrm{tot}}\)|_{\rho_\textrm{mix}}=\frac1{\cal N}\(\Delta\langle
H_{\textrm{tot}}\rangle-\Delta S_{\textrm{tot}}\)|_{\rho_i} + O(\tp_i^3)\,.
 \labell{treetop}
 \ee

Note that the above analysis did not reveal much about the structure of the
quadratic contributions and so we did not actually establish that the shifts in
the entropy in eqs.~\reef{jumbo1} and \reef{jumbo3} are negative. However, the
latter is easily shown by introducing the standard representation of the
logarithm in terms of the resolvent, as follows
\begin{equation}
\log (\rho+\delta \rho)=-\int_0^\infty
d\beta\, \left[\frac{1}{\rho+\delta \rho +\beta }-\frac{1}{\beta+1}\right]\,.
 \labell{resolve}
\end{equation}
The advantage of this representation is that even when $\rho$ and $\delta \rho$
do not commute, it is straightforward to expand the above expression for small
perturbations with
\begin{equation}
\frac{1}{\rho+\delta \rho +\beta}=\frac{1}{\rho+\beta}-\frac{1}{\rho +\beta}
\delta \rho\frac{1}{\rho+\beta}+\frac{1}{\rho +\beta}\delta \rho\frac{1}{\rho+\beta}
\delta \rho\frac{1}{\rho+\beta}+\cdots\,.
 \labell{expandlog}
\end{equation}

Now for any of the pure global states where the excitations appear in one copy
of the QFT, it is straightforward to show
 \bea
\Delta S_{\tot}|_{\rho_i} &=& -\Tr[\rho_i\,\log\rho_i] +\Tr[\rho_0\,\log\rho_0]
 \labell{kapow1}\\
 &=& -\tr[\hr_i\,\log\hr_i] +\Tr[\hr_0\,\log\hr_0]\nonumber\\
&=& -\tr[(\hr_0 + \delta\hr)\,\log(\hr_0 + \delta\hr)] +\Tr[\hr_0\,\log\hr_0]
\,.
 \nonumber
 \eea
That is, as before, the simple tensor product structure of $\rho_i$ and
$\rho_0$ allows us to reduce the calculation of $\Delta S_{\tot}$ to the single
copy of the QFT carrying the excitation. Now we can apply eqs.~\reef{resolve}
and \reef{expandlog} to this expression. Examining the terms linear in
$\delta\hr$, one again finds $\Delta S_{\tot} =\Delta\langle
H_{\textrm{tot}}\rangle$. Hence to leading order, we recover the equality
already found twice above. Now also including the second order terms, we find
 \begin{equation}
\(\Delta \langle H_{\textrm{tot}} \rangle-\Delta S_{\textrm{tot}}\)|_{\rho_i}
=\int_0^\infty d\beta\,\, \beta\,\textrm{tr}\left(\frac{1}{\hr_0 +\beta}
\delta \hr\frac{1}{\hr_0+\beta}\delta \hr\frac{1}{\hr_0+\beta}\right)+\cdots \,.
 \labell{kapow1b}
 \end{equation}
Note the second order term above is explicitly positive since the matrix
$(\hr_0+\beta)^{-1}$ in the center of the integrand is positive definite.
Further, this expression now captures all of the second order terms and so as
required the relative entropy is positive. Of course, given the result in
eq.~\reef{treetop}, the same positivity applies for the mixed state.

As a final comment, let us note that Bekenstein's thought experiment involves a
dynamical process and the exchange of entropy and energy between two systems.
Interpreting the Bekenstein bound in terms of relative entropy, the same
reasoning can also be applied in flat space and for any region, in particular
without referring to black holes. The flat space experiment would involve an
excitation with a modular energy difference $\Delta\langle H \rangle$ with
respect to the vacuum in a region $V$. Under some evolution this modular energy
(the Rindler energy in Bekenstein's experiment) is assumed to be at the same
time conserved but passed to a thermal reservoir \ie being converted into
`heat' in the thermodynamical language (represented by the black hole in
Bekenstein's thought experiment).\footnote{Note that in Bekenstein's
experiment, the initial Rindler energy is conserved along ordinary time
evolution in the form of ordinary energy because it is proportional to the
energy as meassured by asymptotic observers.} This gives $\Delta
S_{\textrm{res}}=\Delta\langle H \rangle$ because for the reservoir, with a
large number of degree of freedom, we are always in the small deviation
scenario (note the temperature here is $T=1$). The increase of the entropy
under this evolution requires $\Delta S_{\textrm{res}}- \Delta S=\Delta\langle
H \rangle-\Delta S\ge 0$. In fact, as shown in section \ref{secondlaw},
positivity of relative entropy can always be interpreted in this way as a
consequence of a second law for specific time evolutions which are CPTP but
nonunitary in the region. A simple example for the present case is given by an
evolution which adds identical and independent field species and mixes the
state in such larger Hilbert space, as described above in this section. This
process may represent for our purposes, the evolution of the initial system
which is finally absorbed by the reservoir. Implicitly, the above discussion
shows that this `evolution' preserves the value of $\Delta\langle H \rangle$.
Also in the limit of a large number of species, we should get $\Delta
S_{\textrm{res}}=\Delta\langle H \rangle$. Here $\Delta S_{\textrm{res}}$ is
the variation of the entropy of the bath due to presence of the probe, which is
now distributed among a large number of field species. Hence, the relative
entropy bound can also be considered a cons
equence of a second law under a CPTP
evolution, in analogy with the derivation of the Bekenstein bound using the
generalized second law.


\end{document}